\documentclass[fleqn,usenatbib]{mnras}

\usepackage{newtxtext,newtxmath}

\usepackage[T1]{fontenc}

\DeclareRobustCommand{\VAN}[3]{#2}
\let\VANthebibliography\thebibliography
\def\thebibliography{\DeclareRobustCommand{\VAN}[3]{##3}\VANthebibliography}

\usepackage{verbatim}

\usepackage{graphicx}	
\usepackage{amsmath}	

\usepackage{arydshln}
\usepackage{subcaption}
\usepackage{orcidlink}

\title[Halo environments of MIGHTEE radio-AGN]{MIGHTEE: The dark matter haloes, duty cycle and mechanical feedback from radio-AGN up to $z \sim 2.5$}

\author[Hamlett et al.]{
Joel Hamlett$^{\orcidlink{0009-0002-9611-1970}}$,$^{1}$\thanks{E-mail: joel.hamlett@physics.ox.ac.uk}
Catherine L. Hale$^{\orcidlink{0000-0001-6279-4772}}$,$^{1,2}$
Matt J. Jarvis$^{\orcidlink{0000-0001-7039-9078}}$,$^{1,3}$
David Alonso$^{\orcidlink{0000-0002-4598-9719}}$,$^{1}$
Natalia Stylianou$^{\orcidlink{0000-0002-6029-005X}}$$^{1}$ and
\newauthor
Imogen H. Whittam$^{\orcidlink{0000-0003-2265-5983}}$$^{1,3}$
\\
\\
$^{1}$Astrophysics, University of Oxford, Denys Wilkinson Building, Keble Road, Oxford, OX1 3RH, UK\\
$^{2}$School of Physics and Astronomy, Institute for Astronomy, University of Edinburgh, Royal Observatory, Blackford Hill, EH9 3HJ Edinburgh, UK\\
$^{3}$Department of Physics and Astronomy, University of the Western Cape, Robert Sobukwe Road, 7535 Bellville, Cape Town, South Africa\\
}

\date{Accepted XXX. Received YYY; in original form ZZZ}

\pubyear{\the\year{}}

\begin{document}
\label{firstpage}
\pagerange{\pageref{firstpage}--\pageref{lastpage}}
\maketitle

\begin{abstract}
Radio-AGN are observed to be more strongly clustered than non-active galaxies, though it is unclear whether this is simply due to their preference for massive host galaxies, or if they reside in distinct environments beyond this mass dependence. 
Using data from three fields covered by the MIGHTEE survey,
we measure the angular two-point cross-correlation functions with a large, stellar mass-limited population of near-infrared selected galaxies, overcoming limitations of previous single-deep-field studies. By fitting halo occupation distribution models, we infer the galaxy bias parameters, $b$, for radio-AGN in three redshift ranges with median redshifts of $z_{\rm{med}}=0.76^{+0.17}_{-0.28}$, $1.25^{+0.14}_{-0.17}$ and $1.75^{+0.44}_{-0.18}$, finding $b=1.94^{+0.07}_{-0.07}$, $2.50^{+0.11}_{-0.18}$ and $3.38^{+0.27}_{-0.38}$, respectively. The typical dark matter halo mass decreases with increasing redshift: $\log_{10}(\langle M_{\rm{h}} \rangle/{\rm{M_\odot}})=13.44^{+0.08}_{-0.08}$, $13.17^{+0.07}_{-0.06}$ and $13.03^{+0.09}_{-0.10}$, which we attribute to the increased abundance of cold gas required to fuel AGN activity at earlier times. The AGN duty cycle is determined to be $\sim5-9\%$, and we estimate that the total energy radiated by radio-jets over $0<z<2.5$ is $\sim10^{53}\ {\rm{J}}$ per halo, which is sufficient to account for the observed excess heating of gas beyond that of gravitational collapse. Comparing the typical dark matter halo masses to the values obtained for the control sample, we find that the halo masses of radio-AGN are $1.54^{+0.47}_{-0.33}$, $1.11^{+0.25}_{-0.20}$ and $1.82^{+1.04}_{-0.57}$ times greater than those of the stellar mass- and redshift-matched galaxies. This difference could arise because AGN feedback suppresses stellar mass growth while leaving halo mass unchanged, or because radio-AGN preferentially reside in earlier forming haloes which are more strongly clustered.
\end{abstract}

\begin{keywords}
galaxies: active -- radio continuum: galaxies -- galaxies: haloes -- large-scale structure of Universe -- cosmology: observations -- dark matter
\end{keywords}



\section{Introduction}\label{sec:introduction}

It is well-established that the environment in which a galaxy resides can significantly affect its evolution, with active galactic nuclei (AGN) playing an important role in this process \citep{bower2006, croton2006, hopkins2006, harrison2018}. AGN are powerful astrophysical phenomena, powered by the accretion of matter onto the supermassive black hole (SMBH) at the centre of a galaxy \citep{padovani2017}. Radio-AGN represent the most massive and energetic AGN in the Universe \citep{best2005} with black hole masses ${\gtrsim}10^8\ {\rm{M_\odot}}$ \citep{jarvis2002, metcalf2006}. These objects emit radio waves through synchrotron radiation produced by their jets of relativistic charged particles \citep{antonucci1993}. The energetic input of AGN through their jets has a significant impact on both their host galaxies \citep[e.g.,][]{rawlings2004, springel2005, Cicone2014, Harrison2014} and their large-scale environment \citep[e.g.,][]{gitti2012, gilli2019, eckert2021}. By injecting energy into the surrounding medium, AGN heat the gas in their environment, preventing it from cooling and collapsing to form stars (see \citealt{mcnamara2012} for a review). First introduced to resolve discrepancies between observation and simulations \citep[e.g.,][]{bower2006, croton2006, mccarthy2010}, this process is known as AGN feedback, and is now recognized as a key mechanism in regulating galaxy evolution \citep{fabian2012, hardcastle2020, gaspari2020}. Investigating the environments of AGN helps to determine whether certain environmental conditions trigger or support AGN activity, and provides an insight into how AGN feedback influences their surroundings. 

Because radio signals are unattenuated by dust, it is possible to conduct studies on radio-AGN out to high redshifts \citep[$z\sim7$; e.g.,][]{Endsley2022}. However, such investigations are often limited by the depths of optical follow-up observations which are necessary to obtain redshifts for these sources \citep[e.g.,][]{Jarvis2009,McAlpine2013,Smolcic2017}. Earlier works on the environments of radio sources cross-matched with optical galaxies from wide-area surveys, allowing the clustering of AGN to be measured out to $z\sim0.5$ \citep{magliocchetti2004, brand2005, wake2008, donoso2010, fine2011}. More recent efforts have extended clustering analyses to higher redshifts using radio surveys that focus on smaller fields with deeper optical survey coverage \citep{hickox2009, lindsay2014a, magliocchetti2017, hale2018, chakraborty2020, mazumder2022}.

Radio-AGN are predominantly hosted by the most massive galaxies, both locally \cite[e.g.,][]{Eales1997,Jarvis2001_Kz,best2005, mauch2007, sabater2019, capetti2022} and at higher redshifts \cite[e.g.,][]{smolcic2009, gurkan2014, magliocchetti2016, uchiyama2022, best2023}. The likelihood of a galaxy hosting an AGN is strongly correlated with its stellar mass \citep{best2005, mauch2007, sabater2019, magliocchetti2020, kondapally2025}, with nearly all of the most massive local elliptical galaxies hosting an AGN \citep{brown2011, capetti2022, grossova2022}. However, despite this strong mass dependence, the wide range of radio luminosities observed among AGN appear to be largely uncorrelated with the stellar masses of their host galaxies \citep{mauch2007}.

In contrast to optical-AGN \cite[e.g.,][]{kauffmann2008, donoso2010, sabater2013, retana2017}, most radio-AGN are located in overdense and cluster-like structures. 
This has been quantified in numerous clustering studies, which consistently show that radio-AGN inhabit dark matter haloes with characteristic masses of $M_{\rm{h}}\sim10^{13}-10^{14}\ {\rm M_\odot}$ across a wide range of radio luminosities, frequencies and redshifts \citep[e.g.,][ see the review by \citealt{magliocchetti2022}]{allison2015, magliocchetti2017, retana2017, hale2018, petter2024}. Radio-AGN have also been found to be more concentrated towards cluster centres, with the brightest cluster galaxy more likely to host a radio-AGN than other galaxies in the same over-density \citep[e.g.,][]{best2007, smolcic2011, hatch2014, mo2018, croston2019}.

Given the strong connection between radio-AGN activity and host galaxy mass, an open question is whether their enhanced clustering is simply due to their preference for massive hosts, or if they occupy distinct environments beyond this mass dependence (i.e., are galaxies that host AGN simply a random subset of massive galaxies?).
Differences in the clustering between AGN and non-active galaxies of the same stellar mass may arise from galaxy-level effects, such as feedback reducing star formation or AGN preferentially residing in galaxies at the centres of haloes, as well as from assembly bias \citep{gao2005, wechsler2006}, if AGN and stellar mass-matched galaxies tend to reside in haloes with different formation histories at fixed halo mass.
Previous works have investigated this by comparing samples of cross-matched AGN to control samples with similar properties (see Section 4.3.2 of the review by \citealt{magliocchetti2022}). A number of studies find that radio-AGN inhabit denser environments and more massive dark matter haloes than stellar mass-matched galaxies, even when controlling for quantities such as velocity dispersion (a proxy for black hole mass) or colour \citep{kauffmann2008, mandelbaum2009}. This environmental enhancement appears to depend on both host-galaxy properties and radio luminosity: passive AGN (hosted by galaxies with little ongoing star formation) are preferentially found in overdense regions, whereas non-passive galaxies hosting AGN do not differ from a control sample \citep{bardelli2010}. At the highest radio luminosities ($L_{1.4{\rm{GHz}}}>10^{24.5}\ {\rm{W\ Hz^{-1}}}$), the enhanced environmental density was found to disappear \citep{malavasi2015}. More recent work finds that radio-AGN largely follow the environments of mass-matched galaxies, except for a higher prevalence of AGN in galaxy clusters \citep{kolwa2019}.

The connection between the radio luminosity of AGN and their environments remains unclear. Previous works have produced conflicting results on whether the clustering of radio-AGN depends on their luminosity, with some reporting that high-luminosity AGN are found in higher-density regions than their lower luminosity counterparts \citep[e.g.,][]{bardelli2010, lindsay2014a, hale2018, croston2019, mo2020}, while others have found no significant dependence of environment on radio luminosity \cite[e.g.,][]{magliocchetti2004, kauffmann2008, wylezalek2013, castignani2014, kolwa2019}. Conversely, a smaller number of studies suggest the opposite trend, with low-luminosity AGN preferentially inhabiting denser regions \citep[e.g.,][]{donoso2010, malavasi2015, uchiyama2022}.

To investigate the clustering of sources, we measure the two-point correlation function (TPCF), which is a widely used measure of the statistical clustering of galaxies. It can be expressed as the real-space correlation function, $\xi(r)$, which measures clustering as a function of three-dimensional separation \citep[e.g.,][]{peebles1980, peebles1983, magliocchetti2004, coil2013}, or the angular correlation function, $\omega(\theta)$, which quantifies clustering in two-dimensional projections on the sky when accurate line-of-sight distances are not available \citep[e.g.,][]{totsuji1969, cress1996, blake2002, overzier2003, wang2013, hatfield2016}.

The clustering of galaxies is closely linked to the underlying distribution of dark matter in the Universe, as galaxies form within dark matter haloes. The efficiency of galaxy formation is enhanced in dense environments, where structures of a given mass collapse earlier due to the higher surrounding density. Consequently, galaxies do not trace the underlying dark matter distribution uniformly but are instead biased tracers \citep{mo1996, peacock1999, cooray2002}. The clustering of galaxies, $\xi_{\rm{gal}}$, and the background dark matter distribution, $\xi_{\rm{DM}}$, are related by the bias parameter, $b$, through
\begin{equation} \label{eqn:bias}
    b(r,z)=\frac{\delta_{\rm{gal}}(r,z)}{\delta_{\rm{DM}}(r,z)}=\sqrt{\frac{\xi_{\rm{gal}}(r,z)}{\xi_{\rm{DM}}(r,z)}},
\end{equation}
where $\delta_{\rm{gal}}$ and $\delta_{\rm{DM}}$ are the local galaxy and dark matter overdensities, respectively \citep{kaiser1984}. In the linear regime, the bias is approximately independent of scale, $b(r,z) \approx b(z)$. The bias parameter depends on halo mass, with more massive haloes being more strongly clustered, and the epoch of halo collapse, with haloes that collapse earlier having higher bias \citep{mo1996}.

The aim of this paper is to investigate whether there is a difference between the environments of radio-selected AGN and a control sample of galaxies matched in both stellar mass and redshift by comparing their projected clustering. The radio data are from the MeerKAT International GHz Tiered Extragalactic Exploration (MIGHTEE; \citealt{jarvis2016}) survey, which targets multiple well-studied fields with rich ancillary data. Using a catalogue of radio sources that have been cross-matched with their host galaxies observed at optical and near-infrared (NIR) wavelengths, a sample of $\sim2000$ AGN covering $\sim7.5\ {\rm{deg}}^2$ and extending to $z=2.5$ is constructed by applying a redshift-dependent luminosity threshold. 
Previous radio-AGN clustering studies have either used wide-area surveys but with higher flux limits \citep[e.g.,][]{mandelbaum2009, allison2015, retana2017, petter2024}, or considered a single deep field with area $\sim2\ {\rm{deg}}^2$ \citep[e.g.,][]{magliocchetti2017, hale2018}. MIGHTEE, by spanning three widely separated fields with optical/NIR overlap at high radio sensitivity, both reduces the impact of cosmic variance on the clustering measurements and allows clustering to be measured out to high redshift.
To quantify the differences in environment, halo occupation models are fit to the measured correlation functions in order to infer the statistical properties of the dark matter haloes that the samples inhabit. 

The paper is organized as follows: In Section~\ref{sec:data} the data, the selection of AGN and the construction of the control sample are discussed. In Section~\ref{sec:methods} we describe the methods that are used to measure clustering and the chosen halo occupation model. The results are presented and discussed in Section~\ref{sec:results}, and our findings are summarized in Section~\ref{sec:conclusion}.

For all calculations in this paper, we assume $\Lambda$ cold dark matter (${\Lambda}{\rm{CDM}}$) cosmology with $\Omega_\Lambda=0.7$, $\Omega_{\rm{m}}=0.3$, $H_0=100h\ {\rm{km\ s^{-1}\ Mpc^{-1}}}$ with $h=0.7$, $\sigma_8=0.81$ and $n_s=0.96$. All magnitudes are given in the AB system \citep{oke1983}.

\section{Data}\label{sec:data}
\subsection{Radio data}\label{ssec:radio_data}

The radio sources in this work are from the Data Release 1 (DR1; \citealt{hale2025}) catalogues of the $\sim1.2-1.3\ \rm{GHz}$ MIGHTEE survey \citep{jarvis2016, heywood2022}. Both higher- ($\sim5\ {\rm{arcsec}}$) and lower-resolution ($\sim8\ {\rm{arcsec}}$) images are available, and we use the catalogues that were derived from the higher-resolution data. The sources span three fields with a wealth of available ancillary data: $4.2\ \rm{deg}^2$ of the Cosmic Evolution Survey (COSMOS) field, $14.4\ \rm{deg}^2$ of the \textit{XMM-Newton} Large Scale Structure (XMM-LSS) field, and $1.5\ \rm{deg}^2$ of the \textit{Chandra Deep Field}-South (CDFS), with central root mean square (r.m.s.) sensitivities of ${\sim}1.2{-}3.6\ \mu\rm{Jy \ beam}^{-1}$. By combining the fields, we increase the cumulative survey area and reduce the impact of cosmic variance on the clustering measurements. The wide separation of the fields further helps to mitigate field-specific systematics that could otherwise affect the results. 

The catalogues account for the confusion limited nature of the MIGHTEE images through an iterative source detection procedure, as described in detail by \cite{hale2025}. The resulting catalogues contain 20\,886, 72\,187 and 21\,152 radio sources in the COSMOS, XMM-LSS and CDFS fields, respectively.

\subsection{Optical/NIR data}\label{ssec:optical_data}

The catalogues of optical/NIR sources that are used in this study have photometry from several surveys. The NIR photometry is provided by the VISTA Deep Extragalactic Observations (VIDEO) survey \citep{jarvis2013} in XMM-LSS and CDFS, and the UltraVISTA survey \citep{mccracken2012} in COSMOS which covers the \textit{YJH$K_{\rm{s}}$} bands. Further NIR coverage in the \textit{grizy} bands for all three fields is provided by the second data release of the Hyper Suprime-Cam Subaru Strategic Program (HSC-SSP; \citealt{aihara2022}). The optical photometry includes data from the Canada-France-Hawaii Telescope Legacy Survey (CFHTLS; \citealt{cuillandre2012}), which provides \textit{ugriz} band data in the COSMOS and XMM-LSS fields, as well as from the VST Optical Imaging of the CDFS and ES1 Fields (VOICE; \citealt{vaccari2016}) survey, which covers the \textit{ugriz} bands in the CDFS field.

Sources were selected in the $K_{\rm{s}}$-band using the source finding software $\textsc{SExtractor}$\footnote{\href{https://www.astromatic.net/software/sextractor/}{https://www.astromatic.net/software/sextractor/}} \citep{bertin1996}. After applying masks to remove areas around bright stars and artifacts, the resulting catalogues contain $419\,015$, $474\,602$ and $367\,090$ sources in the COSMOS, XMM-LSS and CDFS fields, respectively, covering areas of $1.7$, $4.3$ and $3.3\ {\rm{deg}}^2$. Therefore, in the COSMOS and XMM-LSS fields, the area available for clustering analyses is limited by the multi-wavelength coverage, while in CDFS it is limited by the area of the radio observations.

A comprehensive description of the catalogues will be provided by Stylianou et al. (in preparation).

\subsection{Redshift PDFs}\label{ssec:redshift_pdfs}

Spectroscopic redshifts are available for a subset of sources, obtained by cross-matching with the merged spectroscopic catalogues of \cite{vaccari2015} using a $1\ {\rm{arcsec}}$ matching radius. For the regions overlapping between the multi-wavelength and radio data, this results in spectroscopic redshifts for $10\%$, $13\%$ and $4\%$ of sources in the COSMOS, XMM-LSS and CDFS fields, respectively. 

For the remaining sources, less accurate photometric redshifts are used. To account for the uncertainties in these estimates, the redshifts are represented as probability distribution functions ($z$-PDFs). In the CDFS field, the $z$-PDFs were derived using \textsc{LePHARE} \citep{arnouts1999, ilbert2006} which uses a $\chi^2$ minimization approach to determine the most likely redshift by fitting the observed photometry to a suite of galaxy and AGN template spectra. For the COSMOS and XMM-LSS fields, $z$-PDFs were found for each galaxy using the methods and catalogues of \cite{hatfield2022}. They combined two methods for estimating photometric redshifts: template fitting using $\textsc{LePHARE}$, and the machine learning algorithm $\textsc{GPz}$ \citep{almosallam2016, gomes2018, stylianou2022}, where a machine learning model was trained using galaxies with photometric data and spectroscopic redshifts. This produced two distinct estimates for the $z$-PDFs of each source that were combined using a Hierarchical Bayesian model \citep{dahlen2013, duncan2018, duncan2019} to create a single consensus distribution that outperforms each method individually. This is not possible in the CDFS field due to the heterogeneous nature of the optical-wavelength data and the comparative lack of spectroscopic redshift information in this field.

\subsection{Stellar mass completeness}\label{ssec:stellar_mass_completeness}

The stellar mass of each optical/NIR galaxy was estimated using the \textsc{LePHARE} code by fixing the redshift to the peak of the $z$-PDF corresponding to the best-fitting galaxy template, and adopting the \cite{BC03} stellar population synthesis models available in \textsc{LePHARE}. In the CDFS field, when an AGN template provides a better fit, we scale the stellar mass estimate to the peak AGN $z$-PDF redshift using $M_\ast\propto D_{\rm{L}}^2(z)$ where $D_{\rm{L}}(z)$ is the luminosity distance. This approximation is reasonable because the $K_{\rm{s}}$-band observations are relatively insensitive to ongoing star-formation so consistently trace the bulk of the stellar mass \citep{kodama2003}. Therefore, although the spectral energy distribution of the galaxy may evolve with redshift, the resulting variation in stellar mass estimates for fixed observed $K_{\rm{s}}$-band magnitude will be relatively small. The stellar masses of sources with spectroscopic redshifts are similarly scaled.

\begin{figure}
 \includegraphics[width=\columnwidth]{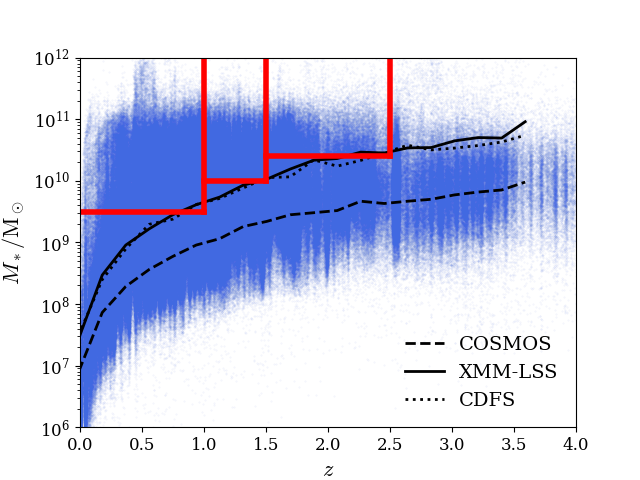}
 \caption{Stellar mass and redshift distribution of NIR-selected galaxies from all three fields (blue points). The banding effect is caused by the discrete redshift bins of \protect \textsc{LePHARE}. The black lines are the 90\% stellar mass completeness limits in each field, found by following the method of \protect \cite{pozetti2010}. The red boxes are our chosen stellar mass and redshift sub-samples.}
 \label{fig:stellar_mass_completeness}
\end{figure}

Stellar mass completeness refers to the stellar mass above which all galaxies are detectable given the flux limit of the survey. Below this limit, an increasing fraction of galaxies are too faint to be observed, leading to an incomplete sample that would bias the clustering results. To determine the completeness limit, the method outlined in \cite{pozetti2010} was adopted. This approach is illustrated in Figure~\ref{fig:stellar_mass_completeness}, which presents the stellar masses of the galaxies as a function of peak $z$-PDF (or spectroscopic) redshift. The lowest value of stellar mass, $M_\ast$, at which a galaxy could be detected is calculated using its observed $2\ {\rm{arcsec}}$ aperture $K_{\rm{s}}$-band magnitude by
\begin{equation}\label{eqn:min_stellar_mass}
    \log_{10}(M_{\rm{lim}})=\log_{10}(M_\ast)+0.4\left(K_{\rm{s}}-K_{\rm{s}}^{\rm{lim}}\right),
\end{equation}
where the completeness limits are $K_{\rm{s}}^{\rm{lim}}=24.8$ in COSMOS and $K_{\rm{s}}^{\rm{lim}}=23.5$ in XMM-LSS and CDFS\footnote{The magnitude limit in COSMOS corresponds to the $5\sigma,2''$ completeness limits from the fifth UltraVISTA data release (DR5). Although Stylianou et al. use the deeper DR6 limits, we adopt the DR5 values as this is the release that was used for cross-matching the radio and optical/NIR catalogues (see Hale et al. in preparation). The magnitude limit in XMM-LSS and CDFS corresponds to $\sim90\%$ completeness in VIDEO at that depth \citep{jarvis2013}.}. For each field, we calculate the minimum masses at which galaxies could be observed and compute the 90th percentile of these data for bins of redshift, shown by the black lines in Figure~\ref{fig:stellar_mass_completeness}. This is used to apply a minimum stellar mass threshold in each redshift bin. The red boxes show the redshift and stellar mass sub-samples that are analysed in subsequent sections.

\subsection{Cross-matching radio sources with host galaxies}\label{ssec:cross_matching}

Cross-matching radio sources with their optical/NIR counterparts associates them with their host galaxies for which redshift and stellar mass estimates have been made. Having redshifts enables the conversion of radio-fluxes to luminosities, which are used to identify AGN, and allows the sample to be divided into redshift bins so that the evolution of clustering over cosmic time can be investigated. Without cross-matching, multiple radio components that belong to the same source, for instance the two lobes of a radio jet, could be misidentified as distinct sources with small angular separations when they are in fact part of a single extended source. When calculating the clustering, this would erroneously enhance the signal at small angular scales \citep[e.g.,][]{cress1996}. Associating radio sources with their true optical/NIR counterparts allows such radio components to be correctly merged into a single catalogue entry, and allows genuinely overlapping sources to be deblended.

To cross-match radio sources with their $K_{\rm{s}}$-band host galaxies, two methods were used: (i) the automated \textit{Likelihood Ratio} matching method \citep[][see e.g., \citealt{mcalpine2012}; \citealt{kondapally2021}; \citealt{whittam2023}]{deruiter1977, sutherland1992}, and (ii) manual cross-matching by MIGHTEE team members via the MIGHTEE zoo, a \textit{Zooniverse} project \citep{lintott2008, fortson2012} internal to the MIGHTEE consortium. A detailed description of the cross-matching procedure will be provided by Hale et al. (in preparation).

Overall, $\sim94\%$, $90\%$ and $88\%$ of radio sources in the area that overlaps with the optical/NIR data were successfully matched to their host galaxy in the COSMOS, XMM-LSS and CDFS fields, respectively, resulting in $12\,987$, $28\,297$ and $18\,180$ cross-matched radio sources.

\subsection{Selecting AGN}\label{ssec:selecting_AGN}

In this work, AGN are selected by their radio luminosities. Since radio-AGN are more dominant than star-forming galaxies at high radio luminosity \citep[e.g.,][]{magliocchetti2002, mauch2007}, it is possible to apply a radio luminosity threshold such that the majority of selected sources are AGN. To convert the radio flux densities of sources to rest-frame luminosities at $\rm{1.4\ GHz}$, we assume that the radio sources exhibit a simple power-law radio spectrum $S_{\nu}\propto{\nu}^{\alpha}$, where $S_\nu$ is the flux density of the source with observed effective frequency $\nu$, and the standard spectral index of the synchrotron power-law, $\alpha=-0.7$, is assumed. The local effective frequency at the location of a source is derived from the effective frequency maps described in Section 2.3 of \cite{hale2025}. These account for variations across the survey field introduced by instrumental factors such as variations in observed frequency due to the primary beam, where decrease in sensitivity across the field, and flagging, which removes data affected by interference or instrument issues \citep{hugo2022}. Including the $k$-correction factor $(1+z)^{-(1+\alpha)}$ to account for the shift in peak flux frequency due to redshift leads to
\begin{equation}\label{eqn:luminosity}
    L_{\rm{1.4GHz}}=\frac{4{\pi}D_{\rm{L}}^2(z)}{(1+z)^{1+\alpha}}\left(\frac{\nu}{\rm{1.4\ GHz}}\right)^{-\alpha}{S_\nu},
\end{equation}
where $z$ is the redshift of the source, and $D_{\rm{L}}(z)$ is the corresponding luminosity distance.

To determine the luminosity threshold that separates AGN from star-forming galaxies, we utilise the classified radio sources from \cite{whittam2022}. Their study successfully cross-matched 5223 out of the 6102 ($86\%$) radio sources from the MIGHTEE Early Science radio continuum data release \citep{whittam2023} with their host galaxies, and classified 88\% of these cross-matched sources as AGN, star-forming galaxies, or probable star-forming galaxies within the central $0.86\ \rm{deg}^2$ of the COSMOS field. The classification was done using a combination of several methods: radio-excess, mid-infrared colours, optical morphology, characteristic X-ray emission and very long baseline interferometry measurements. We bin these sources in redshift bins containing approximately equal numbers of objects, and find the minimum $L_{\rm{1.4GHz}}$ value above which 95\% of sources are classified as AGN. To create a continuous luminosity threshold in redshift, $L_{\rm{AGN}}(z)$, we interpolate in $\log_{10}(L_{\rm{1.4GHz}})$ between the median redshifts of the sources in each bin. The 95\% threshold was chosen as a compromise between providing a high-purity AGN sample while maintaining a sufficiently large population for statistical analysis. Since in this method AGN are selected by their intrinsic properties (luminosity and redshift) the same threshold is assumed to hold for the DR1 data.

\begin{figure}
 \includegraphics[width=\columnwidth]{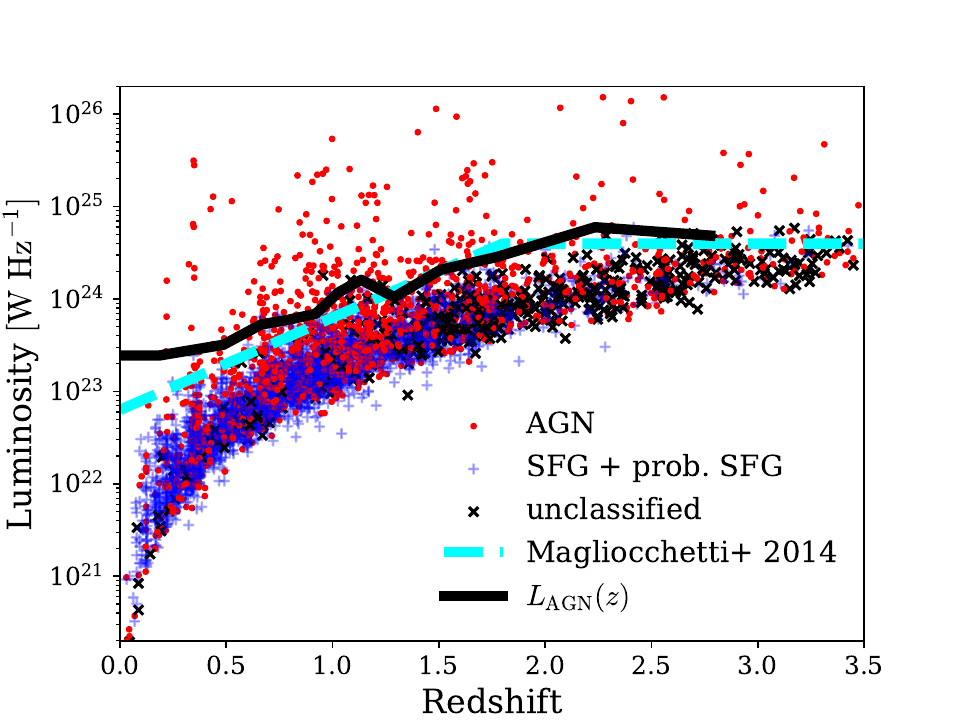}
 \caption{Radio luminosity at 1.4 GHz as a function of redshift for classified sources from \protect \cite{whittam2022}. The black points are the unclassified sources, blue points are those classified as star-forming galaxies or probable star-forming galaxies, and red points are those classified as AGN. The star-forming galaxy points are partially transparent to improve clarity. The solid black line is the radio luminosity threshold, $L_{\rm{AGN}}(z)$, above which 95\% of sources are classified as AGN. This same threshold is applied to the DR1 radio catalogue to create the AGN sample. For comparison, the AGN threshold of \protect \cite{magliocchetti2014} is denoted by the dashed cyan line.}
 \label{fig:early_science_agn_threshold}
\end{figure}

Figure~\ref{fig:early_science_agn_threshold} presents the redshift and luminosity distribution of the classified Early Science sources along with our measurements of $L_{\rm{AGN}}(z)$. For comparison, we also include the AGN luminosity threshold of \cite{magliocchetti2014}, which roughly corresponds to the break of the local radio luminosity function of star-forming galaxies. This threshold has been used in several previous clustering studies \citep[e.g.,][]{magliocchetti2017, chakraborty2020, mazumder2022}, and is similar to ours. 

Of the DR1 radio sources that were not successfully cross-matched with their optical/NIR host galaxies, only 75 have radio flux densities that are high enough to potentially exceed $L_{\rm{AGN}}(z)$ assuming they have $z<2.5$. The incompleteness in cross-matching is therefore not expected to have a significant impact on the clustering measurements.

\subsection{Redshift bins and weights} \label{ssec:weights}

We measure the clustering of galaxies for three bins of redshift: $0 < z < 1$, $1 < z < 1.5$, and $1.5 < z < 2.5$. These ranges were selected to contain a sufficiently large number of AGN while maintaining a redshift width of ${\Delta}z\leq1$ to avoid excessive cosmic evolution within a single bin. The numbers of sources in each bin are given in Table~\ref{tab:hod_results}. 

Since the photometric redshifts are represented as PDFs, the approach of \cite{arnouts2002} is followed and every source is assigned a weight for each redshift bin corresponding to the probability that the true redshift of the source lies within that redshift range. When calculating the clustering, every galaxy appears in every redshift bin, with its contribution weighted accordingly. For the optical/NIR galaxies, the weights are determined by integrating the normalized $z$-PDF of each source, $p_i(z)$, over the redshift range of the bin, $[z_{\rm{min}}, z_{\rm{max}}]$: 
\begin{equation}\label{eqn:weight_optical}
    w_i=\int_{z_{\rm{min}}}^{z_{\rm{max}}}p_i(z)\,{\rm{d}}z.
\end{equation}
For sources with spectroscopic redshifts, $w_i=1$ if $z_{\rm{spec}}$ is between the redshift bin limits and 0 otherwise.

 A radio source will only be classified as an AGN if its luminosity exceeds the redshift-dependent luminosity threshold. To incorporate this into the weighting, for each radio source 1000 random redshift samples are drawn from its $z$-PDF using an inverse cumulative distribution function sampler. At each sampled redshift, the corresponding radio luminosity is computed from the observed flux of the source (scaled to $1.4\ {\rm{GHz}}$), and the sample is retained if the AGN luminosity threshold is exceeded. The weight assigned to the source, $w_{\rm{AGN}}$, is then taken to be the fraction of samples that both fall within the redshift bin and meet the AGN luminosity criterion. For sources with a spectroscopic redshift, $w_{\rm{AGN}}=1$ if its $L_{\rm{1.4GHz}}>L_{\rm{AGN}}(z_{\rm{spec}})$ and 0 otherwise.

\begin{figure*}
 \includegraphics[width=2\columnwidth]{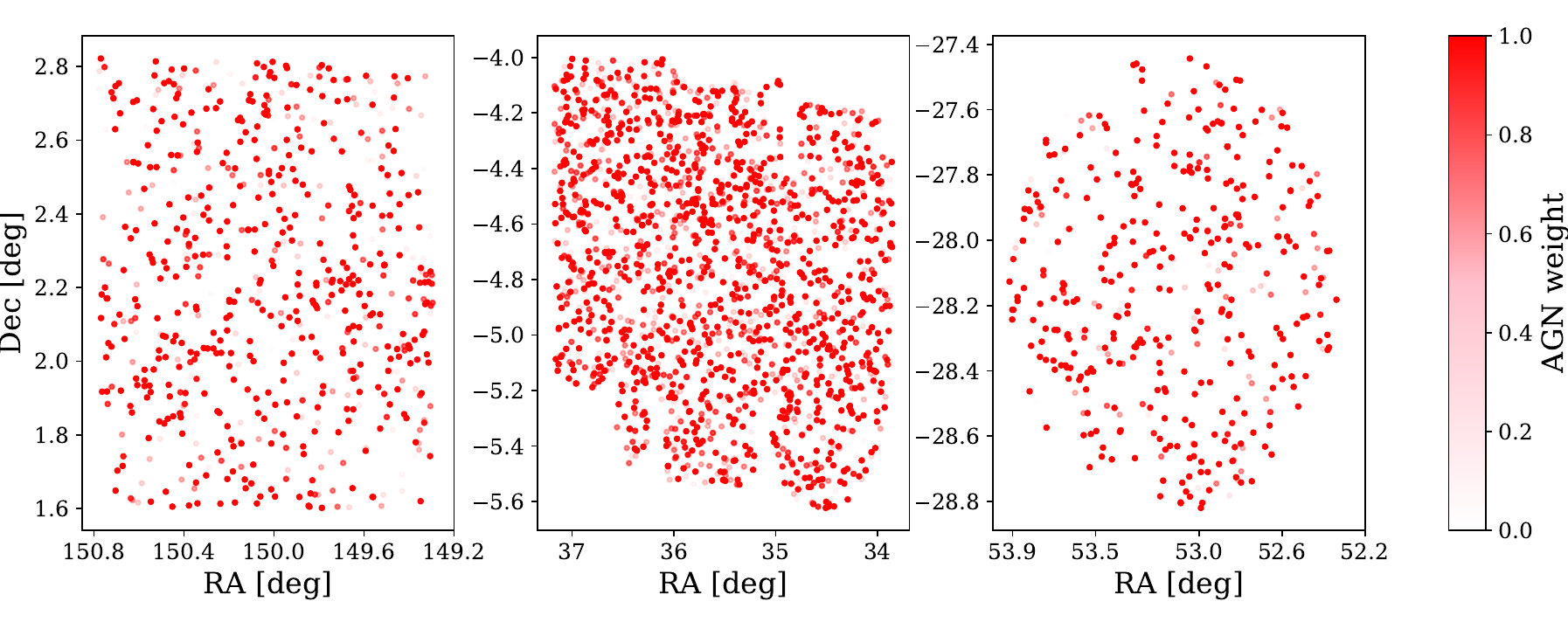}
 \caption{The positions of radio sources on the sky after applying the star mask in each field. The transparency of each point reflects the weight assigned to that source, which represents the probability that the source exceeds the AGN luminosity threshold.} The fields from left to right are COSMOS, XMM-LSS and CDFS. Note that XMM-LSS (middle) is much larger than the other two fields.
 \label{fig:radio_ra_and_dec}
\end{figure*}

In Figure~\ref{fig:radio_ra_and_dec}, we show the distribution of radio sources on the sky in each field. The transparency of each radio source corresponds to the probability that it exceeds the AGN luminosity threshold based on its assigned weight.

\subsection{The matched galaxy sample}\label{ssec:matched_sample}

Since our clustering measurements are based on angular separations (see Section~\ref{ssec:angular_correlation_function}), it is essential that the radio-AGN and comparison galaxy sample have similar redshift distributions; differences in redshift would lead to variations in the physical scales corresponding to a given angular separation. Additionally, it is well-established that galaxies with higher stellar masses reside in more massive haloes which exhibit stronger clustering signals \citep[e.g.,][]{mccracken2015, hatfield2016}. Therefore, to isolate the effects of the presence of an AGN on the clustering signal, we construct a control sample (hereafter the `matched galaxy sample') from the full galaxy population, such that the sum of the $z$-PDFs and the stellar mass distribution of the selected sources closely match those of our AGN sample.

The first step in constructing the matched galaxy sample is to measure the AGN host distribution in redshift and stellar mass. The peak $z$-PDF values and their corresponding stellar masses for radio sources with luminosities that exceed $L_{\rm{AGN}}(z_{\rm{peak}})$ are binned in a regular grid in the $z$ - $\log_{10}M_\ast$ plane. The binning is performed over the ranges $0<z<4$ and $8<\log_{10}(M_\ast / {\rm{M_\odot}})<13$, using bin widths of $\Delta z = 0.1$ and $\Delta \log_{10} M_\ast=0.05$. This produces a two-dimensional histogram that describes the joint redshift-stellar mass distribution of AGN host galaxies. The same binning procedure is applied to the full optical/NIR galaxy sample.

The matched galaxy sample is then constructed by randomly selecting optical/NIR galaxies without replacement on a bin-by-bin basis. In each bin containing at least one AGN, up to twice as many optical/NIR galaxies as AGN are selected, subject to the number of available galaxies in that bin. The resulting numbers of galaxies in the matched galaxy sample in each redshift bin are shown in Table~\ref{tab:hod_results}.

Although the construction of the matched sample was performed using the peak $z$-PDF values, we validate the selection by accounting for the redshift uncertainties using the full $z$-PDFs. To do this, we sample once from the $z$-PDF of each galaxy in the AGN and matched galaxy sample. The stellar masses are scaled to these sampled redshifts using $M_\ast \propto D_{\rm{L}}^2(z)$ where $D_{\rm{L}}(z)$ is the luminosity distance (as in Section~\ref{ssec:stellar_mass_completeness}). We then perform a Kolmogorov–Smirnov (KS) test on the resulting redshift and stellar mass distributions between $0<z<2.5$. Typically, below a $p$-value of $0.05$, the null hypothesis that both samples come from the same parent distribution is rejected. Sampling many times, we find averages of $p=0.10$ and $0.07$ for the resulting redshift and stellar mass distributions, respectively, so they are consistent with being drawn from the same underlying distributions.

\begin{figure}
 \includegraphics[width=1\columnwidth]{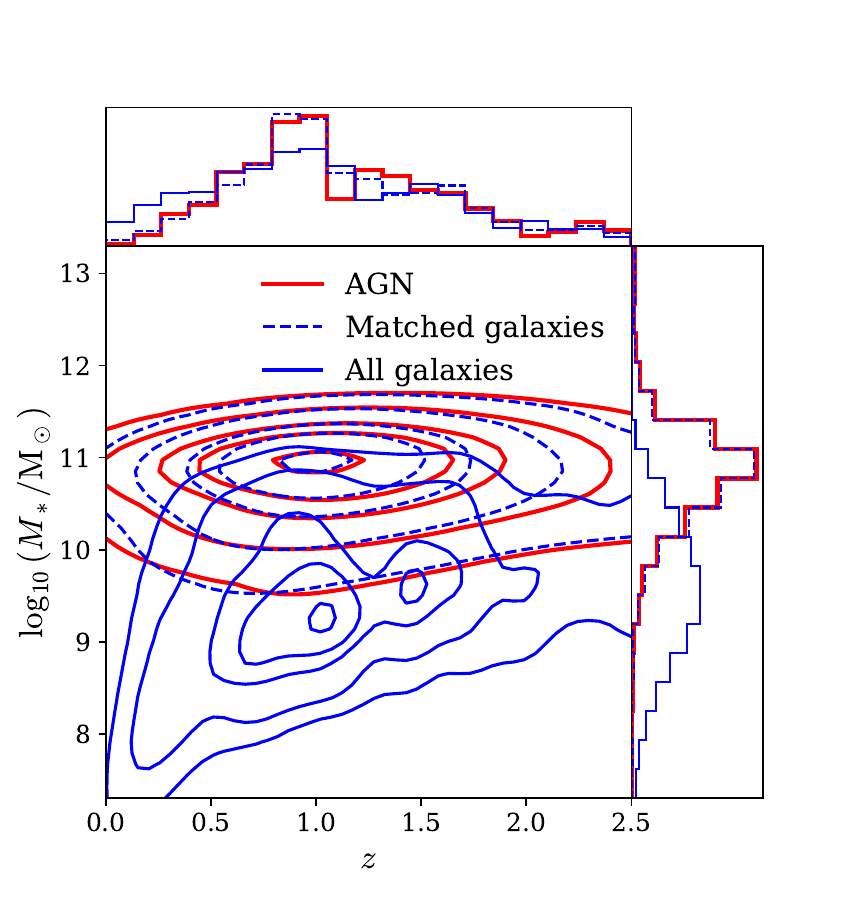}
 \caption{The redshift and stellar mass distributions of galaxies before (solid blue line) and after (dashed blue line) matching to the AGN sample (solid red line), constructed by sampling from the $z$-PDF of each galaxy. The main panel shows the contour plots of these distributions, while the top and right panels display the corresponding normalized histograms for redshift and stellar mass, respectively. The contours are the $10$, $25$, $50$, $68$ and $95\%$ levels.}
 \label{fig:z_and_stellar_mass_pdfs}
\end{figure}

The joint redshift and stellar mass distribution of the full optical/NIR sample before matching, the AGN sample and the matched galaxy sample are shown in Figure~\ref{fig:z_and_stellar_mass_pdfs}.

\section{Methods}\label{sec:methods}

\subsection{Angular two-point correlation function} \label{ssec:angular_correlation_function}

In this work, the projected angular clustering is measured rather than the full three-dimensional spatial clustering because we lack precise spectroscopic redshifts for most sources. Without accurate redshifts, estimates of $\xi(r)$ could be biased due to uncertainties in the line-of-sight distances. $\omega(\theta)$ is defined as the excess probability of finding a pair of galaxies from two populations separated by an angle $\theta$ on the sky, over the probability that would be obtained if the galaxies were distributed randomly:
\begin{equation}\label{eqn:probability}
    {\delta}P=\sigma[1+\omega(\theta)]{\delta\Omega},
\end{equation}
where ${\delta}P$ is the probability of finding two galaxies separated by angle $\theta$, $\sigma$ is the surface density of sources on the sky, and $\delta\Omega$ is the solid angle \citep{peebles1980}. If the separation between pairs of galaxies from the same population is measured, then $\omega(\theta)$ is known as the $\it{auto}$-correlation function, whereas if the two populations are distinct, it is known as the $\it{cross}$-correlation function. 
In the case of the cross-correlation between two populations with surface densities $\sigma_1$ and $\sigma_2$, the excess probability is given by
\begin{equation}\label{eqn:probability_cross}
    {\delta}P_{12}=\sigma_1\sigma_2[1+\omega(\theta)] \delta\Omega_1 \delta\Omega_2.
\end{equation}

\subsubsection{Estimating $\omega(\theta)$ from data} \label{sssec:estimating_tpcf}

The TPCF can be estimated using the data by comparing the pair counts of separations between data sources for bins of angular separation to pair counts of separations between sources in a catalogue of sources with randomly generated positions. We use the estimator of \cite{szapudi1998}:
\begin{equation} \label{eqn:tpcf}
    \omega(\theta)=\frac{\overline{D_nD_m}(\theta)-\overline{D_nR_m}(\theta)-\overline{D_mR_n}(\theta)+\overline{R_nR_m}(\theta)}{\overline{R_nR_m}(\theta)},
\end{equation}
where $\overline{D_nD_m}(\theta)$, $\overline{D_nR_m}(\theta)$ and $\overline{R_nR_m}(\theta)$ are the data-data, data-random and random-random pair counts for separations within $\theta$ to $\theta+\mathrm{d}\theta$ between the two populations, denoted by $n,m=1,2$. For $n=m$, this expression reduces to the auto-correlation estimator of \cite{landy1993}, who showed that the inclusion of the cross term, $\overline{D_nR_m}(\theta)$, in the estimator produces errors that are closer to their expected Poissonian values. When calculating the pair counts, each galaxy contributes only its assigned weight. Since $\mathcal{O}(10)$ times more random positions than data are used in order to minimize statistical noise in pair counts, all pair counts are normalized by the total number of counts over all bins of angular separation so that they sum to unity. 

Since the bin counts are discrete, it may naively be assumed that the uncertainties on $\omega(\theta)$ for the $i^{\rm{th}}$ $\theta$ bin are Poissonian, and given by 
\begin{equation}\label{eqn:poissonian_error}
    \delta\omega_i=\frac{1+\omega_i}{\sqrt{DD_i}},
\end{equation} 
where $DD_i$ are the unnormalized data-data pair counts for that bin. However, this underestimates the uncertainties as it fails to take into account both the fact that the $\theta$ bins are correlated, and the contribution of cosmic variance from large-scale matter fluctuations which dominates the uncertainty on large angular scales. Instead, the uncertainties in $\omega(\theta)$ are calculated using the jackknife sampling method \citep{quenouille1956}. Here, the survey field is split into $N_{\rm{patch}}$ sub-regions of roughly equal area. Each patch is then excluded in turn from the clustering measurements resulting in $N_{\rm{patch}}$ estimates of $\omega_i$. The elements of the covariance matrix are then estimated to be
\begin{equation}\label{eqn:cov_jk}
    {\rm{Cov}}_{ij}(\omega)=\frac{N_{\rm{patch}}-1}{N_{\rm{patch}}}\sum_{k=1}^{N_{\rm{patch}}}\left(\omega_i^k-\overline{\omega}_i\right)\left(\omega_j^k-\overline{\omega}_j\right),
\end{equation}
where $\omega_i^k$ is the value of $\omega_i$ in the $k\rm{th}$ jackknife realisation, and the expectation value, $\overline{\omega}_i$, is the mean value of $\omega_i$ over all $N_{\rm{patch}}$ jackknife estimates, $\overline{\omega}_i=\sum_{k=1}^{N_{\rm{patch}}}\omega_i^k/N_{\rm{patch}}$. The uncertainty on $\omega_i$ is the standard deviation of the individual jackknife realizations from the mean value multiplied by $\sqrt{N_{\rm{patch}}-1}$ which qualitatively accounts for the fact that the individual jackknife samples are not independent \citep{norberg2009}.

To calculate the clustering results in this work, the $\texttt{TreeCorr}$\footnote{\href{https://rmjarvis.github.io/TreeCorr}{https://rmjarvis.github.io/TreeCorr}} python package \citep{jarvis2004} is used, which takes $N_{\rm{patch}}$, and the $\rm{RA}$ and $\rm{Dec}$ of the data and random sources as inputs. We also provide the weight for each data source, as described in Section~\ref{ssec:weights}. The estimator of Equation~\ref{eqn:tpcf} is then used to calculate $\omega(\theta)$. For the jackknife uncertainties, we use $N_{\rm{patch}}=100$ patches spread over the three fields.

\subsubsection{Random Catalogues} \label{sssec:random_catalogue}

As discussed in Section~\ref{sssec:estimating_tpcf}, to estimate the TPCF of the data, a catalogue of randomly distributed sources is required. The selection of objects in these catalogues should mirror the selection of the data to ensure that the physical clustering of sources is measured, rather than observational selection effects which will affect the observed density of sources.

To generate sources for the radio random catalogues, the approach of \cite{hale2018} is followed. First, random points in $\rm{RA}$ and $\cos(\rm{Dec})$ are generated, and any points that lie outside of the r.m.s. map of the fields are removed. This is repeated until there are $\sim2\times10^7$ points in each field. Next, each point is randomly assigned a flux density and redshift from an entry in the Square Kilometre Array Design Study (SKADS) $100\ \rm{deg}^2$ Simulated Sky simulations component catalogue \citep{wilman2008, wilman2010}. This flux is scaled to 1.4 GHz using the maps of local effective frequencies (as discussed in Section~\ref{ssec:selecting_AGN}) and assuming the power-law relation $S_\nu\propto\nu^\alpha$.

The noise across each field is not uniform due to the variation in beam sensitivity and artifacts around bright sources. This introduces deviations to the observed fluxes, causing some faint sources to appear to fall below the detection threshold, whereas others may be boosted above it. This affects the completeness of the catalogue and the distribution of sources, especially near the edges of the fields where the noise is greatest. To replicate these effects in the random catalogue, a noise term is added to the assigned flux of each point. This is sampled from a Gaussian distribution with a mean of zero and standard deviation equal to the corresponding r.m.s. value at the location of the point. The total flux of each point is then calculated as the sum of the assigned flux and this noise, and a point is included in the random catalogue only if its total flux is greater than a $5\sigma$ threshold, where $\sigma$ is the r.m.s. at the location of the point. For the remaining points, their $1.4\ {\rm{GHz}}$ luminosities are calculated using their scaled fluxes and assigned redshifts, and the AGN luminosity threshold of Section~\ref{ssec:selecting_AGN} is applied.

More advanced processes exist to account for other factors that affect the selection of sources, such as extended sources having lower signal-to-noise ratios (SNR), and smearing due to time and bandwidth averaging, calibration errors and ionospheric distortion reducing peak flux densities \citep{hale2024}. However, $99\%$ of sources with $w_{\rm{AGN}}>0.1$ have a $\rm{SNR}>28\sigma$, so these effects will be minimal.

After repeating this process for each of the three fields, a star mask is applied to exclude random sources located near bright stars where photometric measurements in the optical/NIR data are not possible. Finally, we randomly select ten times as many random sources as there are data sources in each field. Using a significantly larger number of random sources than data reduces the statistical noise in the TPCF estimation.

\subsection{HOD modelling} \label{ssec:hod_modelling}

To connect the AGN clustering measurements to the statistical properties of the dark matter haloes they inhabit, we make use of a Halo Occupation Distribution \citep[HOD;][]{cooray2002, berlind2002, zehavi2005} model. The HOD framework links the typical number of galaxies in a halo with its mass, providing a statistical description of how galaxies occupy dark matter haloes. This relies on knowledge of both the halo profile and the halo mass function, which must be calibrated using numerical simulations. By assuming a model of how the bias varies with halo mass, the correlation function of galaxies can be predicted and compared to observational data. The parameters of the HOD model, such as the minimum halo mass required to host a central galaxy and how the number of satellite galaxies increases with mass, can then be constrained by fitting the correlation function predicted by the model to the observed clustering measurements \citep{zehavi2005, zheng2005}. 

In HOD modelling, each dark matter halo is assumed to be populated by either zero or one central galaxy with a number of orbiting satellite galaxies. The occupation of galaxies depends solely on halo mass, with no additional dependence on its environment or formation history. Central galaxies are present in haloes with sufficiently large mass, and satellite galaxies only occupy haloes that already host a central galaxy. The number of satellites then increases monotonically with halo mass \citep{zheng2005}. 

For a choice of concentration-mass relation (how the concentration of dark matter haloes depends on their mass), halo mass function (the number density of haloes as a function of mass), and halo bias model (how dark matter haloes are distributed relative to the overall matter distribution of the Universe), the HOD model can be used to predict the galaxy power-spectrum, and then projected to give the angular correlation function \citep{kin1999, chon2004}.

Within the HOD framework, the TPCF can be decomposed into two contributions: the \textit{one}-halo term and the \textit{two}-halo term \citep{berlind2002}. The one-halo term describes the clustering signal from pairs of galaxies that reside within the same dark matter halo. This term dominates at small angular scales, typically corresponding to physical scales of less than $\sim 1\ \rm{Mpc}$ \citep{cooray2002}, and reflects the internal structure of haloes, including the distribution of satellite galaxies around the central galaxy. In contrast, the two-halo term accounts for the clustering of galaxies that reside in separate dark matter haloes. It becomes significant at larger angular scales and is governed by the large-scale distribution of dark matter haloes, tracing the underlying cosmic web.

We note, however, that HOD modelling tends to underpredict the clustering signal in the `quasi-linear' transitional regime between the one- and two-halo terms. This is due to halo-exclusion effects (haloes cannot overlap) and the breakdown of linear perturbation theory on these scales \citep[e.g.,][]{fedeli2014, mead2015}. While there is no simple correction, this caveat should be kept in mind when interpreting our results.

\subsubsection{The HOD model} \label{sssec:hod_model}

For the HOD modelling in this work, we use the LSST Dark Energy Science Collaboration's Python Core Cosmology Library ($\texttt{CCL}$)\footnote{\href{https://github.com/LSSTDESC/CCL}{https://github.com/LSSTDESC/CCL}} python package \citep{chisari2019}. We assume the \cite{duffy2008} concentration-mass relation, the \cite{despali2016} halo mass function, and the \cite{tinker2010} halo bias model, which were calibrated using simulations. It is also assumed that the radial distribution of dark matter within haloes has a Navarro-Frenk-White profile \citep*[NFW;][]{navarro1995}, and that the distribution of satellite galaxies directly follows this profile. Halo masses are taken to be the virial mass, defined with respect to the critical density following \citet{bryan1998}.

The HOD model used by \texttt{CCL} draws from several papers, including \cite{zheng2005}, \cite{ando2017} and \cite{nicola2020}, and has the following five parameters of interest:
\begin{enumerate}
    \item $M_{\rm{min}}$, the characteristic halo mass at which half of haloes host a central galaxy;
    \item $M_0$, the minimum halo mass required for a halo to host a satellite galaxy;
    \item $M_1$, the characteristic halo mass at which a halo typically contains a single satellite galaxy;
    \item $\sigma_{{\ln}M}$ describes the smoothness of the transition in halo mass for central galaxy formation;\footnote{Note that $\sigma_{{\ln}M}$ is defined such that all logarithms of mass entering $\overline{N}_{\rm{c}}(M_{\rm{h}})$ are natural logarithms. This differs from the convention in some papers where $\log_{10}$ is used.} and
    \item $\alpha_{\rm{s}}$, the power-law index governing how the number of satellites scales with the halo mass.
\end{enumerate}

For a given halo mass, $M_{\rm{h}}$, the average number of central galaxies is given by an expression that smoothly transitions from 0 to 1:
\begin{equation} \label{eqn:num_centrals}
    \overline{N}_{\rm{c}}(M_{\rm{h}})=\frac{1}{2}\left[1+{\rm{erf}}\left(\frac{\ln{M_{\rm{h}}}-\ln{M_{\rm{min}}}}{\sigma_{{\ln}{M}}}\right)\right],
\end{equation}
where the error function
\begin{equation} \label{eqn:erfl}
    {\rm{erf}}(x)=\frac{2}{\sqrt{\pi}}\int_0^x{\rm{d}}t\,e^{-t^2}.
\end{equation}
Above the minimum mass required to host a satellite galaxy, $M_0$, the average number of satellites in a halo that contains a central galaxy increases as a power-law:
\begin{equation} \label{eqn:num_satellites}
    \overline{N}_{\rm{s}}(M_{\rm{h}})=\Theta(M_{\rm{h}}-M_0)\left(\frac{M_{\rm{h}}-M_0}{M_1}\right)^{\alpha_{\rm{s}}},
\end{equation}
where $\Theta(x)$ is the Heaviside step function. 

The average total number of galaxies in a halo of given halo mass is then,
\begin{equation} \label{eqn:num_total}
    \overline{N}_{\rm{tot}}(M_{\rm{h}})=\overline{N}_{\rm{c}}(M_{\rm{h}})\times\left[1+\overline{N}_{\rm{s}}(M_{\rm{h}})\right].
\end{equation}
Note that $\overline{N}_{\rm{tot}}$ is defined so that a halo must have a central galaxy in order to have a satellite contribution to the total number of galaxies.

Once the best-fitting HOD parameters have been determined, the following values can be derived: the mean halo mass,
\begin{equation} \label{eqn:mean_halo_mass}
    \langle M_{\rm{h}} \rangle(z)=\int{\rm{d}}M\,M\,n(M,z)\frac{\overline{N}_{\rm{tot}}(M)}{\overline{n}_{\rm{g}}(z)},
\end{equation}
the fraction of galaxies that are satellites,
\begin{equation} \label{eqn:satellite_fraction}
    f_{\rm{sat}}(z)=1-f_{\rm{cen}}(z)=1-\int{\rm{d}}M\,n(M,z)\frac{\overline{N}_{\rm{c}}(M)}{\overline{n}_{\rm{g}}(z)},
\end{equation}
and the mean galaxy bias,
\begin{equation} \label{eqn:mean_bias}
    b(z)=\int{\rm{d}}M\,b_{\rm{h}}(M,z)n(M,z)\frac{\overline{N}_{\rm{tot}}(M)}{\overline{n}_{\rm{g}}(z)},
\end{equation}
where $n(M, z)$ is the halo mass function, $b_{\rm{h}}(M,z)$ is the halo bias model, and the integrals are normalized by the mean number density of galaxies,
\begin{equation} \label{eqn:mean_number_density}
    \overline{n}_{\rm{g}}(z)=\int{\rm{d}}M\,n(M,z)\overline{N}_{\rm{tot}}(M).
\end{equation}

One additional consideration when modelling the cross-correlation functions is the overlap between the samples. In other clustering studies, it is usually assumed that the two populations being cross-correlated are independent, or that the overlap between the two samples is small enough that it can be ignored. However, our AGN are a subset of the optical/NIR galaxies, and the matched galaxy sample is drawn from the full optical/NIR population. This introduces additional self-pairing terms when measuring the cross-correlation that correspond to galaxies that are present in both samples. To account for these effects in the model, we modified the $\texttt{CCL}$ code, with the theory and changes described in Appendix~\ref{sec:cross_correlation_model}.

\subsubsection{Model fitting and integral constraints} \label{sssec:model_fitting}

Since the observed fields have finite sizes, the maximum possible angular separation of two sources is limited. The observed TPCF, $\omega_{\rm obs}$, will therefore be underestimated at large angular scales and negatively offset from the true TPCF, $\omega_{\rm true}$:
\begin{equation}\label{eqn:obs_tpcf_offset}
    \omega_{\rm{obs}}(\theta)=\omega_{\rm{true}}(\theta)-\sigma^2_{\rm{IC}},
\end{equation}
where the offset, $\sigma^2_{\rm{IC}}$, is known as the integral constraint. Its value is given analytically by the expression from \cite{groth1977},
\begin{equation}\label{eqn:integral_constraint_analytic}
    \sigma^2_{\rm{IC}}=\frac{1}{\Omega^2}\iint\omega_{\rm{true}}(\theta)\,\rm{d}\Omega_1\rm{d}\Omega_2,
\end{equation}
where $\rm{d}\Omega_1\rm{d}\Omega_2$ denotes integrating over the field solid angle twice. This can be estimated numerically using the random-random pair counts \citep{roche1999}:
\begin{equation}\label{eqn:integral_constraint_numeric}
    \sigma^2_{\rm{IC}}=\frac{\sum_i\overline{R_nR_m}(\theta_i)\,\omega_i^{\rm{true}}}{\sum_i\overline{R_nR_m}(\theta_i)}.
\end{equation}

To fit our model to the observed TPCF, we use the Markov chain Monte Carlo (MCMC) sampling method provided by the python package $\texttt{emcee}$\footnote{\href{https://emcee.readthedocs.io/en/stable/}{https://emcee.readthedocs.io/en/stable/}} \citep{foreman-mackey2013}. The starting positions of the walkers are drawn randomly from uniform priors over 
$11<\log_{10}(M_{\rm{min}}/{\rm{M_\odot}})<15$; 
$8<\log_{10}(M_0/{\rm{M_\odot}})<\log_{10}(M_1/{\rm{M_\odot}})$; 
$0<\sigma_{\ln{M}}<1.4$;
$\log_{10}(M_{\rm{min}}/{\rm{M_\odot}})<\log_{10}(M_1/{\rm{M_\odot}})<16$;
and $0.4<\alpha_{\rm{s}}<1.6$. Although the parameters are fit separately in each of the redshift bins, they are assumed to be constant within each bin.

The two parameters, $M_{\rm{min}}$ and $\sigma_{\ln{M}}$, that control the shape of $\overline{N}_{\rm{c}}(M_{\rm{h}})$, are highly degenerate since both have qualitatively similar effects on $\omega(\theta)$ \citep{salcedo2020}. For the optical/NIR galaxy TPCFs, this degeneracy is broken using $\overline{n}_{\rm{g}}$ by fitting simultaneously to both the measured $\omega(\theta)$ using the covariance matrix defined in Equation~\ref{eqn:cov_jk}, and the observed number density of galaxies. A Gaussian likelihood is defined, $\mathcal{L}=\exp(-\frac{1}{2}\chi^2)$, where the total $\chi^2$ is the sum of these two contributions:
\begin{equation}\label{eqn:chi2}
\begin{split}
    \chi^2=&\sum_{i}\sum_{j}\left(\omega_i^{\rm{true}}-\omega_i^{\rm{model}}\right){\rm{Cov}}^{-1}_{ij}\left(\omega_j^{\rm{true}}-\omega_j^{\rm{model}}\right)\\
    &+\frac{\left(N_{\rm{g}}^{\rm{obs}}-N_{\rm{g}}^{\rm{mod}}\right)^2}{\sigma_{\rm{gal}}^2}.
\end{split}
\end{equation}
In the first term, $\omega_i^{\rm{model}}$ is the value of the TPCF from the HOD model for the $i^{\rm{th}}$ $\theta$ bin, $\omega_i^{\rm{true}}$ is related to the measured TPCF by Equation~\ref{eqn:obs_tpcf_offset}, and ${\rm{Cov}}_{ij}^{-1}$ are the components of the inverse of the covariance matrix. In the second term, the observed number of galaxies, $N_{\rm{g}}^{\rm{obs}}$, is the sum of the assigned weights for galaxies above the stellar mass threshold (i.e., $N_{\rm{g}}^{\rm{obs}}=\sum_iw_i$), and the number of galaxies predicted by a given model is found by multiplying the predicted number density by the comoving volume covered by the fields:
\begin{equation}\label{eqn:num_galaxies_model}
    N_{\rm{g}}^{\rm{mod}}=\overline{n}_{\rm{g}}\times \frac{\Omega}{4\pi} \times\frac{4}{3}\pi\left[d^3(z_{\rm{max}})-d^3(z_{\rm{min}})\right],
\end{equation} 
where $d(z)$ is the comoving distance to redshift $z$, and $\Omega$ is the solid angle covered by the fields in steradians.  The uncertainty on the number of galaxies, $\sigma^2_{\rm{gal}}=\sigma^2_{\rm{Pois}}+\sigma^2_{\rm{cv}}$, contains contributions from Poisson noise, $\sigma_{\rm{Pois}}=\sqrt{N^{\rm{mod}}_{\rm{g}}}$, and cosmic variance, $\sigma_{\rm{cv}}=b\,\sigma_{\rm{dm}}N^{\rm{mod}}_{\rm{g}}$. The galaxy bias is calculated using Equation~\ref{eqn:mean_bias} at each step in the MCMC chain, and the fractional dark matter root cosmic variance, $\sigma_{\rm{dm}}$, in each field is estimated using the cosmic variance cookbook of \cite{moster2011} which scales with the width of the redshift bin as $\sigma_{\rm{dm}}\propto{\Delta z}^{-1/2}$. The total cosmic variance for the combined fields is then given by a volume weighted sum,
\begin{equation}\label{eqn:total_cosmic_variance}
    \sigma_{\rm{cv}}^2=\frac{\sum_i V_i^2 \sigma_i^2}{(\sum_iV_i)^2},
\end{equation} 
where $V_i$ is the comoving volume of each field between the redshift bin limits, calculated as in Equation~\ref{eqn:num_galaxies_model}.

For the AGN and matched galaxy sample TPCFs, the model is fitted solely to $\omega(\theta)$ (the first term of Equation~\ref{eqn:chi2}) with the $N_{\rm{gal}}$ terms omitted. This is because the observed AGN population is not complete, since not all radio sources are successfully cross-matched with their host galaxies, which will decrease the inferred number density. Additionally, the number of galaxies in the matched sample is chosen arbitrarily, and so cannot be used in the model fitting. When fitting to the TPCF of the optical/NIR galaxies, we include both terms.

Since the number of AGN is relatively small, the uncertainties on the AGN auto-correlation are large, leading to weak constraints on the HOD parameters when fitting to it alone. For this reason, we fit to the cross-correlation between the AGN (or matched galaxy) sample and the optical/NIR galaxies. This requires knowledge of the best-fitting HOD parameters for the optical/NIR sample. To properly propagate uncertainties, in each redshift bin, a joint fitting is performed to the optical/NIR auto-correlation, the AGN cross-correlation, and the matched galaxy cross-correlation, simultaneously. The total likelihood is the product of the individual likelihoods:
\begin{equation}\label{eqn:total_likelihood}
\begin{split}
    &\mathcal{L}_{\rm{tot}}(\phi_{\rm{AGN}}, \phi_{\rm{mat}}, \phi_{\rm{opt}}) \\ &= \mathcal{L}_{\rm{opt}}(\phi_{\rm{opt}}) \mathcal{L}_{\rm{AGN}\times\rm{opt}}(\phi_{\rm{AGN}},\phi_{\rm{opt}}) \mathcal{L}_{\rm{mat}\times{opt}}(\phi_{\rm{mat}},\phi_{\rm{opt}}),
\end{split}
\end{equation}
where $\phi_{\rm{opt}}$, $\phi_{\rm{AGN}}$ and $\phi_{\rm{mat}}$ are the HOD parameters for the optical/NIR, AGN and matched galaxy samples, respectively. This assumes that the uncertainties on the cross-correlations are independent of those on the optical/NIR auto-correlation. Given that the optical/NIR sample is over an order of magnitude larger than the AGN and matched galaxy samples, the uncertainties on the cross-correlation measurements will be dominated by the shot noise of the latter, rather than by fluctuations in the optical/NIR population. The covariance between the uncertainties will therefore be small, and we find that fitting to the auto-correlation function alone gives the same best-fitting parameters as those obtained when fitting to the three sets of measurements simultaneously.

When fitting to the optical/NIR auto-correlation functions, all five HOD parameters are fit. However, for the AGN and matched galaxy samples, the fit is restricted to only three free parameters: $M_{\rm{min}}$, $M_1$ and $\alpha_{\rm{s}}$, while fixing $M_0=M_{\rm{min}}$ and $\sigma_{\ln{M}}=1$, chosen to be consistent with the values found for high-luminosity radio-galaxies by \cite{petter2024}. We find that allowing $M_0$ to vary has a negligible impact on the resulting model correlation function and the best-fitting values of the other parameters. Fixing $\sigma_{\ln{M}}$ is necessary because, for these samples, number density constraints cannot be reliably used to break its degeneracy with $M_{\rm{min}}$. 

We use 60 walkers (5 per model parameter) with a 500 step burn-in phase and 2500 iterations. The fitting for the $1.5<z<2.5$ bin was also repeated with twice as many iterations which produced almost identical best-fitting HOD parameters and uncertainties, so we are confident that the number of iterations here is sufficient to reach convergence. The quoted upper and lower uncertainties for the HOD parameters are the $16^{\rm{th}}$ and $84^{\rm{th}}$ percentiles of the posterior distributions.

\section{Results and Discussion} \label{sec:results}

\subsection{Auto-correlation functions of AGN and matched galaxies}\label{ssec:auto_correlations}

\begin{figure*}
 \includegraphics[width=2\columnwidth]{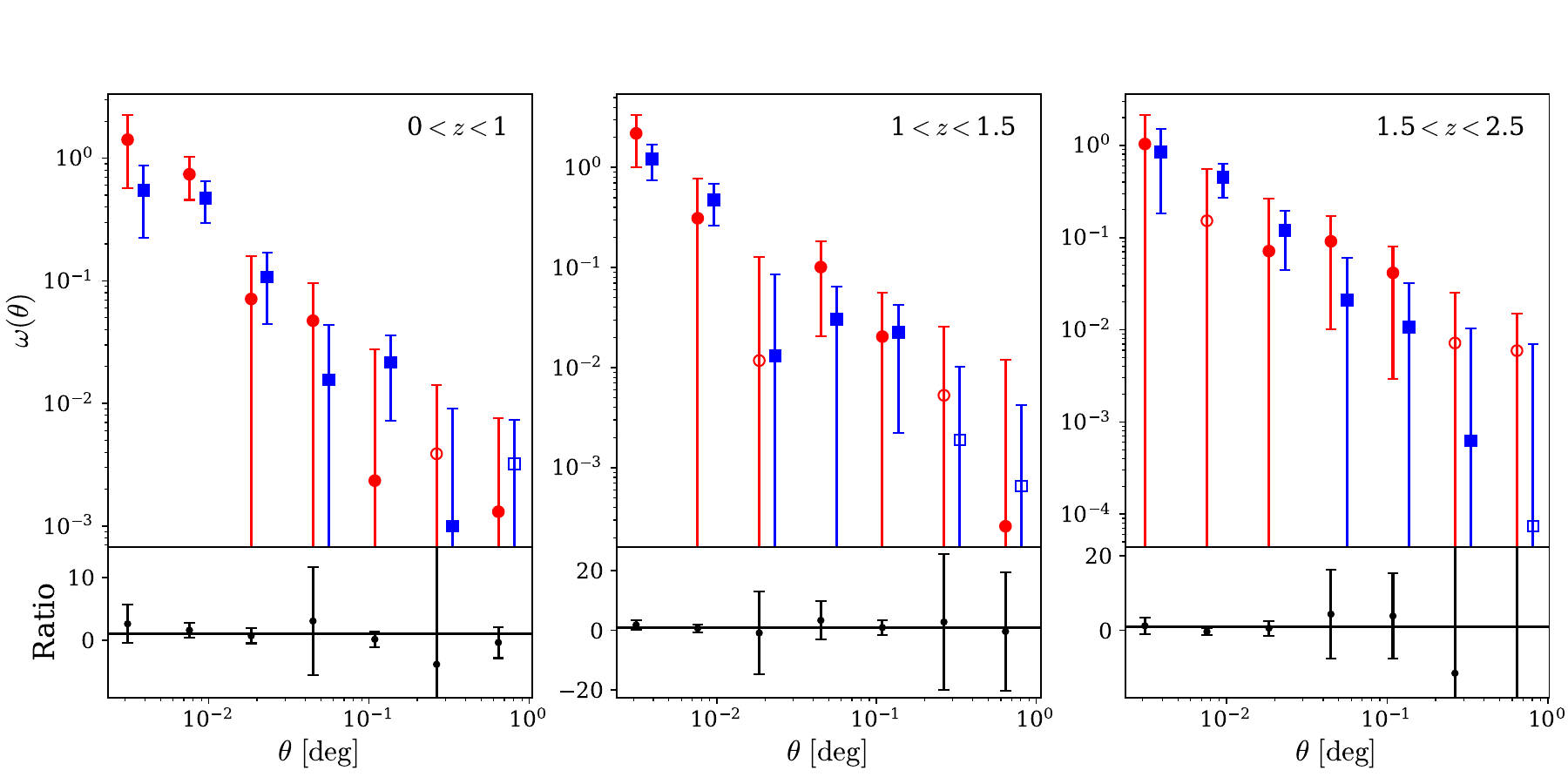}
 \caption{The measured angular two-point auto-correlation functions, $\omega(\theta)$, for AGN (red circles) and the stellar mass-matched galaxies sample (blue squares) as functions of angular separation, $\theta$, for each redshift bin. Left to right, these are: $0<z<1$, $1<z<1.5$ and $1.5<z<2.5$. Open markers are negative values. For clarity, the matched galaxy correlations are offset by $+0.1$ dex in $\theta$. The bottom panels give the ratio of the two correlation functions, with the black horizontal line at $\omega_{\rm{AGN}}/\omega_{\rm{mat}}=1$.}
 \label{fig:auto_correlations}
\end{figure*}

Figure~\ref{fig:auto_correlations} presents the measured auto-correlation functions of the AGN and matched galaxy samples for each redshift bin. The bottom panels in each plot are the ratio of the two correlation functions. The uncertainties are found using jackknife patches, as discussed in Section~\ref{sssec:estimating_tpcf}. We use $\theta$ bins ranging from $\theta=2\times10^{-3}-1^\circ$, producing 7 values of $\omega_i$ that are uniformly spaced in $\log{\theta}$ space. The lower $\theta$ limit was chosen to be approximately twice the $5\ {\rm{arcsec}}$ resolution of the radio data, below which radio source blending may become an issue, while the upper limit was chosen to be slightly greater than the angular scale where the measurements drop to zero due to the finite sizes of the fields.

All of the individual points of the measured auto-correlation function for the two samples overlap within the uncertainties. To assess whether the differences between the two curves are statistically significant, we perform a $\chi^2$ test, accounting for correlations between $\theta$ bins using the covariance matrices that were calculated with Equation~\ref{eqn:cov_jk}. We take $\chi^2=\boldsymbol{\Delta}^{\rm{T}}\left[{\rm{Cov}}(\Delta)\right]^{-1}\boldsymbol{\Delta}$, where the total covariance matrix, ${\rm{Cov}}(\Delta)$, is the sum of the individual matrices for the two samples, and $\boldsymbol{\Delta} = \boldsymbol{\omega}_{\rm{AGN}} - \boldsymbol{\omega}_{\rm{mat}}$ is the difference between the two curves. 
The corresponding $p$-value is computed using the $\chi^2$ cumulative distribution function, which quantifies the probability of obtaining the observed difference under the null hypothesis that the two samples have the same underlying clustering. 

From the lowest to highest redshift bin, we obtain $p$-values of 0.91, 0.99 and 0.82, which are all much higher than the significance threshold. There is therefore no strong evidence for significant difference between the auto-correlations of the two samples. This motivates our use of the cross-correlation functions with the full galaxy population.

\subsection{Cross-correlation functions of AGN and matched galaxies with optical/NIR galaxies}\label{ssec:cross_correlations}

\begin{figure*}
 \includegraphics[width=2\columnwidth]{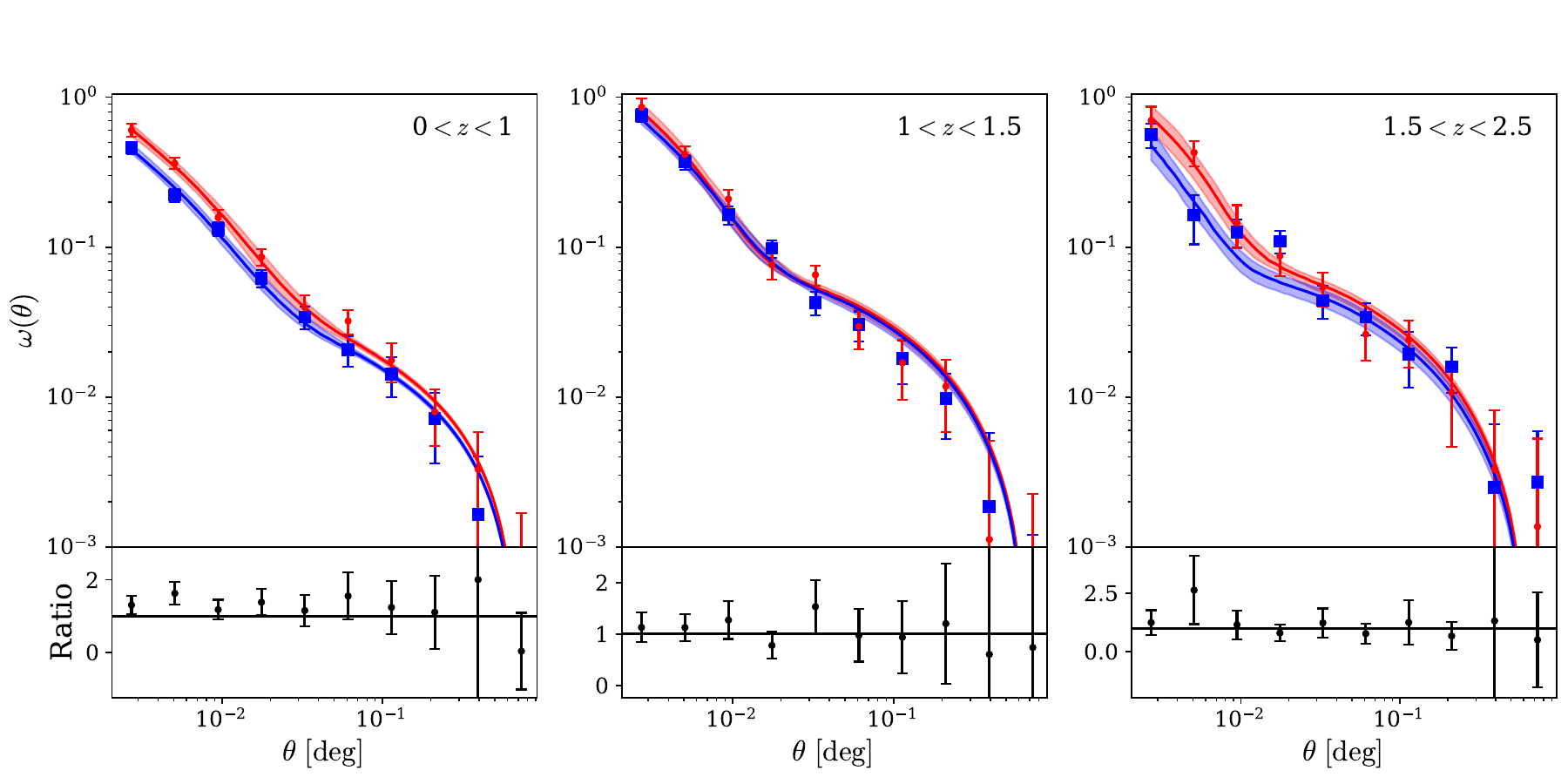}
 \caption{The measured angular two-point cross-correlation functions, $\omega(\theta)$, between AGN and optical/NIR galaxies (red circles), and the cross-correlation of stellar mass-matched galaxies with optical/NIR galaxies (blue squares) as functions of angular separation, $\theta$, for each redshift bin. Left to right, these are: $0<z<1$, $1<z<1.5$ and $1.5<z<2.5$. The solid lines are the best-fitting HOD models and the shaded regions are the $1\sigma$ confidence intervals. The bottom panels gives the ratio of the two correlation functions, with the black horizontal line at $\omega_{\rm{AGN}}/\omega_{\rm{mat}}=1$.}
 \label{fig:cross_correlations}
\end{figure*}

To further investigate the environmental differences between AGN and the matched galaxy sample, we measure their cross-correlations with the full galaxy population. Since the optical/NIR galaxy catalogue is significantly larger than the AGN sample, computing the cross-correlation greatly increases the number of pair counts. This reduces statistical uncertainties and increases the likelihood of detecting significant differences in clustering between the two samples. The smaller error bars will also produce tighter constraints on the best-fitting HOD parameters. As before, the measurements are limited by the resolution of the radio data, so we adopt the same $\theta$ range as for the auto-correlation measurements while increasing the number of bins to 10.

Figure~\ref{fig:cross_correlations} shows the measured cross-correlations of AGN with optical/NIR galaxies and the cross-correlations of the matched galaxy sample with optical/NIR galaxies as functions of angular separation for each of our redshift bins and stellar mass thresholds. The uncertainties are found using the jackknife method described in Section~\ref{ssec:angular_correlation_function}. 

For the $0<z<1$ redshift bin at small angular separations, the cross-correlation signal for AGN is stronger than that of the matched galaxy sample, suggesting that galaxies are more densely clustered around AGN within the same halo than non-active galaxies of similar stellar mass. The ratio plots shows that for $0<z<1$, the AGN clustering signal is $\sim1.5$ times greater on scales $\lesssim0.1^\circ$. For higher redshifts, the difference between the AGN and matched galaxy signals is less pronounced; though, the AGN measurements tend to lie consistently above those of the matched sample.

For larger angular separations, however, the cross-correlations for AGN and the matched galaxies are indistinguishable within the uncertainties in all redshift bins. On large scales where the clustering of galaxies is well described by linear perturbation theory, the auto-correlation of AGN is proportional to the square of the bias parameter, $b^2_{\rm{AGN}}$, whereas the cross-correlation scales as the product of the biases of the two populations, $b_{\rm{AGN}}b_{\rm{opt}}$ \citep{cooray2002}. As a result, the relative difference in bias between the AGN and matched galaxy sample is less pronounced in the two-halo term of the cross-correlation.

To test whether the differences between the two curves are statistically significant, we perform the same $\chi^2$ test as described in Section~\ref{ssec:auto_correlations}. Since in each bin the AGN and matched galaxy samples are cross-correlated with the same galaxy sample, their measurements are not independent, and this must be accounted for in the covariance matrix. The total covariance matrix is
\begin{equation}
\begin{split}
   &{\rm{Cov}}(\omega_{\rm{AGN}}-\omega_{\rm{mat}})={\rm{Cov}}(\omega_{\rm{AGN}})+{\rm{Cov}}(\omega_{\rm{mat}})\\
   &-\big[{\rm{Cov}}(\omega_{\rm{AGN}},\omega_{\rm{mat}})+{\rm{Cov}}^{\rm{T}}(\omega_{\rm{AGN}},\omega_{\rm{mat}})\big],
\end{split}
\end{equation}
where ${\rm{Cov}}(\omega_{\rm{AGN}})$ and ${\rm{Cov}}(\omega_{\rm{mat}})$ are the covariance matrices for the AGN and matched galaxy cross-correlations estimated using Equation~\ref{eqn:cov_jk}, and the elements of the cross-covariance matrix are given by \citep{norberg2009}
\begin{equation}
\begin{split}
    &{\rm{Cov}}_{ij}(\omega_{\rm{AGN}},\omega_{\rm{mat}})\\
    &=\frac{N_{\rm{patch}}-1}
    {N_{\rm{patch}}}\sum_{k=1}^{N_{\rm{patch}}}\left(\boldsymbol{\omega}_{\rm{AGN}}^k-\overline{\boldsymbol{\omega}}_{\rm{AGN}}\right)_i\left(\boldsymbol{\omega}_{\rm{mat}}^k-\overline{\boldsymbol{\omega}}_{\rm{mat}}\right)_j,
\end{split}
\end{equation}
with ${\rm{Cov}}^{\rm{T}}_{ij}(\omega_{\rm{AGN}},\omega_{\rm{mat}})={\rm{Cov}}_{ji}(\omega_{\rm{AGN}},\omega_{\rm{mat}})$ its transpose. Performing the test, we find a $p$-value of $7.9\times10^{-3}$ for the $0<z<1$ redshift bin which is significantly below the $p=0.05$ significance threshold, so the correlation functions differ significantly. However, for the $1<z<1.5$ and $1.5<z<2.5$ bins, we find $p=0.23$ and $0.06$, respectively, so there is no significant difference between the two clustering signals.

 The enhanced clustering in the lowest redshift bin implies that AGN are more biased tracers of the underlying dark matter distribution and, in turn, reside in more massive dark matter haloes compared to non-active galaxies of the same stellar masses, in agreement with the findings of previous works \cite[e.g.,][]{kauffmann2008, mandelbaum2009, bardelli2010, malavasi2015}.

 In addition to measuring the clustering of the AGN and matched sample, we also investigated whether the cross-correlations of AGN with the optical/NIR galaxy sample depend on their radio luminosity by dividing the sample into low- and high-luminosity sub-samples. Since the AGN selection threshold is redshift-dependent, a corresponding minimum luminosity was applied in each bin so that the sub-samples were drawn from comparable redshift distributions. New AGN weights were recomputed for each radio source following a similar procedure to that in Section~\ref{ssec:weights}, giving the probability that the source both exceeds the luminosity threshold and falls within the luminosity range of the sub-sample. Applying the same significance test as above, no significant differences were found between the clustering of low- and high-luminosity AGN. While this null result may indicate that the environments of AGN are genuinely independent of radio luminosity, it is more likely a consequence of the large statistical uncertainties resulting from the small sample sizes. Therefore, a larger dataset is required to establish this conclusively.

\subsection{HOD modelling results}\label{ssec:hod_results}

Having determined that galaxies hosting AGN are significantly more clustered than non-active galaxies matched in stellar mass, we now present the results of HOD modelling to infer the properties of the dark matter haloes that host our galaxy samples. We first present the best-fitting HOD parameters for the optical/NIR auto-correlations function for each redshift and stellar mass bin, which must be known to measure the HOD parameters of the AGN and matched galaxy samples from their cross-correlations. We then derive the best-fitting AGN HOD parameters and compare to previous results in the literature. Finally, we compare the best-fitting HOD parameters of the AGN and matched galaxy samples.

Table~\ref{tab:hod_results} presents the best-fitting HOD parameters for the optical/NIR, matched galaxy and AGN samples for each stellar mass and redshift bin along with the number of objects in each sample and the median redshifts.

{\renewcommand{\arraystretch}{1.2}
\begin{table*}
 \caption{The number counts ($N$), median redshift ($z_{\rm{med}}$) and best-fitting HOD parameters for the full optical/NIR samples (`All'), the AGN sample and the matched galaxy sample, along with the corresponding reduced $\chi^2_\nu=\chi^2/{\rm{d.o.f.}}$, for each stellar mass threshold and redshift bin. All masses are given as $\log_{10}$ values in units of $\rm{M_\odot}$. Note that the number counts are actually the sums of the weights in each sub-sample. The quoted uncertainties are the $16^{\rm{th}}$ and $84^{\rm{th}}$ percentiles of the posteriors.}
 \begin{tabular}{lccccccccr}
  \hline
  $M_\ast$ threshold & Sample & $N$ & $z_{\rm{med}}$ & $M_{\rm{min}}$ & $M_0$ & $\sigma_{\ln{M}}$ & $M_1$ & $\alpha_{\rm{s}}$ & $\chi^2_\nu$\\
  \hline
  $0\,<\,z<1$\\
  \hline
    9.5&All&$186\,954$&$0.71^{+0.21}_{-0.31}$&$11.57^{+0.06}_{-0.02}$&$9.50^{+1.06}_{-1.06}$&$0.46^{+0.42}_{-0.33}$&$12.75^{+0.02}_{-0.03}$&$1.07^{+0.03}_{-0.04}$&2.66\\
&AGN&$952$&$0.76^{+0.17}_{-0.28}$&$12.81^{+0.09}_{-0.13}$&$-$&1&$13.62^{+0.35}_{-0.52}$&$0.93^{+0.40}_{-0.33}$&0.93\\
&Matched&$1\,720$&$0.78^{+0.15}_{-0.30}$&$12.52^{+0.12}_{-0.21}$&$-$&1&$13.46^{+0.42}_{-0.79}$&$0.84^{+0.42}_{-0.25}$&0.84\\
  \hline
  $1\,<\,z<1.5$\\
  \hline
  10&All&$69\,518$&$1.21^{+0.21}_{-0.16}$&$12.03^{+0.08}_{-0.05}$&$9.91^{+1.19}_{-1.28}$&$0.68^{+0.40}_{-0.42}$&$13.20^{+0.02}_{-0.02}$&$1.24^{+0.03}_{-0.04}$&0.34\\
&AGN&$567$&$1.25^{+0.14}_{-0.17}$&$12.88^{+0.11}_{-0.35}$&$-$&1&$14.11^{+1.24}_{-1.16}$&$0.89^{+0.45}_{-0.33}$&1.50\\
&Matched&$1\,229$&$1.22^{+0.19}_{-0.16}$&$12.64^{+0.24}_{-0.50}$&$-$&1&$13.42^{+1.06}_{-0.76}$&$1.05^{+0.30}_{-0.37}$&0.99\\
  \hline
  $1.5\,<\,z<2.5$\\
  \hline
    10.4&All&$36\,207$&$1.85^{+0.38}_{-0.26}$&$12.41^{+0.07}_{-0.06}$&$10.17^{+1.67}_{-1.51}$&$0.74^{+0.31}_{-0.45}$&$13.52^{+0.03}_{-0.03}$&$1.48^{+0.08}_{-0.18}$&1.99\\
&AGN&$498$&$1.75^{+0.44}_{-0.18}$&$12.87^{+0.18}_{-0.32}$&$-$&1&$13.93^{+1.39}_{-0.75}$&$0.99^{+0.42}_{-0.42}$&1.13\\
&Matched&$990$&$1.76^{+0.44}_{-0.19}$&$12.51^{+0.25}_{-0.42}$&$-$&1&$13.97^{+1.31}_{-0.90}$&$0.90^{+0.46}_{-0.37}$&3.43\\
  \hline
 \end{tabular}
 \label{tab:hod_results}
\end{table*}
}

\subsubsection{Optical/NIR sample HOD results}\label{sssec:optical_hod_results}

{\renewcommand{\arraystretch}{1.2}
\begin{table*}
 \caption{Characteristic halo masses, $M_{\rm{min}}$ and $M_1$, of optical/NIR galaxies from previous works for comparable redshift ranges and stellar mass thresholds to the three used in this study.}
 \begin{tabular}{lccccr}
  \hline
  Survey & Reference & Redshift & $M_\ast$ threshold & $\log_{10}(M_{\rm{min}}/{\rm{M_\odot}})$ & $\log_{10}(M_1/{\rm{M_\odot}})$ \\
  \hline
  VIDEO & \cite{hatfield2016} & $0.5<z<0.75$ & $9.6$ & $11.80^{+0.09}_{-0.08}$ & $12.80^{+0.22}_{-0.16}$ \\
  & & $1<z<1.25$ & $10.1$ & $12.00^{+0.06}_{-0.08}$ & $13.20^{+0.14}_{-0.24}$ \\ 
  \hline
  UltraVISTA & \cite{mccracken2015} & $1.5<z<2.5$ & $10.4$ & $\sim12.3$ & $\sim13.5$\\
  &(see their Figure 9)\\
  \hline
 \end{tabular}
 \label{tab:previous_optical_hod_params}
\end{table*}
}

\begin{figure*}
 \includegraphics[width=2\columnwidth]{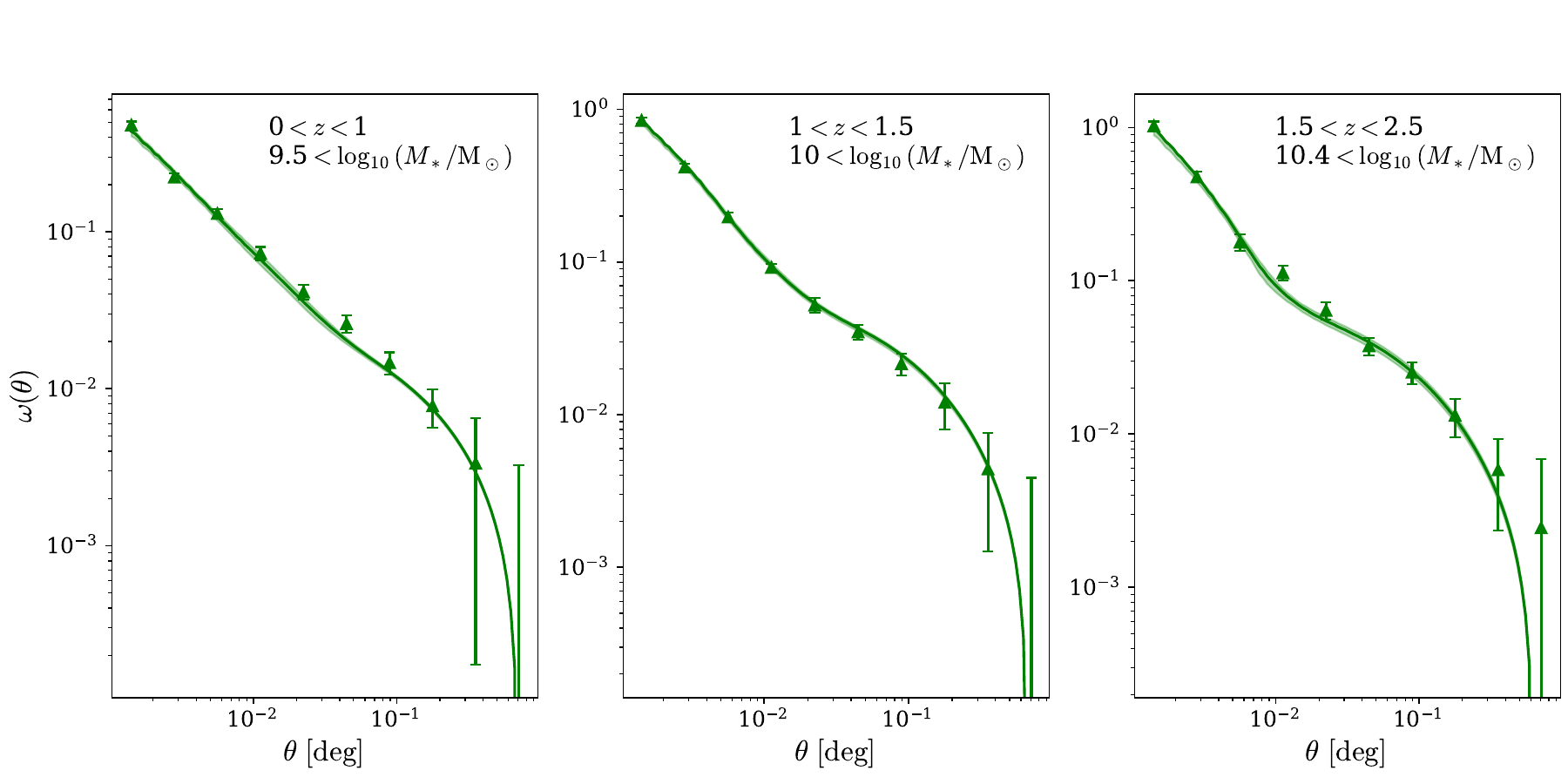}
 \caption{The measured angular two-point auto-correlation functions, $\omega(\theta)$, for the full optical/NIR galaxy populations (upwards pointing green triangles) as functions of angular separation, $\theta$, for each stellar mass and redshift bin. Left to right, these are: $0<z<1$, $1<z<1.5$ and $1.5<z<2.5$ with stellar mass thresholds of $\log_{10}(M_\ast/M_\odot)>9.5$, $10$ and $10.4$, respectively. The solid lines are the best-fitting HOD models and the shaded regions are the $1\sigma$ confidence intervals. Note that the model is fitted to the observed number densities of galaxies in addition to the clustering measurements shown in this figure.}
 \label{fig:auto_all}
\end{figure*}

In this section, the best-fitting HOD parameters for the auto-correlations of the optical/NIR samples are discussed and compared to the values found by previous works. The measured auto-correlation functions and best-fitting models are shown in Figure~\ref{fig:auto_all}.

For the best-fitting HOD parameters, the reduced $\chi^2_\nu=2.66$, $0.34$ and $1.99$ where obtained for the $0<z<1$, $1<z<1.5$ and $1.5<z<2.5$ redshift bins, respectively. The number of degrees of freedom is calculated as the number of $\theta$ bins (10) plus one (for fitting to $\overline{n}_{\rm{g}}$), minus the number of model parameters (5).

Table~\ref{tab:previous_optical_hod_params} presents the characteristic halo masses, $M_{\rm{min}}$ and $M_1$, from previous studies that use the same five parameter HOD model. We include the results for their closest matching redshift ranges and stellar mass thresholds to the three used in this work. In particular, \cite{hatfield2016} measured the auto-correlation function of galaxies from the first data release of the VIDEO survey out to $z\sim1.7$ over an area of $1.5\ {\rm{deg}}^2$, while \cite{mccracken2015} measured the clustering of galaxies from UltraVISTA to $z\sim2$ over $1\ {\rm{deg}}^2$.

Overall, we find that our measurements of $M_{\rm{min}}$ and $M_1$ are broadly consistent with those studies across the full redshift range considered. In the lowest redshift bin, we find a slightly lower value of $\log_{10}(M_{\rm{min}}/{\rm{M_\ast}})=11.57^{+0.06}_{-0.02}$ compared to the $\log_{10}(M_{\rm{min}}/{\rm{M_\ast}})=11.80^{+0.09}_{-0.08}$ reported by \cite{hatfield2016}, although their result corresponds to a stellar mass threshold that is higher by $0.1\ {\rm{dex}}$. The overall agreement in the inferred HOD parameters, together with the acceptable reduced $\chi^2$ values, improves confidence in the reliability of the AGN HOD parameters derived from the cross-correlations.

\subsubsection{AGN sample HOD results}\label{sssec:agn_hod_results}

This section discusses the best-fitting HOD parameters for the AGN cross-correlation functions presented in Table~\ref{tab:hod_results}, and compares the derived typical halo masses and bias values to previous results in the literature.

Good fits to the AGN cross-correlation were obtained for all redshift bins with reduced chi-squared values of $\chi^2_\nu=0.93$, $1.50$ and $1.13$, for the $0<z<1$, $1<z<1.5$ and $1.5<z<2.5$ redshift bins, respectively, for 7 degrees of freedom.

The best constrained parameter in terms of relative uncertainty is $M_{\rm{min}}$, which remains consistent across the redshift bins. It ranges from $\log_{10}(M_{\rm{min}}/{\rm{M_\odot}})=12.81^{+0.09}_{-0.13}$ in the $0<z<1$ bin to $\log_{10}(M_{\rm{min}}/{\rm{M_\odot}})=12.87^{+0.18}_{-0.32}$ in the $1.5<z<2.5$ bin. This suggests that the characteristic halo mass required to host a radio-AGN does not evolve strongly over the redshift range considered. In the intermediate redshift bin, $1<z<1.5$, with $z_{\rm{med}}=1.25^{+0.14}_{-0.17}$, we obtain $\log_{10}(M_{\rm{min}}/{\rm{M_\odot}})=12.88^{+0.11}_{-0.35}$. This is lower than the value reported by \cite{magliocchetti2017} of $\log_{10}(M_{\rm{min}}/{\rm{M_\odot}})=13.6^{+0.3}_{-0.6}$ for a Very Large Array (VLA) radio sample in COSMOS at $z\approx1.25$. They assumed that all radio-AGN are central galaxies (i.e., one AGN per halo) and fit only to the two-halo term, for which angular correlation scales with the bias as $\omega_{\rm{2h}}{\propto}b^2$. Satellites preferentially reside in higher mass haloes (typically around $M_1$) that have larger $b_{\rm{h}}(M_{\rm{h}}, z)$. Including satellites therefore raises the mean bias above what would be predicted for centrals alone. To reproduce the observed two-halo amplitude with central galaxies only, $M_{\rm{min}}$ must be driven to higher values so that galaxies occupy more massive, and hence more biased, haloes. This may explain our lower $M_{\rm{min}}$ value, although we note the large uncertainty associated with the measurement of \cite{magliocchetti2017} due to the limited areal coverage within just the COSMOS field.

The parameters that determine the number of AGN satellite galaxies, $M_1$ and $\alpha_{\rm{s}}$, are poorly constrained with large uncertainties relative to those of the full optical/NIR sample. For increasing redshift bins, we find $\log_{10}(M_1/{\rm{M_\odot}})=13.62^{+0.35}_{-0.52}$, $14.11^{+1.24}_{-1.16}$ and $13.93^{+1.39}_{-0.75}$. If the fraction of AGN hosts that are satellite galaxies is small, variations in $M_1$ or $\alpha_{\rm{s}}$ will have minimal effect on the modelled correlation function (see the flat prior distributions in Appendix~\ref{sec:corner_plots}). Evaluating Equation~\ref{eqn:satellite_fraction}, we derive satellite fractions of $f_{\rm{sat}}=0.18^{+0.16}_{-0.09}$, $0.06^{+0.21}_{-0.05}$ and $0.06^{+0.09}_{-0.05}$. This aligns with observations that the vast majority of radio-loud AGN are located at the centres of their haloes \citep[e.g.,][]{hatch2014, mo2018, croston2019}. 

Table~\ref{tab:derived_params} presents the values of the derived parameters, $\langle{M_{\rm{h}}}\rangle$, $f_{\rm{sat}}$ and $b$, for the AGN and matched galaxy samples. These were evaluated by sampling from the posterior distributions of the best-fitting HOD parameters, and a single value for each redshift bin was found by marginalizing $\langle{M_{\rm{h}}}(z)\rangle$, $f_{\rm{sat}}(z)$ and $b(z)$ over the redshift distribution of the sample. Although all three of the AGN HOD parameters are consistent between the redshift bins, the corresponding values of $\langle{M_{\rm{h}}}\rangle$ vary due to the evolving halo mass function. For $1.5<z<2.5$, we find $\log_{10}(\langle{M_{\rm{h}}}\rangle/{\rm{M_\odot}})=13.03^{+0.09}_{-0.10}$ which increases to $\log_{10}(\langle{M_{\rm{h}}}\rangle/{\rm{M_\odot}})=13.44^{+0.08}_{-0.08}$ by $0<z<1$. This suggests that radio-AGN in our sample typically reside in galaxy groups, which have halo masses of $M_{\rm{h}}\sim10^{11}-10^{14}\ {\rm{M_\odot}}$, rather than larger galaxy clusters with $M_{\rm{h}}\gtrsim10^{14}\ {\rm{M_\odot}}$ \citep{lim2017, calderon2019}. This is in agreement with studies that investigate galaxy densities around radio-galaxies \citep[e.g.,][]{best2004, croston2019}. 

At low redshift, our estimate of $\langle{M_{\rm{h}}}\rangle$ for $0<z<1$ is in line with the findings of previous works. For example, \cite{hickox2009} investigated the clustering of radio sources with $L_{\rm{1.4\ GHz}}>10^{23.8}\ {\rm{W\ Hz^{-1}}}$ and $0.25<z<0.8$ using the Westerbork Synthesis Radio Telescope \citep{devries2002}, and found radio-AGN are strongly clustered with halo masses $\log_{10}(M_{\rm{h}}/{\rm{M_\odot}})\approx13.3$ (converted to our cosmology using $h=0.7$). \cite{mandelbaum2009} measured the clustering and lensing of radio-AGN with $z\lesssim0.3$ from the Faint Images of the Radio Sky at Twenty-Centimetres radio survey \citep[FIRST;][]{becker1995}, and found that they reside in less massive haloes with $\log_{10}(M_{\rm{h}}/{\rm{M_\odot}})\approx13.0$.

Our estimates of $\langle{M_{\rm{h}}}\rangle$ at higher redshift are similar to the findings of previous clustering studies, though they lie at the lower end of the reported range. For example, \cite{allison2015} cross-correlated radio-AGN from FIRST with lensing of the cosmic microwave background and found a typical halo mass of $\log_{10}(\langle{M_{\rm{h}}}\rangle/{\rm{M_\odot}})=13.6^{+0.3}_{-0.4}$ at a redshift of $z\approx1.5$. This is consistent with our value at $z_{\rm{med}}=1.25^{+0.14}_{-0.17}$ of $\log_{10}(\langle{M_{\rm{h}}}\rangle/{\rm{M_\odot}})=13.17^{+0.07}_{-0.06}$. \cite{retana2017} measured the projected correlation function of radio-loud quasars from FIRST over the redshift range $0.3<z<2.3$, and found a higher average dark matter halo masses of $\log(\langle{M_{\rm{h}}}\rangle/{\rm{M_\odot}})\sim13.5$ at $z\sim1.5$. 

One possible explanation for our slightly lower halo mass estimates is the difference in flux limits across surveys. Both of these studies adopted a flux density threshold of $S_{\rm{1.4GHz}}>1\ {\rm{mJy}}$. Similarly, \cite{magliocchetti2017} imposed a $S_{\rm{1.4GHz}}>0.15\ {\rm{mJy}}$ flux density limit. The minimum fluxes selected by our luminosity threshold for increasing redshift bins are $S_{\rm{1.4GHz}}>0.18$, $0.13$ and $0.17\ \rm{mJy}$. Higher flux thresholds will preferentially select more luminous radio-AGN, especially at high redshift, than the deeper MIGHTEE sample (compare Figure~\ref{fig:early_science_agn_threshold} to Figure 3 of \citealt{retana2017}). Previous works have found that the clustering strength is correlated with luminosity \citep[e.g.,][]{bardelli2010, lindsay2014a, hale2018, croston2019, mo2020}, implying that more luminous AGN occupy more massive haloes. In support of this explanation, \cite{hale2018}, who measured the angular auto-correlation of radio sources from VLA observations of the COSMOS field at a lower flux limit of $\sim13\ {\mu}{\rm{Jy\ beam^{-1}}}$ at $3\ {\rm{GHz}}$ ($0.02\ {\rm{mJy\ beam^{-1}}}$ at $1.4\ {\rm{GHz}}$ assuming $\alpha=-0.7$) also found lower halo mass estimates of $\log_{10}(M_{\rm{h}}/{\rm{M_\odot}})\sim13.1-13.3$ for AGN with $L_{\rm{3GHz}}>10^{23}\ {\rm{W\ Hz^{-1}}}$. This would also explain the lower mass estimates of \cite{mandelbaum2009} whose sample extended to $L_{\rm{1.4GHz}}\approx10^{23}\ {\rm{W\ Hz^{-1}}}$ compared to our threshold of $L_{\rm{AGN}}(z=0)\approx10^{23.4}\ {\rm{W\ Hz^{-1}}}$.

Our finding that the typical host halo mass of radio-AGN increases at later times is an important result of this work. This trend may reflect the greater abundance of cold gas for accretion at earlier epochs which would allow for radio-AGN activity to occur in lower-mass haloes. Such a scenario would account for the strong evolution in the comoving space number density of powerful radio sources \citep{dunlop1990, jarvis2001, rigby2011}. Previous clustering studies have found little evolution in the environments of radio-AGN out to $z\sim2.5$ \cite[see Section 3.2 of][]{magliocchetti2017}. Though \cite{hale2018} observed a flattening of the bias parameter with increasing redshift which would suggest a decrease in typical host halo mass.

\begin{figure*}
 \includegraphics[width=2\columnwidth]{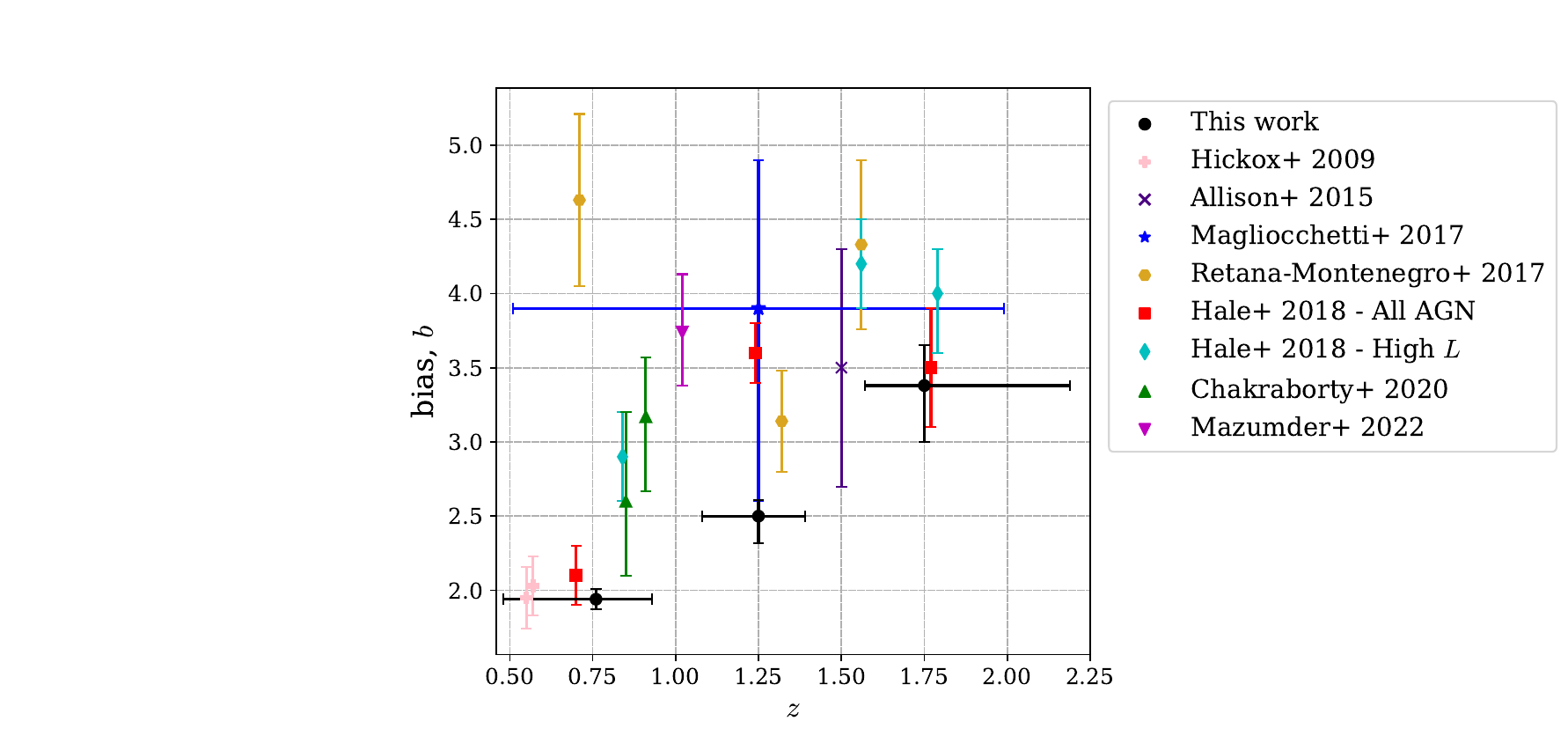}
 \caption{Radio-AGN bias estimates from this work (black circles) and previous clustering studies as a function of redshift. These include bias estimates from: \citealt{hickox2009} (pink plusses); \citealt{allison2015} (purple cross); \citealt{magliocchetti2017} (blue star); \citealt{retana2017} (gold hexagons); the full AGN population from \citealt{hale2018} (red squares), and high-luminosity AGN ($L_{\rm{3GHz}}>10^{23}\ {\rm{W\ Hz^{-1}}}$; cyan diamonds); \citealt{chakraborty2020} (green triangles); and \citealt{mazumder2022} (downwards-pointing magenta triangles).}
 \label{fig:bias}
\end{figure*}

Figure~\ref{fig:bias} presents our estimates of the AGN bias parameter as a function of redshift, showing a clear increase in bias towards earlier times. Included in the plot are results obtained by previous radio-AGN clustering studies at a range of observing frequencies. These include the results from \cite{hickox2009}, \cite{allison2015}, \cite{magliocchetti2017}, \cite{retana2017} and \cite{hale2018}. Also included are the bias estimates of \cite{chakraborty2020}, who investigated the clustering of radio-selected AGN at $400\ {\rm{MHz}}$ and $612\ {\rm{MHz}}$ using archival data from the upgraded Giant Metrewave Radio Telescope \citep[uGMRT;][]{chakraborty2019}; and \cite{mazumder2022}, who investigated the clustering of radio sources at $325\ {\rm{MHz}}$ using archival GMRT data \citep{swarup1991}. Excluded from the plot are the anomalously high bias values of \cite{lindsay2014a} who cross-correlated VLA radio sources with optical and NIR galaxies and found $b(z=1.35)=7.62\pm1.27$, $b(z=1.55)=9.91\pm2.48$ and $b(z=1.77)=11.14\pm3.01$ for luminosity thresholds of $L_{\rm{1.4GHz}}>10^{23}$, $10^{23.5}$ and $10^{24}\ {\rm{W\ Hz^{-1}}}$, respectively. The most directly comparable study to this work is that of \cite{hale2018}, which used a source sample of similar depth and comparable r.m.s. radio sensitivity to that of the MIGHTEE dataset.

For our lowest redshift bin, we find a bias of $b=1.94^{+0.07}_{-0.07}$ at $z_{\rm{med}}=0.76^{+0.17}_{-0.28}$, in agreement with the value reported by \cite{hale2018} of $b=2.1\pm0.2$ for AGN with $z<1$ at a comparable median redshift of $z_{\rm{med}}=0.70$.
\citeauthor{hale2018} also report a bias of $b=2.9^{+0.3}_{-0.3}$ at $z_{\rm{med}}=0.84$ for a high-luminosity subsample selected with $L_{\rm{3GHz}}>10^{23}\ {\rm{W\ Hz^{-1}}}$, equivalent to $L_{\rm{1.4GHz}}>10^{23.2}\ {\rm{W\ Hz^{-1}}}$. That luminosity cut lies below our $L_{\rm{AGN}}(z)$ threshold over the redshift range of the bin, so this is the population that is most similar to the sample used in this work at this redshift, and we find our measured bias to be lower than theirs.
In the intermediate redshift range, $1<z<1.5$, we find $b=2.50^{+0.11}_{-0.18}$ at $z_{\rm{med}}=1.25^{+0.14}_{-0.17}$ which is significantly lower than the $b=3.6\pm0.2$ at $z_{\rm{med}}=1.24$ reported by \citeauthor{hale2018} for their full AGN sample across their whole redshift range ($\Delta z \approx 4$). 
For our highest redshift bin, $1.5<z<2.5$, we obtain $b=3.38^{+0.27}_{-0.38}$ at $z_{\rm{med}}=1.75^{+0.44}_{-0.18}$, in agreement with their $b=3.5\pm0.4$ at $z_{\rm{med}}=1.77$ for AGN with $z\ge1$. Our result is also consistent with their high luminosity sample bias value of $b=4.0^{+0.3}_{-0.4}$ at $z_{\rm{med}}=1.79$.

The lower bias values found in this work may stem from several factors. First, we used a narrower redshift bins which reduces the blending of potentially evolving clustering signals over cosmic time. Since radio-AGN are selected by their luminosity alone, another possibility is that the sample is contaminated with star-forming galaxies, which are known to be less clustered than AGN \citep[e.g.,][]{magliocchetti2017, hale2018, chakraborty2020}. To test this, we experimented with increasing the luminosity threshold used to select AGN from the minimum luminosity above which $95\%$ of sources in \cite{whittam2022} were classified as AGN, to the luminosity above which $100\%$ of sources were classified as AGN (see Section~\ref{ssec:selecting_AGN}). No significant increase in the bias parameter was found.

\cite{hale2018}, \cite{chakraborty2020} and \cite{mazumder2022} all obtain their bias estimates by fitting a power-law to the angular auto-correlation function of the form $\omega(\theta)\propto\theta^{-0.8}$. The most likely explanation for the lower bias values is that, instead of fitting a simple power-law to the entire clustering signal, in this work the one- and two-halo term contributions are modelled separately, resulting in a more accurate representation of the true clustering signal. Fitting a power-law (i.e., a straight line in $\log\omega$-$\log\theta$) does not account for the dip in clustering amplitude at intermediate scales where the one- and two-halo regimes meet. Because the majority of the data points lie near the peaks of the one- and two-halo contributions, the power-law fit will be pulled up to match those measurements. This would result in a poorer fit at intermediate scales and an overestimation of the clustering amplitude, leading to a higher inferred bias.

{\renewcommand{\arraystretch}{1.2}
\begin{table}
\centering
 \caption{The values of the typical halo mass, $\langle M_{\rm{h}} \rangle$, the satellite fraction, $f_{\rm{sat}}$, and the bias parameter, $b$, for the AGN and matched galaxy samples in each redshift bin. To obtain a single value in each bin, we marginalize $\langle M_{\rm{h}} \rangle (z)$, $f_{\rm{sat}}(z)$, and $b(z)$ over the respective redshift distributions. Uncertainties are obtained by sampling from the posteriors of the HOD parameters and taking the $16^{\rm{th}}$ and $84^{\rm{th}}$ percentiles.}
 \begin{tabular}{lccr}
  \hline
  Sample &  $\log_{10}(\langle M_{\rm{h}} \rangle / {\rm{M_\odot}})$ & $f_{\rm{sat}}$ & $b$\\
  \hline
  $0\,<\,z<1$\\
  \hline
AGN&$13.44^{+0.08}_{-0.08}$&$0.18^{+0.16}_{-0.09}$&$1.94^{+0.07}_{-0.07}$\\
Matched&$13.25^{+0.08}_{-0.09}$&$0.18^{+0.18}_{-0.10}$&$1.69^{+0.07}_{-0.06}$\\
  \hline
  $1\,<\,z<1.5$\\
  \hline
  AGN&$13.17^{+0.07}_{-0.06}$&$0.06^{+0.21}_{-0.05}$&$2.50^{+0.11}_{-0.18}$\\
Matched&$13.12^{+0.06}_{-0.06}$&$0.15^{+0.16}_{-0.12}$&$2.33^{+0.14}_{-0.19}$\\
  \hline
  $1.5\,<\,z<2.5$\\
  \hline
  AGN&$13.03^{+0.09}_{-0.10}$&$0.06^{+0.09}_{-0.05}$&$3.38^{+0.27}_{-0.38}$\\
Matched&$12.77^{+0.14}_{-0.17}$&$0.05^{+0.08}_{-0.05}$&$2.79^{+0.32}_{-0.38}$\\
  \hline
 \end{tabular}
 \label{tab:derived_params}
\end{table}
}

\subsubsection{AGN duty cycle}\label{sssec:duty_cycle}

In addition to constraining the halo mass and bias from clustering, the HOD framework can be used to estimate the AGN  duty cycle. This is the fraction of time during which a SMBH is observable as a radio-AGN given the luminosity threshold. At any redshift, only a fraction $f_{\rm{DC}}$ of galaxies hosting a sufficiently massive SMBH will be radio-loud and included in the sample. Assuming that the fraction of active SMBHs is random, and not correlated with halo mass, this intermittency can be implemented in the HOD framework by rescaling the central galaxy occupation \citep[e.g.,][]{miyaji2011, krumpe2015}:
\begin{equation}
    \overline{N}_{\rm{c}}(M_{\rm{h}}) \rightarrow f_{\rm{DC}} \overline{N}_{\rm{c}}(M_{\rm{h}}).
\end{equation}
However, introducing a mass-independent multiplicative factor simply renormalizes both the model pair counts and the mean number density. Pair counts in the one- and two-halo terms scale with the products of the occupation numbers (for example, $\overline{N}_{\rm{c}}\times\overline{N}_{\rm{c}}$ or $\overline{N}_{\rm{c}}\times\overline{N}_{\rm{s}}$; see Appendix~\ref{sec:cross_correlation_model}), so they pick up the same overall factor of $f_{\rm{DC}}$ as the mean galaxy number density (Equation~\ref{eqn:mean_number_density}). Because the TPCF is normalized by the mean density \citep{nicola2020}, this global factor cancels out, and the normalized clustering signal is unchanged. Consequently, a mass-independent $f_{\rm{DC}}$ cannot be constrained by fitting to the clustering measurements alone.

Instead, $f_{\rm{DC}}$ can be estimated from the number densities. A straightforward approach is to compare the ratio of the number density of observed AGN to that of potential host galaxies (i.e., galaxies in haloes massive enough to host such AGN). The observed number is found by summing the weights, while the number density of potential host galaxies is found by marginalizing Equation~\ref{eqn:mean_number_density} for the best-fitting AGN HOD parameters over the redshift distribution, such that
\begin{equation}
    f_{\rm{DC}}=\frac{\sum_i w_{{\rm{AGN}},i}}{\overline{n}^{\rm{mod}}_{\rm{AGN}}V_{\rm{tot}}},
\end{equation}
where $V_{\rm{tot}}$ is the total comoving volume covered by the fields between the limits of the redshift bin. The observed AGN number density can be directly taken from the weights provided the selection is volume-limited for the chosen luminosity threshold and redshift bin. In this case, the sample is complete across the bin and no AGN are missed purely due to sensitivity. If the selection of AGN was flux-limited, the completeness would vary with redshift, and the direct counts would underestimate the true number density. As evident in Figure~\ref{fig:early_science_agn_threshold}, the selection threshold, $L_{\rm{AGN}}(z)$, remains above the survey flux limit at all redshifts in each bin.

One caveat of this method is the fact that approximately $10\%$ of the radio sources were not successfully cross-matched with their optical counterparts and are excluded from the sample, slightly lowering the duty cycle estimates. These unmatched sources are predominantly low-SNR detections, while the luminosity threshold primarily selects high-SNR sources (see Section~\ref{sssec:random_catalogue}), so the impact of this incompleteness is minimal.

Sampling from the HOD parameter posteriors, we find $f_{\rm{DC}}=0.09^{+0.03}_{-0.03}$, $0.08^{+0.04}_{-0.06}$ and $0.05^{+0.04}_{-0.03}$ for the $0<z<1$, $1<z<1.5$ and $1.5<z<2.5$ redshift bins, respectively. These estimates are consistent with the results of \cite{petter2024}, who calculated the duty cycle of low-frequency radio-galaxies with $L_{\rm{150MHz}}\gtrsim10^{25.25}\ {\rm{W\ Hz^{-1}}}$ (equivalent to $\gtrsim10^{24.6}\ {\rm{W\ Hz^{-1}}}$ at $1.4\ {\rm{GHz}}$) which is similar to $L_{\rm{AGN}}(z)$ in the $1.5<z<2.5$ redshift bin. They found that the duty cycle increases from $\sim5\%$ at $z\sim1.5$ up to $\sim10-20\%$ for $z\lesssim1$.

The fraction of galaxies that host radio-AGN is a strong function of galaxy mass, increasing as $\propto M_\ast^{2.5}$ \citep[e.g.,][]{best2005} in the local Universe. It is therefore important to quote host galaxy stellar masses alongside duty cycle estimates. Sampling from the $z$-PDFs of radio sources and retaining only samples where the corresponding luminosity exceeds $L_{\rm{AGN}}(z)$, we find median stellar masses for increasing redshift bins of $\log_{10}(M_{\ast,{\rm med}}/{\rm M_\odot})=10.95^{+0.29}_{-0.52}$, $10.82^{+0.34}_{-0.47}$ and $10.82^{+0.35}_{-0.52}$.

Given this strong stellar mass dependence, an effective duty cycle can also be inferred by calculating the ratio of the number of radio-AGN to the total number of galaxies within stellar mass ranges. At the median stellar masses of our samples, we find that this alternative approach yields duty cycle estimates that are lower by a factor of $\sim2-4$, similar to the AGN fractions reported in previous studies at these masses \citep[e.g.,][]{best2005, sabater2019, kondapally2025}. In Section~\ref{sssec:compare_agn_and_matched_hod_results}, we show that stellar mass-matched control galaxies have different HOD parameters than AGN hosts, so AGN halo occupation is not determined by stellar mass alone. The denominator in the stellar mass-based duty cycle therefore includes galaxies in a larger range of halo environments, whereas the HOD estimate is restricted to the subset of haloes consistent with the clustering signal, leading to a higher inferred duty cycle.

The duty cycle can be used to calculate a characteristic timescale, $\tau$, of the observable AGN phase. It can be estimated by multiplying the measured duty cycle by the cosmic time interval spanned by the redshift bin limits, $\Delta t$: $\tau=f_{\rm{DC}} \Delta t$. This represents the total time that an AGN would be radiating above the luminosity threshold between the redshift limits of the bin, not the duration of a single episode of activity. We find $\tau\approx670\ {\rm{Myr}}$ for $0<z<1$ (which spans $\Delta t=7.7\ {\rm{Gyr}}$), $\tau\approx130\ {\rm{Myr}}$ for $1<z<1.5$ ($\Delta t=1.6\ {\rm{Gyr}}$) and $\tau\approx76\ {\rm{Myr}}$ for $1.5<z<2.5$ ($\Delta t=1.6\ {\rm{Gyr}}$). The sum of these characteristic time scales of $\sim880\ {\rm{Myr}}$ over $0<z<2.5$ is similar to the value of $\sim1\ {\rm{Gyr}}$ derived by \cite{magliocchetti2017} for $0<z<2.3$ and a similar luminosity threshold. Given that the lifetimes of radio-bright AGN are limited to a few $\times10\ {\rm{Myr}}$ \citep{blundell1999}, this suggests that AGN undergo multiple radio-loud episodes during this period. Evidence for such recurrent activity is observed in `restarting' AGN, where multiple generations of radio jets are seen within the same source \citep[e.g.,][]{lara1999, schoenmakers2000, saikia2009, nandi2012, mahatma2023}.

\subsubsection{Energy deposited by jet heating}\label{sssec:heating}

We now quantify the energy deposited by radio-AGN into the surrounding gas via heating from their jets. The total kinetic energy emitted per halo over the time period $\Delta t$ is roughly estimated as:
\begin{equation}
    \Delta E_{\rm{kin}}=\frac{\int_{L_{\rm{AGN}}(z)}^\infty {\rm{d}}\log{L} \, \Psi(L,z)}{\int_0^\infty {\rm{d}}M \, n(M,z) \overline{N}_{\rm{c}}(M)} \times \Delta t,
\end{equation}
where $\Psi(L,z)$ is the kinetic heating rate density for radio-excess AGN from \cite{kondapally2023}, derived by combining the evolving luminosity functions of \cite{kondapally2022} with the $1.4\ {\rm{GHz}}$ to kinetic jet power relation of \cite{heckman2014}. The latter is based on the results of \cite{birzan2008} and \cite{cavagnolo2010}, and was calibrated by considering the work required to inflate X-ray cavities. The numerator in the fraction represents the total jet power density of AGN radiating above the luminosity selection threshold $L_{\rm{AGN}}(z)$, while the denominator gives the comoving number density of haloes capable of hosting such AGN. Since the kinetic heating is dominated by sources with $L_{\rm{1.4GHz}}\gtrsim10^{25.5}\ {\rm{W\ Hz^{-1}}}$ \citep{kondapally2023} which is well above $L_{\rm{AGN}}(z)$, essentially all of the heating power is captured.

For our increasing redshift bins, we find $\Delta E_{\rm{kin}}\approx4\times10^{53}$, $1\times10^{53}$ and $2\times10^{53}\ {\rm{J}}$ per halo. Reproducing the observed entropy and thermal properties of the intra-cluster medium requires additional heating of  $0.5-1\ {\rm{keV}}$ per gas particle beyond that of gravitational collapse \citep{borgani2002, kravtsov2012}. Assuming a halo mass of $10^{13.3}\ {\rm{M_\odot}}$, a gas mass fraction of $5\%$ \citep{popesso2024}, and  $N_{\rm{gas}}=M_{\rm{gas}}/(\mu_{\rm{p}} m_{\rm{p}})$ with mean molecular mass $\mu_{\rm{p}}=0.6$ for ionized gas \citep[e.g.,][]{wu1999} and proton mass $m_{\rm{p}}$, this corresponds to $2-3\times10^{53}\ {\rm{J}}$. Therefore, the energy provided by jets over $0<z<2.5$ is sufficient to account for this additional heating.

\subsubsection{Comparison of the AGN and matched galaxy sample best-fitting HOD parameters}\label{sssec:compare_agn_and_matched_hod_results}

In this section, we compare the best-fitting HOD parameters and derived quantities between the AGN and matched galaxy samples. For the matched galaxy sample cross-correlations, our best-fitting models yield reduced chi-squared values of $\chi^2_\nu=0.84$, $0.99$ and $3.43$ for increasing redshift bins. We consider the higher $\chi^2_\nu$ in the highest redshift bin acceptable given that we use a simplified three-parameter model, and fit a single, non-evolving set of HOD parameters over a broad redshift range, ${\Delta}z=1$.

The only statistically significant difference in the individual HOD parameters between the two samples is in their $M_{\rm{min}}$ values for the $0<z<1$ redshift bin. For the AGN sample, $\log_{10}(M_{\rm{min}}/{\rm{M_\odot}})=12.81^{+0.09}_{-0.13}$ compared to $\log_{10}(M_{\rm{min}}/{\rm{M_\odot}})=12.52^{+0.12}_{-0.21}$ for the matched galaxy sample. Due to their large uncertainties, the satellite parameters, $M_1$ and $\alpha_{\rm{s}}$, are consistent within the uncertainties between the AGN and matched galaxy sample for all redshift bins. 

Despite the overlap in the uncertainties of the individual HOD parameters, the values of the derived parameters given in Table~\ref{tab:derived_params} exhibit significant differences. This is because much of the uncertainty in $M_{\rm{min}}$ and $M_1$ is a result of their degeneracy, as evidenced by the elongated contours in the corner plots (see Appendix~\ref{sec:corner_plots}). The largest difference in $\langle{M_{\rm{h}}}\rangle$ between the AGN and matched galaxy samples is in the highest redshift bin, $1.5<z<2.5$: for AGN, $\log_{10}(\langle{M_{\rm{h}}}\rangle/{\rm{M_\odot}})=13.03^{+0.09}_{-0.10}$ compared to $\log_{10}(\langle{M_{\rm{h}}}\rangle/{\rm{M_\odot}})=12.77^{+0.14}_{-0.17}$ for the matched galaxy sample. This means that the typical halo mass hosting galaxies with radio-AGN is approximately $1.82^{+1.04}_{-0.57}$ times greater than that of non-active galaxies with similar stellar masses. The significant difference in $\langle{M_{\rm{h}}}\rangle$ persists for the lowest redshift bin, $0<z<1$. Here, the AGN sample gives $\log_{10}(\langle{M_{\rm{h}}}\rangle/{\rm{M_\odot}})=13.44^{+0.08}_{-0.08}$ compared to $\log_{10}(\langle{M_{\rm{h}}}\rangle/{\rm{M_\odot}})=13.25^{+0.08}_{-0.09}$ for the matched galaxy sample, differing by a factor of $1.54^{+0.47}_{-0.33}$. This is consistent with the results of \cite{mandelbaum2009} who found that the dark matter halo masses of low redshift ($z<0.3$), radio-loud AGN were about twice as massive as those of a control sample matched in stellar mass. However, for the intermediate redshift bin, $1<z<1.5$, the two values of $\langle{M_{\rm{h}}}\rangle$ overlap within the uncertainties: $\log_{10}(\langle{M_{\rm{h}}}\rangle/{\rm{M_\odot}})=13.17^{+0.07}_{-0.06}$ for AGN and $\log_{10}(\langle{M_{\rm{h}}}\rangle/{\rm{M_\odot}})=13.12^{+0.06}_{-0.06}$ for the matched galaxy sample, giving a ratio of $1.11^{+0.25}_{-0.20}$. Though, it is worth noting that all three ratios are statistically consistent with each other within the uncertainties. 

It is unclear whether the halo mass excess seen in AGN hosts is a driver of AGN activity, or a consequence of it. On one hand, radio-AGN activity may be enhanced in overdense structures. More massive haloes have deeper gravitational potential wells which will more efficiently channel gas toward the central galaxy, providing fuel for the SMBH. Such environments are also more likely to host mechanisms that promote gas inflows, such as galaxy interactions \citep[e.g.,][]{alonso2007, sabater2013, goulding2018} and mergers \citep[e.g.,][]{fanidakis2010, treister2012, gao2020, pierce2022} which are known from simulations to be efficient at driving gas towards galaxy centres \citep{springel2005}, triggering AGN activity. 

On the other hand, the observed halo mass excess could be due to the cumulative effect of feedback; by suppressing star formation, jet-mode feedback will reduce stellar mass while leaving halo mass unchanged \citep{croton2006, bower2006, heckman2014, hardcastle2020, scharre2024}. As a result, AGN host galaxies would be assigned to a lower stellar mass bin despite residing in more massive halos. At fixed stellar mass, this effect would naturally produce a clustering excess relative to non-active galaxies. A potential caveat is that the active lifetimes of radio-AGN are on the order of a few $\times10\ {\rm{Myr}}$ \citep{blundell1999} whereas galaxy quenching occurs on longer timescales of $\sim500\ {\rm{Myr}}$ \citep{lian2016}. However, our estimate of the cumulative active lifetime of radio-AGN of $\sim1\ {\rm{Gyr}}$ over $0<z<2.5$ suggests that repeated cycles of activity could be sufficient to sustain this quenching process. 

The higher derived halo masses may also reflect assembly bias, whereby clustering depends on properties beyond halo mass. In standard HOD modelling, it is assumed that the clustering of galaxies depends solely on halo mass, but in reality haloes of the same mass can differ in clustering strength depending on their formation history \citep{gao2005, wechsler2006, croton2007, gao2007}. At fixed mass, haloes that formed earlier have a higher clustering amplitude than haloes that formed more recently \citep{navarro1995, gao2005, wechsler2006}. If AGN preferentially reside in such early-forming haloes, they would appear both more clustered and have had a longer period over which their central black holes could grow.

Distinguishing between these scenarios is challenging within the standard HOD framework, which assumes that clustering depends only on halo mass, and it is likely that a combination of these effects contributes to the observed difference.

An additional possibility is that differences in black hole mass contribute to the enhanced clustering of radio-AGN. There is some evidence that AGN powered by more massive black holes cluster more strongly \citep[e.g.,][]{hatch2014, retana2017}. Since the fraction of galaxies that are radio-loud increases with black hole mass \citep[e.g.,][]{mclure2004, best2005, metcalf2006, whittam2022, Jackson2026}, and our AGN sample is selected by radio luminosity, the observed clustering excess could therefore reflect systematic differences in black hole mass. However, because black hole mass correlates with host galaxy stellar mass \citep[e.g.,][]{magorrian1998, haring2004, mcconnell2013, kormendy2013}, our stellar mass matching should largely remove this effect. Consequently, black hole mass would only play an additional role if jet power depends more strongly on black hole mass than black hole mass does on stellar mass.

We also note that it is likely that many of the galaxies in the matched sample also underwent radio-loud AGN activity at earlier epochs. This would also dilute any differences we are able to measure in the clustering of current radio-loud AGN and the radio-dormant AGN in the control sample. The true difference in the halo masses of stellar mass-matched galaxies with and without AGN may therefore be greater.

\section{Conclusion} \label{sec:conclusion}

In this work, we compared the environments of radio-AGN to those of control galaxies matched in both stellar mass and redshift. Although it is well established that radio-AGN reside in the most massive dark matter haloes \citep{magliocchetti2004, mandelbaum2009, allison2015, magliocchetti2017, retana2017, hale2018}, our aim was to determine whether this is solely a consequence of their preference for massive host galaxies, or whether radio-AGN activity is linked to some additional environmental preference.

Using radio sources from the MIGHTEE radio survey that had been cross-matched with their optical/NIR counterparts from VISTA and HSC, we constructed a sample of $\sim2000$ AGN selected by their radio luminosities. A control sample of galaxies matched in both stellar mass and redshift was drawn from the optical/NIR galaxy catalogue, accounting for the full photometric redshift PDFs. These samples were divided into three redshift ranges: $0<z<1$, $1<z<1.5$ and $1.5<z<2.5$.

Measurements of the angular two-point auto-correlation functions showed no significant difference between the AGN and matched galaxy samples. This motivated our use of the cross-correlations with the full galaxy population, where the increased pair counts reduces statistical uncertainties, and we found that AGN are more clustered than the control sample.

By fitting a HOD model to the measured clustering signals, the statistical properties of the dark matter haloes that our samples inhabit were investigated. AGN were found to occupy dark matter haloes with masses $\log_{10}(M_{\rm{h}}/{\rm{M_\odot}})\gtrsim12.8$. Unlike some previous studies that report no significant evolution in the environmental properties of radio-AGN \citep[e.g.,][]{magliocchetti2017}, we find a clear decrease in the typical halo mass of radio-AGN with redshift: for increasing redshift bins, $\log_{10}(\langle M_{\rm{h}} \rangle/{\rm{M_\odot}})=13.44^{+0.08}_{-0.08}$, $13.17^{+0.07}_{-0.06}$ and $13.03^{+0.09}_{-0.10}$. We interpret this trend as a consequence of the higher abundance of cold gas at earlier epochs, which could more readily fuel AGN activity. This would also explain the strong evolution in the comoving space number density of powerful radio sources \citep{dunlop1990, jarvis2001, rigby2011}. The masses we obtain are broadly consistent with previous works \citep{hickox2009, mandelbaum2009, allison2015, magliocchetti2017, retana2017, hale2018}, though lie at the lower end of mass estimates. This may be because MIGHTEE has a lower flux limit than previous surveys which would preferentially select lower-luminosity radio-AGN. Our halo masses suggest that radio-AGN typically reside in galaxy groups rather than higher mass clusters \citep{lim2017, calderon2019}. 

For increasing redshift bins, we find bias parameters for radio-AGN of $b=1.94^{+0.07}_{-0.07}$, $2.50^{+0.11}_{-0.18}$ and $3.38^{+0.27}_{-0.38}$. Comparing these bias estimates with those found by \cite{hale2018}, our measurements in the lowest and highest redshift bins are consistent with their values within the uncertainties. However, for the intermediate redshift bin, our value is significantly lower. We attribute this to the use of full HOD modelling, which explicitly models the one- and two-halo contributions to the clustering signal, rather than using a simple power-law fit.

By comparing the ratio of the number of observed radio-AGN to the number of galaxies in haloes massive enough to host a SMBH predicted by the HOD model, a duty cycle of $\sim5-9\%$ is inferred. Given that the active lifetime of radio-AGN is limited to a few $\times10\ {\rm{Myr}}$ \citep{blundell1999}, this implies that each AGN undergoes multiple periods of radio-loud activity over the redshift ranges considered. We estimate that the cumulative energy deposited into the intergalactic medium by radio jets since $z=2.5$ is $\sim10^{53}\ {\rm{J}}$. This is sufficient to account for the observed excess energy in the gas of galaxy groups assuming $0.5-1\ {\rm{keV}}$ of additional heating per gas particle above that of gravitational collapse.

Comparing the typical halo masses that we derive for the AGN and matched galaxy samples, we find that for $0<z<1$, the typical dark matter halo mass of radio-AGN is $1.54^{+0.47}_{-0.33}$ times greater than that of galaxies matched in stellar mass and redshift. A similarly significant excess is seen at $1.5<z<2.5$, with a ratio of $1.82^{+1.04}_{-0.57}$. The halo mass excess of radio-AGN may be a result of AGN feedback which would suppress star formation and result in lower stellar masses for the same halo mass. Alternatively, it could indicate that galaxies hosting radio-AGN preferentially reside in earlier-forming and more clustered haloes, which have had more time to accrete gas and grow their central black hole.

The main limitation of this work is the number of AGN, which leads to large statistical uncertainties in the measured correlation functions. A larger sample would allow for tighter constraints on the HOD parameters and the use of narrower redshift bins to better trace the evolution of radio-AGN environments. It would also allow for a robust comparison between low- and high-luminosity AGN populations. Radio clustering studies have been limited by the lack of overlapping multi-wavelength data that are required to obtain the redshifts of host galaxies. Current and upcoming wide-area radio surveys such as with the Low-Frequency Array \citep[LOFAR;][]{haarlem2005}, the Australian Square Kilometre Array Pathfinder radio telescope \citep[ASKAP;][]{norris2011} and ultimately the Square Kilometre Array Observatory\footnote{\href{https://www.skao.int/en}{https://www.skao.int/en}} (SKAO) will produce vastly increased source counts, with photometric redshift information provided by the next generation of surveys with LSST \citep{lsst2009} and Euclid \citep{euclid2025} together with spectroscopic programmes like DESI \citep{levi2013} and WEAVE-LOFAR \citep{smith2016}.

\section*{Acknowledgements}
JH acknowledges funding from the Science and Technology Facilities Council (STFC) [grant code ST/Y509474/1]. MJJ, CLH and IHW acknowledge support from the Hintze Family Charitable Foundation through the Oxford Hintze Centre for Astrophysical Surveys. CLH also acknowledges support from STFC [ST/Y000951/1]. MJJ acknowledges the support of the STFC consolidated grant [ST/S000488/1] and [ST/W000903/1]. MJJ and NS acknowledge funding from a UKRI Frontiers Research Grant [EP/X026639/1]. DA acknowledges support from the Beecroft Trust. We would like to thank the anonymous referee for their useful comments that have helped to improve this paper.

The MeerKAT telescope is operated by the South African Radio Astronomy Observatory, which is a facility of the National Research Foundation, an agency of the Department of Science and Innovation. We acknowledge the use of the ilifu cloud computing facility – www.ilifu.ac.za, a partnership between the University of Cape Town, the University of the Western Cape, Stellenbosch University, Sol Plaatje University and the Cape Peninsula University of Technology. The Ilifu facility is supported by contributions from the Inter-University Institute for Data Intensive Astronomy (IDIA – a partnership between the University of Cape Town, the University of Pretoria and the University of the Western Cape, the Computational Biology division at UCT and the Data Intensive Research Initiative of South Africa (DIRISA). The authors acknowledge the Centre for High Performance Computing (CHPC), South Africa, for providing computational resources to this research project.

This work is based on data products from observations made with ESO Telescopes at the La Silla Paranal Observatory under ESO programme ID 179.A-2005 (Ultra-VISTA) and ID 179.A-2006(VIDEO) and on data products produced by CALET and the Cambridge Astronomy Survey Unit on behalf of the Ultra-VISTA and VIDEO consortia.

\section*{Data Availability}

The MIGHTEE continuum DR1 data used in this work are accessible through \href{https://doi.org/10.48479/7msw-r692}{\texttt{https://doi.org/10.48479/7msw-r692}} with information given in \cite{hale2025}. The cross-matched catalogues will be made available by Hale et al. (in preparation), and the multi-wavelength data catalogues will be made available by Stylianau et al. (in preparation). The results presented in this work can be obtained through a reasonable request to the author.



\bibliographystyle{mnras}
\bibliography{references} 

@ARTICLE{Jackson2026,
       author = {{Jackson}, Charlotte L. and {Matthews}, James H. and {Whittam}, Imogen H. and {Jarvis}, Matt J. and {Temple}, Matthew J. and {Rankine}, Amy L. and {Hewett}, Paul C.},
        title = "{Exploring the quasar disc-wind-jet connection with LoTSS and SDSS}",
      journal = {\mnras},
         year = 2026,
        month = jan,
          doi = {10.1093/mnras/stag065},
archivePrefix = {arXiv},
       eprint = {2510.25833},
 primaryClass = {astro-ph.GA},
       adsurl = {https://ui.adsabs.harvard.edu/abs/2026MNRAS.tmp...59J}
}

@ARTICLE{BC03,
       author = {{Bruzual}, G. and {Charlot}, S.},
        title = "{Stellar population synthesis at the resolution of 2003}",
      journal = {\mnras},
         year = 2003,
        month = oct,
       volume = {344},
       number = {4},
        pages = {1000-1028},
          doi = {10.1046/j.1365-8711.2003.06897.x},
archivePrefix = {arXiv},
       eprint = {astro-ph/0309134},
 primaryClass = {astro-ph},
       adsurl = {https://ui.adsabs.harvard.edu/abs/2003MNRAS.344.1000B}
}

@ARTICLE{Eales1997,
       author = {{Eales}, Stephen and {Rawlings}, Steve and {Law-Green}, Duncan and {Cotter}, Garret and {Lacy}, Mark},
        title = "{A first sample of faint radio sources with virtually complete redshifts - I. Infrared images, the Hubble diagram and the alignment effect}",
      journal = {\mnras},
         year = 1997,
        month = nov,
       volume = {291},
       number = {4},
        pages = {593-615},
          doi = {10.1093/mnras/291.4.593},
archivePrefix = {arXiv},
       eprint = {astro-ph/9701023},
 primaryClass = {astro-ph},
       adsurl = {https://ui.adsabs.harvard.edu/abs/1997MNRAS.291..593E}
}

@ARTICLE{Jarvis2001_Kz,
       author = {{Jarvis}, Matt J. and {Rawlings}, Steve and {Eales}, Steve and {Blundell}, Katherine M. and {Bunker}, Andrew J. and {Croft}, Steve and {McLure}, Ross J. and {Willott}, Chris J.},
        title = "{A sample of 6C radio sources designed to find objects at redshift z>4 - III. Imaging and the radio galaxy K-z relation}",
      journal = {\mnras},
         year = 2001,
        month = oct,
       volume = {326},
       number = {4},
        pages = {1585-1600},
          doi = {10.1111/j.1365-2966.2001.04730.x},
archivePrefix = {arXiv},
       eprint = {astro-ph/0106130},
 primaryClass = {astro-ph},
       adsurl = {https://ui.adsabs.harvard.edu/abs/2001MNRAS.326.1585J}
}

@ARTICLE{Endsley2022,
       author = {{Endsley}, Ryan and {Stark}, Daniel P. and {Fan}, Xiaohui and {Smit}, Renske and {Wang}, Feige and {Yang}, Jinyi and {Hainline}, Kevin and {Lyu}, Jianwei and {Bouwens}, Rychard and {Schouws}, Sander},
        title = "{Radio and far-IR emission associated with a massive star-forming galaxy candidate at z ≃ 6.8: a radio-loud AGN in the reionization era?}",
      journal = {\mnras},
         year = 2022,
        month = may,
       volume = {512},
       number = {3},
        pages = {4248-4261},
          doi = {10.1093/mnras/stac737},
archivePrefix = {arXiv},
       eprint = {2108.01084},
 primaryClass = {astro-ph.GA},
       adsurl = {https://ui.adsabs.harvard.edu/abs/2022MNRAS.512.4248E}
}

@ARTICLE{Smolcic2017,
       author = {{Smol{\v{c}}i{\'c}}, V. and {Delvecchio}, I. and {Zamorani}, G. and {Baran}, N. and {Novak}, M. and {Delhaize}, J. and {Schinnerer}, E. and {Berta}, S. and {Bondi}, M. and {Ciliegi}, P. and {Capak}, P. and {Civano}, F. and {Karim}, A. and {Le Fevre}, O. and {Ilbert}, O. and {Laigle}, C. and {Marchesi}, S. and {McCracken}, H.~J. and {Tasca}, L. and {Salvato}, M. and {Vardoulaki}, E.},
        title = "{The VLA-COSMOS 3 GHz Large Project: Multiwavelength counterparts and the composition of the faint radio population}",
      journal = {\aap},
         year = 2017,
        month = jun,
       volume = {602},
          eid = {A2},
        pages = {A2},
          doi = {10.1051/0004-6361/201630223},
archivePrefix = {arXiv},
       eprint = {1703.09719},
 primaryClass = {astro-ph.GA},
       adsurl = {https://ui.adsabs.harvard.edu/abs/2017A&A...602A...2S}
}

@ARTICLE{McAlpine2013,
       author = {{McAlpine}, K. and {Jarvis}, M.~J. and {Bonfield}, D.~G.},
        title = "{Evolution of faint radio sources in the VIDEO-XMM3 field}",
      journal = {\mnras},
         year = 2013,
        month = dec,
       volume = {436},
       number = {2},
        pages = {1084-1095},
          doi = {10.1093/mnras/stt1638},
archivePrefix = {arXiv},
       eprint = {1309.0358},
 primaryClass = {astro-ph.CO},
       adsurl = {https://ui.adsabs.harvard.edu/abs/2013MNRAS.436.1084M}
}

@ARTICLE{Jarvis2009,
       author = {{Jarvis}, Matt J. and {Teimourian}, Hanifa and {Simpson}, Chris and {Smith}, Daniel J.~B. and {Rawlings}, Steve and {Bonfield}, David},
        title = "{The discovery of a typical radio galaxy at z = 4.88}",
      journal = {\mnras},
         year = 2009,
        month = sep,
       volume = {398},
       number = {1},
        pages = {L83-L87},
          doi = {10.1111/j.1745-3933.2009.00715.x},
archivePrefix = {arXiv},
       eprint = {0907.1447},
 primaryClass = {astro-ph.CO},
       adsurl = {https://ui.adsabs.harvard.edu/abs/2009MNRAS.398L..83J}
}

@ARTICLE{Harrison2014,
       author = {{Harrison}, C.~M. and {Alexander}, D.~M. and {Mullaney}, J.~R. and {Swinbank}, A.~M.},
        title = "{Kiloparsec-scale outflows are prevalent among luminous AGN: outflows and feedback in the context of the overall AGN population}",
      journal = {\mnras},
         year = 2014,
        month = jul,
       volume = {441},
       number = {4},
        pages = {3306-3347},
          doi = {10.1093/mnras/stu515},
archivePrefix = {arXiv},
       eprint = {1403.3086},
 primaryClass = {astro-ph.GA},
       adsurl = {https://ui.adsabs.harvard.edu/abs/2014MNRAS.441.3306H}
}

@ARTICLE{springel2005,
       author = {{Springel}, Volker and {Di Matteo}, Tiziana and {Hernquist}, Lars},
        title = "{Black Holes in Galaxy Mergers: The Formation of Red Elliptical Galaxies}",
      journal = {\apjl},
         year = 2005,
        month = feb,
       volume = {620},
       number = {2},
        pages = {L79-L82},
          doi = {10.1086/428772},
archivePrefix = {arXiv},
       eprint = {astro-ph/0409436},
 primaryClass = {astro-ph},
       adsurl = {https://ui.adsabs.harvard.edu/abs/2005ApJ...620L..79S}
}

@ARTICLE{Cicone2014,
       author = {{Cicone}, C. and {Maiolino}, R. and {Sturm}, E. and {Graci{\'a}-Carpio}, J. and {Feruglio}, C. and {Neri}, R. and {Aalto}, S. and {Davies}, R. and {Fiore}, F. and {Fischer}, J. and {Garc{\'\i}a-Burillo}, S. and {Gonz{\'a}lez-Alfonso}, E. and {Hailey-Dunsheath}, S. and {Piconcelli}, E. and {Veilleux}, S.},
        title = "{Massive molecular outflows and evidence for AGN feedback from CO observations}",
      journal = {\aap},
         year = 2014,
        month = feb,
       volume = {562},
          eid = {A21},
        pages = {A21},
          doi = {10.1051/0004-6361/201322464},
archivePrefix = {arXiv},
       eprint = {1311.2595},
 primaryClass = {astro-ph.CO},
       adsurl = {https://ui.adsabs.harvard.edu/abs/2014A&A...562A..21C}
}

@BOOK{peebles1980,
       author = {{Peebles}, P.~J.~E.},
        title = "{The large-scale structure of the universe}",
         year = 1980,
       adsurl = {https://ui.adsabs.harvard.edu/abs/1980lssu.book.....P}
}

@ARTICLE{landy1993,
       author = {{Landy}, Stephen D. and {Szalay}, Alexander S.},
        title = "{Bias and Variance of Angular Correlation Functions}",
      journal = {\apj},
         year = 1993,
        month = jul,
       volume = {412},
        pages = {64},
          doi = {10.1086/172900},
       adsurl = {https://ui.adsabs.harvard.edu/abs/1993ApJ...412...64L}
}

@article{hale2018,
    author = {Hale, C L and Jarvis, M J and Delvecchio, I and Hatfield, P W and Novak, M and Smolčić, V and Zamorani, G},
    title = {The clustering and bias of radio-selected AGN and star-forming galaxies in the COSMOS field},
    journal = {\mnras},
    volume = {474},
    number = {3},
    pages = {4133-4150},
    year = {2018},
    month = {11},
    issn = {0035-8711},
    doi = {10.1093/mnras/stx2954},
    url = {https://doi.org/10.1093/mnras/stx2954},
    eprint = {https://academic.oup.com/mnras/article-pdf/474/3/4133/22989439/stx2954.pdf},
}

@INPROCEEDINGS{jarvis2016,
       author = {{Jarvis}, M. and {Taylor}, R. and {Agudo}, I. and {Allison}, J.~R. and {Deane}, R.~P. and {Frank}, B. and {Gupta}, N. and {Heywood}, I. and {Maddox}, N. and {McAlpine}, K. and {Santos}, M. and {Scaife}, A.~M.~M. and {Vaccari}, M. and {Zwart}, J.~T.~L. and {Adams}, E. and {Bacon}, D.~J. and {Baker}, A.~J. and {Bassett}, B.~A. and {Best}, P.~N. and {Beswick}, R. and {Blyth}, S. and {Brown}, M.~L. and {Bruggen}, M. and {Cluver}, M. and {Colafrancesco}, S. and {Cotter}, G. and {Cress}, C. and {Dav{\'e}}, R. and {Ferrari}, C. and {Hardcastle}, M.~J. and {Hale}, C.~L. and {Harrison}, I. and {Hatfield}, P.~W. and {Klockner}, H.~R. and {Kolwa}, S. and {Malefahlo}, E. and {Marubini}, T. and {Mauch}, T. and {Moodley}, K. and {Morganti}, R. and {Norris}, R.~P. and {Peters}, J.~A. and {Prandoni}, I. and {Prescott}, M. and {Oliver}, S. and {Oozeer}, N. and {Rottgering}, H.~J.~A. and {Seymour}, N. and {Simpson}, C. and {Smirnov}, O. and {Smith}, D.~J.~B.},
        title = "{The MeerKAT International GHz Tiered Extragalactic Exploration (MIGHTEE) Survey}",
    booktitle = {MeerKAT Science: On the Pathway to the SKA},
         year = 2016,
        month = jan,
          eid = {6},
        pages = {6},
          doi = {10.22323/1.277.0006},
archivePrefix = {arXiv},
       eprint = {1709.01901},
 primaryClass = {astro-ph.GA},
       adsurl = {https://ui.adsabs.harvard.edu/abs/2016mks..confE...6J}
}

@article{heywood2022,
    author = {Heywood, I and Jarvis, M J and Hale, C L and Whittam, I H and Bester, H L and Hugo, B and Kenyon, J S and Prescott, M and Smirnov, O M and Tasse, C and Afonso, J M and Best, P N and Collier, J D and Deane, R P and Frank, B S and Hardcastle, M J and Knowles, K and Maddox, N and Murphy, E J and Prandoni, I and Randriamampandry, S M and Santos, M G and Sekhar, S and Tabatabaei, F and Taylor, A R and Thorat, K},
    title = {MIGHTEE: total intensity radio continuum imaging and the COSMOS/XMM-LSS Early Science fields},
    journal = {\mnras},
    volume = {509},
    number = {2},
    pages = {2150-2168},
    year = {2022},
    month = {10},
    issn = {0035-8711},
    doi = {10.1093/mnras/stab3021},
    url = {https://doi.org/10.1093/mnras/stab3021},
    eprint = {https://academic.oup.com/mnras/article-pdf/509/2/2150/41192382/stab3021.pdf},
}

@article{hale2025,
    author = {Hale, C L and Heywood, I and Jarvis, M J and Whittam, I H and Best, P N and An, Fangxia and Bowler, R A A and Harrison, I and Matthews, A and Smith, D J B and Taylor, A R and Vaccari, M},
    title = {MIGHTEE: the continuum survey Data Release 1},
    journal = {\mnras},
    volume = {536},
    number = {3},
    pages = {2187-2211},
    year = {2025},
    month = {11},
    issn = {0035-8711},
    doi = {10.1093/mnras/stae2528},
    url = {https://doi.org/10.1093/mnras/stae2528},
    eprint = {https://academic.oup.com/mnras/article-pdf/536/3/2187/60580400/stae2528.pdf},
}

@ARTICLE{wilman2008,
       author = {Wilman, R. J. and {Miller}, L. and {Jarvis}, M.~J. and {Mauch}, T. and {Levrier}, F. and {Abdalla}, F.~B. and {Rawlings}, S. and {Kl{\"o}ckner}, H. -R. and {Obreschkow}, D. and {Olteanu}, D. and {Young}, S.},
        title = "{A semi-empirical simulation of the extragalactic radio continuum sky for next generation radio telescopes}",
      journal = {\mnras},
         year = 2008,
        month = aug,
       volume = {388},
       number = {3},
        pages = {1335-1348},
          doi = {10.1111/j.1365-2966.2008.13486.x},
archivePrefix = {arXiv},
       eprint = {0805.3413},
 primaryClass = {astro-ph},
       adsurl = {https://ui.adsabs.harvard.edu/abs/2008MNRAS.388.1335W}
}

@article{wilman2010,
    author = {Wilman, R. J. and Jarvis, M. J. and Mauch, T. and Rawlings, S. and Hickey, S.},
    title = {An infrared–radio simulation of the extragalactic sky: from the Square Kilometre Array to Herschel},
    journal = {\mnras},
    volume = {405},
    number = {1},
    pages = {447-461},
    year = {2010},
    month = {06},
    issn = {0035-8711},
    doi = {10.1111/j.1365-2966.2010.16453.x},
    url = {https://doi.org/10.1111/j.1365-2966.2010.16453.x},
    eprint = {https://academic.oup.com/mnras/article-pdf/405/1/447/3374829/mnras0405-0447.pdf},
}

@article{hatfield2022,
    author = {Hatfield, P W and Jarvis, M J and Adams, N and Bowler, R A A and Häußler, B and Duncan, K J},
    title = {Hybrid photometric redshifts for sources in the COSMOS and XMM-LSS fields},
    journal = {\mnras},
    volume = {513},
    number = {3},
    pages = {3719-3733},
    year = {2022},
    month = {04},
    issn = {0035-8711},
    doi = {10.1093/mnras/stac1042},
    url = {https://doi.org/10.1093/mnras/stac1042},
    eprint = {https://academic.oup.com/mnras/article-pdf/513/3/3719/43693703/stac1042.pdf},
}

@ARTICLE{arnouts1999,
       author = {{Arnouts}, S. and {Cristiani}, S. and {Moscardini}, L. and {Matarrese}, S. and {Lucchin}, F. and {Fontana}, A. and {Giallongo}, E.},
        title = "{Measuring and modelling the redshift evolution of clustering: the Hubble Deep Field North}",
      journal = {\mnras},
         year = 1999,
        month = dec,
       volume = {310},
       number = {2},
        pages = {540-556},
          doi = {10.1046/j.1365-8711.1999.02978.x},
archivePrefix = {arXiv},
       eprint = {astro-ph/9902290},
 primaryClass = {astro-ph},
       adsurl = {https://ui.adsabs.harvard.edu/abs/1999MNRAS.310..540A}
}

@article{almosallam2016,
    author = {Almosallam, Ibrahim A. and Jarvis, Matt J. and Roberts, Stephen J.},
    title = { GPz: non-stationary sparse Gaussian processes for heteroscedastic uncertainty estimation in photometric redshifts},
    journal = {\mnras},
    volume = {462},
    number = {1},
    pages = {726-739},
    year = {2016},
    month = {07},
    issn = {0035-8711},
    doi = {10.1093/mnras/stw1618},
    url = {https://doi.org/10.1093/mnras/stw1618},
    eprint = {https://academic.oup.com/mnras/article-pdf/462/1/726/18470400/stw1618.pdf},
}

@ARTICLE{gomes2018,
       author = {{Gomes}, Zahra and {Jarvis}, Matt J. and {Almosallam}, Ibrahim A. and {Roberts}, Stephen J.},
        title = "{Improving photometric redshift estimation using GPZ: size information, post processing, and improved photometry}",
      journal = {\mnras},
         year = 2018,
        month = mar,
       volume = {475},
       number = {1},
        pages = {331-342},
          doi = {10.1093/mnras/stx3187},
archivePrefix = {arXiv},
       eprint = {1712.02256},
 primaryClass = {astro-ph.GA},
       adsurl = {https://ui.adsabs.harvard.edu/abs/2018MNRAS.475..331G}
}

@ARTICLE{ilbert2006,
       author = {{Ilbert}, O. and {Arnouts}, S. and {McCracken}, H.~J. and {Bolzonella}, M. and {Bertin}, E. and {Le F{\`e}vre}, O. and {Mellier}, Y. and {Zamorani}, G. and {Pell{\`o}}, R. and {Iovino}, A. and {Tresse}, L. and {Le Brun}, V. and {Bottini}, D. and {Garilli}, B. and {Maccagni}, D. and {Picat}, J.~P. and {Scaramella}, R. and {Scodeggio}, M. and {Vettolani}, G. and {Zanichelli}, A. and {Adami}, C. and {Bardelli}, S. and {Cappi}, A. and {Charlot}, S. and {Ciliegi}, P. and {Contini}, T. and {Cucciati}, O. and {Foucaud}, S. and {Franzetti}, P. and {Gavignaud}, I. and {Guzzo}, L. and {Marano}, B. and {Marinoni}, C. and {Mazure}, A. and {Meneux}, B. and {Merighi}, R. and {Paltani}, S. and {Pollo}, A. and {Pozzetti}, L. and {Radovich}, M. and {Zucca}, E. and {Bondi}, M. and {Bongiorno}, A. and {Busarello}, G. and {de La Torre}, S. and {Gregorini}, L. and {Lamareille}, F. and {Mathez}, G. and {Merluzzi}, P. and {Ripepi}, V. and {Rizzo}, D. and {Vergani}, D.},
        title = "{Accurate photometric redshifts for the CFHT legacy survey calibrated using the VIMOS VLT deep survey}",
      journal = {\aap},
         year = 2006,
        month = oct,
       volume = {457},
       number = {3},
        pages = {841-856},
          doi = {10.1051/0004-6361:20065138},
archivePrefix = {arXiv},
       eprint = {astro-ph/0603217},
 primaryClass = {astro-ph},
       adsurl = {https://ui.adsabs.harvard.edu/abs/2006A&A...457..841I}
}

@ARTICLE{duncan2018,
       author = {{Duncan}, Kenneth J. and {Brown}, Michael J.~I. and {Williams}, Wendy L. and {Best}, Philip N. and {Buat}, Veronique and {Burgarella}, Denis and {Jarvis}, Matt J. and {Ma{\l}ek}, Katarzyna and {Oliver}, S.~J. and {R{\"o}ttgering}, Huub J.~A. and {Smith}, Daniel J.~B.},
        title = "{Photometric redshifts for the next generation of deep radio continuum surveys - I. Template fitting}",
      journal = {\mnras},
         year = 2018,
        month = jan,
       volume = {473},
       number = {2},
        pages = {2655-2672},
          doi = {10.1093/mnras/stx2536},
archivePrefix = {arXiv},
       eprint = {1709.09183},
 primaryClass = {astro-ph.GA},
       adsurl = {https://ui.adsabs.harvard.edu/abs/2018MNRAS.473.2655D}
}

@article{duncan2019,
	author = {{Duncan} and {Sabater, J.} and {Röttgering, H. J. A.} and {Jarvis, M. J.} and {Smith, D. J. B.} and {Best, P. N.} and {Callingham, J. R.} and {Cochrane, R.} and {Croston, J. H.} and {Hardcastle, M. J.} and {Mingo, B.} and {Morabito, L.} and {Nisbet, D.} and {Prandoni, I.} and {Shimwell, T. W.} and {Tasse, C.} and {White, G. J.} and {Williams, W. L.} and {Alegre, L.} and {Chyży, K. T.} and {Gürkan, G.} and {Hoeft, M.} and {Kondapally, R.} and {Mechev, A. P.} and {Miley, G. K.} and {Schwarz, D. J.} and {van Weeren, R. J.}},
	title = {The LOFAR Two-metre Sky Survey - IV. First Data Release: Photometric redshifts and rest-frame magnitudes⋆⋆⋆},
	DOI= "10.1051/0004-6361/201833562",
	url= "https://doi.org/10.1051/0004-6361/201833562",
	journal = {A&A},
	year = 2019,
	volume = 622,
	pages = "A3",
}

@ARTICLE{dahlen2013,
       author = {{Dahlen}, Tomas and {Mobasher}, Bahram and {Faber}, Sandra M. and {Ferguson}, Henry C. and {Barro}, Guillermo and {Finkelstein}, Steven L. and {Finlator}, Kristian and {Fontana}, Adriano and {Gruetzbauch}, Ruth and {Johnson}, Seth and {Pforr}, Janine and {Salvato}, Mara and {Wiklind}, Tommy and {Wuyts}, Stijn and {Acquaviva}, Viviana and {Dickinson}, Mark E. and {Guo}, Yicheng and {Huang}, Jiasheng and {Huang}, Kuang-Han and {Newman}, Jeffrey A. and {Bell}, Eric F. and {Conselice}, Christopher J. and {Galametz}, Audrey and {Gawiser}, Eric and {Giavalisco}, Mauro and {Grogin}, Norman A. and {Hathi}, Nimish and {Kocevski}, Dale and {Koekemoer}, Anton M. and {Koo}, David C. and {Lee}, Kyoung-Soo and {McGrath}, Elizabeth J. and {Papovich}, Casey and {Peth}, Michael and {Ryan}, Russell and {Somerville}, Rachel and {Weiner}, Benjamin and {Wilson}, Grant},
        title = "{A Critical Assessment of Photometric Redshift Methods: A CANDELS Investigation}",
      journal = {\apj},
         year = 2013,
        month = oct,
       volume = {775},
       number = {2},
          eid = {93},
        pages = {93},
          doi = {10.1088/0004-637X/775/2/93},
archivePrefix = {arXiv},
       eprint = {1308.5353},
 primaryClass = {astro-ph.CO},
       adsurl = {https://ui.adsabs.harvard.edu/abs/2013ApJ...775...93D}
}

@article{stylianou2022,
doi = {10.1088/1538-3873/ac59bf},
url = {https://dx.doi.org/10.1088/1538-3873/ac59bf},
year = {2022},
month = {apr},
publisher = {The Astronomical Society of the Pacific},
volume = {134},
number = {1034},
pages = {044501},
author = {Natalia Stylianou and Alex I. Malz and Peter Hatfield and John Franklin Crenshaw and Julia Gschwend},
title = {The Sensitivity of GPz Estimates of Photo-z Posterior PDFs to Realistically Complex Training Set Imperfections},
journal = {Publications of the Astronomical Society of the Pacific}
}

@article{norberg2009,
    author = {Norberg, P. and Baugh, C. M. and Gaztañaga, E. and Croton, D. J.},
    title = {Statistical analysis of galaxy surveys – I. Robust error estimation for two-point clustering statistics},
    journal = {\mnras},
    volume = {396},
    number = {1},
    pages = {19-38},
    year = {2009},
    month = {06},
    issn = {0035-8711},
    doi = {10.1111/j.1365-2966.2009.14389.x},
    url = {https://doi.org/10.1111/j.1365-2966.2009.14389.x},
    eprint = {https://academic.oup.com/mnras/article-pdf/396/1/19/4066956/mnras0396-0019.pdf},
}

@ARTICLE{jarvis2004,
       author = {{Jarvis}, M. and {Bernstein}, G. and {Jain}, B.},
        title = "{The skewness of the aperture mass statistic}",
      journal = {\mnras},
         year = 2004,
        month = jul,
       volume = {352},
       number = {1},
        pages = {338-352},
          doi = {10.1111/j.1365-2966.2004.07926.x},
archivePrefix = {arXiv},
       eprint = {astro-ph/0307393},
 primaryClass = {astro-ph},
       adsurl = {https://ui.adsabs.harvard.edu/abs/2004MNRAS.352..338J}
}

@ARTICLE{becker1995,
       author = {{Becker}, Robert H. and {White}, Richard L. and {Helfand}, David J.},
        title = "{The FIRST Survey: Faint Images of the Radio Sky at Twenty Centimeters}",
      journal = {\apj},
         year = 1995,
        month = sep,
       volume = {450},
        pages = {559},
          doi = {10.1086/176166},
       adsurl = {https://ui.adsabs.harvard.edu/abs/1995ApJ...450..559B}
}

@ARTICLE{groth1977,
       author = {{Groth}, E.~J. and {Peebles}, P.~J.~E.},
        title = "{Statistical analysis of catalogs of extragalactic objects. VII. Two- and three-point correlation functions for the high-resolution Shane-Wirtanen catalog of galaxies.}",
      journal = {\apj},
         year = 1977,
        month = oct,
       volume = {217},
        pages = {385-405},
          doi = {10.1086/155588},
       adsurl = {https://ui.adsabs.harvard.edu/abs/1977ApJ...217..385G}
}

@article{roche1999,
    author = {Roche, Nathan and Eales, Stephen A.},
    title = {The angular correlation function and hierarchical moments of ∼70 000 faint galaxies to R=23.5},
    journal = {\mnras},
    volume = {307},
    number = {3},
    pages = {703-721},
    year = {1999},
    month = {08},
    issn = {0035-8711},
    doi = {10.1046/j.1365-8711.1999.02652.x},
    url = {https://doi.org/10.1046/j.1365-8711.1999.02652.x},
    eprint = {https://academic.oup.com/mnras/article-pdf/307/3/703/2850280/307-3-703.pdf},
}

@article{szapudi1998,
doi = {10.1086/311146},
url = {https://dx.doi.org/10.1086/311146},
year = {1998},
month = {jan},
publisher = {},
volume = {494},
number = {1},
pages = {L41},
author = {Szapudi, István and Szalay, Alexander S.},
title = {A New Class of Estimators for the N-Point Correlations},
journal = {The Astrophysical Journal}
}

@article{magliocchetti2002,
    author = {Magliocchetti, Manuela and Maddox, Steve J. and Jackson, Carole A. and Bland-Hawthorn, Joss and Bridges, Terry and Cannon, Russell and Cole, Shaun and Colless, Matthew and Collins, Chris and Couch, Warrick and Dalton, Gavin and De Propris, Roberto and Driver, Simon P. and Efstathiou, George and Ellis, Richard S. and Frenk, Carlos S. and Glazebrook, Karl and Lahav, Ofer and Lewis, Ian and Lumsden, Stuart and Peacock, John A. and Peterson, Bruce A. and Sutherland, Will and Taylor, Keith},
    title = {The 2dF Galaxy Redshift Survey: the population of nearby radio galaxies at the 1-mJy level},
    journal = {\mnras},
    volume = {333},
    number = {1},
    pages = {100-120},
    year = {2002},
    month = {06},
    issn = {0035-8711},
    doi = {10.1046/j.1365-8711.2002.05386.x},
    url = {https://doi.org/10.1046/j.1365-8711.2002.05386.x},
    eprint = {https://academic.oup.com/mnras/article-pdf/333/1/100/4010034/333-1-100.pdf},
}

@article{mauch2007,
    author = {Mauch, Tom and Sadler, Elaine M.},
    title = {Radio sources in the 6dFGS: local luminosity functions at 1.4 GHz for star-forming galaxies and radio-loud AGN},
    journal = {\mnras},
    volume = {375},
    number = {3},
    pages = {931-950},
    year = {2007},
    month = {01},
    issn = {0035-8711},
    doi = {10.1111/j.1365-2966.2006.11353.x},
    url = {https://doi.org/10.1111/j.1365-2966.2006.11353.x},
    eprint = {https://academic.oup.com/mnras/article-pdf/375/3/931/40677396/mnras\_375\_3\_931.pdf},
}

@article{whittam2022,
    author = {Whittam, I H and Jarvis, M J and Hale, C L and Prescott, M and Morabito, L K and Heywood, I and Adams, N J and Afonso, J and An, Fangxia and Ao, Y and Bowler, R A A and Collier, J D and Deane, R P and Delhaize, J and Frank, B and Glowacki, M and Hatfield, P W and Maddox, N and Marchetti, L and Matthews, A M and Prandoni, I and Randriamampandry, S and Randriamanakoto, Z and Smith, D J B and Taylor, A R and Thomas, N L and Vaccari, M},
    title = {MIGHTEE: the nature of the radio-loud AGN population},
    journal = {\mnras},
    volume = {516},
    number = {1},
    pages = {245-263},
    year = {2022},
    month = {08},
    issn = {0035-8711},
    doi = {10.1093/mnras/stac2140},
    url = {https://doi.org/10.1093/mnras/stac2140},
    eprint = {https://academic.oup.com/mnras/article-pdf/516/1/245/45512548/stac2140.pdf},
}

@article{zheng2005,
doi = {10.1086/466510},
url = {https://dx.doi.org/10.1086/466510},
year = {2005},
month = {nov},
publisher = {},
volume = {633},
number = {2},
pages = {791},
author = {Zheng, Zheng and Berlind, Andreas A. and Weinberg, David H. and Benson, Andrew J. and Baugh, Carlton M. and Cole, Shaun and Davé, Romeel and Frenk, Carlos S. and Katz, Neal and Lacey, Cedric G.},
title = {Theoretical Models of the Halo Occupation Distribution: Separating Central and Satellite Galaxies},
journal = {The Astrophysical Journal}
}

@article{ando2017,
    author = {Ando, Shin'ichiro and Benoit-Lévy, Aurélien and Komatsu, Eiichiro},
    title = {Angular power spectrum of galaxies in the 2MASS Redshift Survey},
    journal = {\mnras},
    volume = {473},
    number = {4},
    pages = {4318-4325},
    year = {2017},
    month = {10},
    issn = {0035-8711},
    doi = {10.1093/mnras/stx2634},
    url = {https://doi.org/10.1093/mnras/stx2634},
    eprint = {https://academic.oup.com/mnras/article-pdf/473/4/4318/21907252/stx2634.pdf},
}

@article{nicola2020,
doi = {10.1088/1475-7516/2020/03/044},
url = {https://dx.doi.org/10.1088/1475-7516/2020/03/044},
year = {2020},
month = {mar},
publisher = {},
volume = {2020},
number = {03},
pages = {044},
author = {Nicola, Andrina and Alonso, David and Sánchez, Javier and Slosar, Anže and Awan, Humna and Broussard, Adam and Dunkley, Jo and Gawiser, Eric and Gomes, Zahra and Mandelbaum, Rachel and Miyatake, Hironao and Newman, Jeffrey A. and Sevilla-Noarbe, Ignacio and Skinner, Sarah and Wagoner, Erika L.},
title = {Tomographic galaxy clustering with the Subaru Hyper Suprime-Cam first year public data release},
journal = {Journal of Cosmology and Astroparticle Physics}
}

@ARTICLE{navarro1995,
       author = {{Navarro}, Julio F. and {Frenk}, Carlos S. and {White}, Simon D.~M.},
        title = "{The Structure of Cold Dark Matter Halos}",
      journal = {\apj},
         year = 1996,
        month = may,
       volume = {462},
        pages = {563},
          doi = {10.1086/177173},
archivePrefix = {arXiv},
       eprint = {astro-ph/9508025},
 primaryClass = {astro-ph},
       adsurl = {https://ui.adsabs.harvard.edu/abs/1996ApJ...462..563N}
}

@article{chisari2019,
doi = {10.3847/1538-4365/ab1658},
url = {https://dx.doi.org/10.3847/1538-4365/ab1658},
year = {2019},
month = {may},
publisher = {The American Astronomical Society},
volume = {242},
number = {1},
pages = {2},
author = {Chisari, Nora Elisa and Alonso, David and Krause, Elisabeth and Leonard, C. Danielle and Bull, Philip and Neveu, Jérémy and Villarreal, Antonio and Singh, Sukhdeep and McClintock, Thomas and Ellison, John and Du, Zilong and Zuntz, Joe and Mead, Alexander and Joudaki, Shahab and Lorenz, Christiane S. and Tröster, Tilman and Sanchez, Javier and Lanusse, Francois and Ishak, Mustapha and Hlozek, Renée and Blazek, Jonathan and Campagne, Jean-Eric and Almoubayyed, Husni and Eifler, Tim and Kirby, Matthew and Kirkby, David and Plaszczynski, Stéphane and Slosar, Anže and Vrastil, Michal and Wagoner, Erika L. and (LSST Dark Energy Science Collaboration)},
title = {Core Cosmology Library: Precision Cosmological Predictions for LSST},
journal = {The Astrophysical Journal Supplement Series}
}

@ARTICLE{foreman-mackey2013,
       author = {{Foreman-Mackey}, Daniel and {Hogg}, David W. and {Lang}, Dustin and {Goodman}, Jonathan},
        title = "{emcee: The MCMC Hammer}",
      journal = {\pasp},
         year = 2013,
        month = mar,
       volume = {125},
       number = {925},
        pages = {306},
          doi = {10.1086/670067},
archivePrefix = {arXiv},
       eprint = {1202.3665},
 primaryClass = {astro-ph.IM},
       adsurl = {https://ui.adsabs.harvard.edu/abs/2013PASP..125..306F}
}

@article{hatfield2016,
    author = {Hatfield, P. W. and Lindsay, S. N. and Jarvis, M. J. and Häußler, B. and Vaccari, M. and Verma, A.},
    title = {The galaxy–halo connection in the VIDEO survey at 0.5 \&lt; z \&lt; 1.7},
    journal = {\mnras},
    volume = {459},
    number = {3},
    pages = {2618-2631},
    year = {2016},
    month = {04},
    issn = {0035-8711},
    doi = {10.1093/mnras/stw769},
    url = {https://doi.org/10.1093/mnras/stw769},
    eprint = {https://academic.oup.com/mnras/article-pdf/459/3/2618/8106475/stw769.pdf},
}

@article{tinker2010,
doi = {10.1088/0004-637X/724/2/878},
url = {https://dx.doi.org/10.1088/0004-637X/724/2/878},
year = {2010},
month = {nov},
publisher = {The American Astronomical Society},
volume = {724},
number = {2},
pages = {878},
author = {Tinker, Jeremy L. and Robertson, Brant E. and Kravtsov, Andrey V. and Klypin, Anatoly and Warren, Michael S. and Yepes, Gustavo and Gottlöber, Stefan},
title = {THE LARGE-SCALE BIAS OF DARK MATTER HALOS: NUMERICAL CALIBRATION AND MODEL TESTS},
journal = {The Astrophysical Journal}
}

@article{duffy2008,
    author = {Duffy, Alan R. and Schaye, Joop and Kay, Scott T. and Dalla Vecchia, Claudio},
    title = {Dark matter halo concentrations in the Wilkinson Microwave Anisotropy Probe year 5 cosmology},
    journal = {\mnras: Letters},
    volume = {390},
    number = {1},
    pages = {L64-L68},
    year = {2008},
    month = {10},
    issn = {1745-3925},
    doi = {10.1111/j.1745-3933.2008.00537.x},
    url = {https://doi.org/10.1111/j.1745-3933.2008.00537.x},
    eprint = {https://academic.oup.com/mnrasl/article-pdf/390/1/L64/54682040/mnrasl\_390\_1\_l64.pdf},
}

@article{whittam2023,
    author = {Whittam, I H and Prescott, M and Hale, C L and Jarvis, M J and Heywood, I and An, Fangxia and Glowacki, M and Maddox, N and Marchetti, L and Morabito, L K and Adams, N J and Bowler, R A A and Hatfield, P W and Varadaraj, R G and Collier, J and Frank, B and Taylor, A R and Santos, M G and Vaccari, M and Afonso, J and Ao, Y and Delhaize, J and Knowles, K and Kolwa, S and Randriamampandry, S M and Randriamanakoto, Z and Smirnov, O and Smith, D J B and White, S V},
    title = {MIGHTEE: Multi-wavelength counterparts in the COSMOS field},
    journal = {\mnras},
    volume = {527},
    number = {2},
    pages = {3231-3245},
    year = {2024},
    month = {10},
    issn = {0035-8711},
    doi = {10.1093/mnras/stad3307},
    url = {https://doi.org/10.1093/mnras/stad3307},
    eprint = {https://academic.oup.com/mnras/article-pdf/527/2/3231/56343751/stad3307.pdf},
}

@article{sutherland1992,
    author = {Sutherland, Will and Saunders, Will},
    title = {On the likelihood ratio for source identification},
    journal = {\mnras},
    volume = {259},
    number = {3},
    pages = {413-420},
    year = {1992},
    month = {12},
    issn = {0035-8711},
    doi = {10.1093/mnras/259.3.413},
    url = {https://doi.org/10.1093/mnras/259.3.413},
    eprint = {https://academic.oup.com/mnras/article-pdf/259/3/413/4109075/mnras259-0413.pdf},
}

@article{malavasi2015,
	author = {{Malavasi}, N. and {Bardelli, S.} and {Ciliegi, P.} and {Ilbert, O.} and {Pozzetti, L.} and {Zucca, E.}},
	title = {The environment of radio sources in the VLA-COSMOS survey field⋆},
	DOI= "10.1051/0004-6361/201425155",
	url= "https://doi.org/10.1051/0004-6361/201425155",
	journal = {A&A},
	year = 2015,
	volume = 576,
	pages = "A101",
	month = "",
}

@misc{hugo2022,
      title={Tricolour: an optimized SumThreshold flagger for MeerKAT}, 
      author={Benjamin V. Hugo and Simon Perkins and Bruce Merry and Tom Mauch and Oleg M. Smirnov},
      year={2022},
      eprint={2206.09179},
      archivePrefix={arXiv},
      primaryClass={astro-ph.IM},
      url={https://arxiv.org/abs/2206.09179}, 
}

@article{magliocchetti2017,
    author = {Magliocchetti, M. and Popesso, P. and Brusa, M. and Salvato, M. and Laigle, C. and McCracken, H. J. and Ilbert, O.},
    title = {The clustering properties of radio-selected AGN and star-forming galaxies up to redshifts z ∼ 3},
    journal = {\mnras},
    volume = {464},
    number = {3},
    pages = {3271-3280},
    year = {2017},
    month = {10},
    issn = {0035-8711},
    doi = {10.1093/mnras/stw2541},
    url = {https://doi.org/10.1093/mnras/stw2541},
    eprint = {https://academic.oup.com/mnras/article-pdf/464/3/3271/18517206/stw2541.pdf},
}

@ARTICLE{oke1983,
       author = {{Oke}, J.~B. and {Gunn}, J.~E.},
        title = "{Secondary standard stars for absolute spectrophotometry.}",
      journal = {\apj},
         year = 1983,
        month = mar,
       volume = {266},
        pages = {713-717},
          doi = {10.1086/160817},
       adsurl = {https://ui.adsabs.harvard.edu/abs/1983ApJ...266..713O}
}

@article{alonso2007,
    author = {Alonso, M. Sol and Lambas, Diego G. and Tissera, Patricia and Coldwell, Georgina},
    title = {Active galactic nuclei and galaxy interactions},
    journal = {\mnras},
    volume = {375},
    number = {3},
    pages = {1017-1024},
    year = {2007},
    month = {01},
    issn = {0035-8711},
    doi = {10.1111/j.1365-2966.2007.11367.x},
    url = {https://doi.org/10.1111/j.1365-2966.2007.11367.x},
    eprint = {https://academic.oup.com/mnras/article-pdf/375/3/1017/3938846/mnras0375-1017.pdf},
}

@article{treister2012,
doi = {10.1088/2041-8205/758/2/L39},
url = {https://dx.doi.org/10.1088/2041-8205/758/2/L39},
year = {2012},
month = {oct},
publisher = {The American Astronomical Society},
volume = {758},
number = {2},
pages = {L39},
author = {Treister, E. and Schawinski, K. and Urry, C. M. and Simmons, B. D.},
title = {MAJOR GALAXY MERGERS ONLY TRIGGER THE MOST LUMINOUS ACTIVE GALACTIC NUCLEI},
journal = {The Astrophysical Journal Letters}
}

@article{sabater2013,
    author = {Sabater, J. and Best, P. N. and Argudo-Fernández, M.},
    title = {Effect of the interactions and environment on nuclear activity},
    journal = {\mnras},
    volume = {430},
    number = {1},
    pages = {638-651},
    year = {2013},
    month = {01},
    issn = {0035-8711},
    doi = {10.1093/mnras/sts675},
    url = {https://doi.org/10.1093/mnras/sts675},
    eprint = {https://academic.oup.com/mnras/article-pdf/430/1/638/13761369/sts675.pdf},
}

@article{fanidakis2010,
    author = {Fanidakis, N. and Baugh, C. M. and Benson, A. J. and Bower, R. G. and Cole, S. and Done, C. and Frenk, C. S.},
    title = {Grand unification of AGN activity in the ΛCDM cosmology},
    journal = {\mnras},
    volume = {410},
    number = {1},
    pages = {53-74},
    year = {2010},
    month = {12},
    issn = {0035-8711},
    doi = {10.1111/j.1365-2966.2010.17427.x},
    url = {https://doi.org/10.1111/j.1365-2966.2010.17427.x},
    eprint = {https://academic.oup.com/mnras/article-pdf/410/1/53/18439155/mnras0410-0053.pdf},
}

@article{mccracken2015,
    author = {McCracken, H. J. and Wolk, M. and Colombi, S. and Kilbinger, M. and Ilbert, O. and Peirani, S. and Coupon, J. and Dunlop, J. and Milvang-Jensen, B. and Caputi, K. and Aussel, H. and Béthermin, M. and Le Fèvre, O.},
    title = {Probing the galaxy–halo connection in UltraVISTA to z ∼ 2},
    journal = {\mnras},
    volume = {449},
    number = {1},
    pages = {901-916},
    year = {2015},
    month = {03},
    issn = {0035-8711},
    doi = {10.1093/mnras/stv305},
    url = {https://doi.org/10.1093/mnras/stv305},
    eprint = {https://academic.oup.com/mnras/article-pdf/449/1/901/4143490/stv305.pdf},
}

@article{jarvis2013,
    author = {Jarvis, Matt J. and Bonfield, D. G. and Bruce, V. A. and Geach, J. E. and McAlpine, K. and McLure, R. J. and González-Solares, E. and Irwin, M. and Lewis, J. and Yoldas, A. Kupcu and Andreon, S. and Cross, N. J. G. and Emerson, J. P. and Dalton, G. and Dunlop, J. S. and Hodgkin, S. T. and Le, Fèvre O. and Karouzos, M. and Meisenheimer, K. and Oliver, S. and Rawlings, S. and Simpson, C. and Smail, I. and Smith, D. J. B. and Sullivan, M. and Sutherland, W. and White, S. V. and Zwart, J. T. L.},
    title = {The VISTA Deep Extragalactic Observations (VIDEO) survey★},
    journal = {\mnras},
    volume = {428},
    number = {2},
    pages = {1281-1295},
    year = {2013},
    month = {10},
    issn = {0035-8711},
    doi = {10.1093/mnras/sts118},
    url = {https://doi.org/10.1093/mnras/sts118},
    eprint = {https://academic.oup.com/mnras/article-pdf/428/2/1281/3240562/sts118.pdf},
}

@article{padovani2017,
    author = "Padovani, P. and others",
    title = "{Active galactic nuclei: what\textquoteright{}s in a name?}",
    eprint = "1707.07134",
    archivePrefix = "arXiv",
    primaryClass = "astro-ph.GA",
    doi = "10.1007/s00159-017-0102-9",
    journal = "Astron. Astrophys. Rev.",
    volume = "25",
    number = "1",
    pages = "2",
    year = "2017"
}

@article{hardcastle2020,
title = {Radio galaxies and feedback from AGN jets},
journal = {New Astronomy Reviews},
volume = {88},
pages = {101539},
year = {2020},
issn = {1387-6473},
doi = {https://doi.org/10.1016/j.newar.2020.101539},
url = {https://www.sciencedirect.com/science/article/pii/S1387647320300166},
author = {M.J. Hardcastle and J.H. Croston}
}

@article{eckert2021,
   title={Feedback from Active Galactic Nuclei in Galaxy Groups},
   volume={7},
   ISSN={2218-1997},
   url={http://dx.doi.org/10.3390/universe7050142},
   DOI={10.3390/universe7050142},
   number={5},
   journal={Universe},
   publisher={MDPI AG},
   author={Eckert, Dominique and Gaspari, Massimo and Gastaldello, Fabio and Le Brun, Amandine M. C. and O’Sullivan, Ewan},
   year={2021},
   month=may, pages={142} }

@ARTICLE{fabian2012,
       author = {{Fabian}, A.~C.},
        title = "{Observational Evidence of Active Galactic Nuclei Feedback}",
      journal = {\araa},
         year = 2012,
        month = sep,
       volume = {50},
        pages = {455-489},
          doi = {10.1146/annurev-astro-081811-125521},
archivePrefix = {arXiv},
       eprint = {1204.4114},
 primaryClass = {astro-ph.CO},
       adsurl = {https://ui.adsabs.harvard.edu/abs/2012ARA&A..50..455F}
}

@ARTICLE{scharre2024,
       author = {{Scharr{\'e}}, Lucie and {Sorini}, Daniele and {Dav{\'e}}, Romeel},
        title = "{The effects of stellar and AGN feedback on the cosmic star formation history in the SIMBA simulations}",
      journal = {\mnras},
         year = 2024,
        month = oct,
       volume = {534},
       number = {1},
        pages = {361-383},
          doi = {10.1093/mnras/stae2098},
archivePrefix = {arXiv},
       eprint = {2404.07252},
 primaryClass = {astro-ph.GA},
       adsurl = {https://ui.adsabs.harvard.edu/abs/2024MNRAS.534..361S}
}

@ARTICLE{croton2006,
       author = {{Croton}, Darren J. and {Springel}, Volker and {White}, Simon D.~M. and {De Lucia}, G. and {Frenk}, C.~S. and {Gao}, L. and {Jenkins}, A. and {Kauffmann}, G. and {Navarro}, J.~F. and {Yoshida}, N.},
        title = "{The many lives of active galactic nuclei: cooling flows, black holes and the luminosities and colours of galaxies}",
      journal = {\mnras},
         year = 2006,
        month = jan,
       volume = {365},
       number = {1},
        pages = {11-28},
          doi = {10.1111/j.1365-2966.2005.09675.x},
archivePrefix = {arXiv},
       eprint = {astro-ph/0508046},
 primaryClass = {astro-ph},
       adsurl = {https://ui.adsabs.harvard.edu/abs/2006MNRAS.365...11C}
}

@ARTICLE{gaspari2020,
       author = {{Gaspari}, Massimo and {Tombesi}, Francesco and {Cappi}, Massimo},
        title = "{Linking macro-, meso- and microscales in multiphase AGN feeding and feedback}",
      journal = {Nature Astronomy},
         year = 2020,
        month = jan,
       volume = {4},
        pages = {10-13},
          doi = {10.1038/s41550-019-0970-1},
archivePrefix = {arXiv},
       eprint = {2001.04985},
 primaryClass = {astro-ph.GA},
       adsurl = {https://ui.adsabs.harvard.edu/abs/2020NatAs...4...10G}
}

@ARTICLE{cooray2002,
       author = {{Cooray}, Asantha and {Sheth}, Ravi},
        title = "{Halo models of large scale structure}",
      journal = {\physrep},
         year = 2002,
        month = dec,
       volume = {372},
       number = {1},
        pages = {1-129},
          doi = {10.1016/S0370-1573(02)00276-4},
archivePrefix = {arXiv},
       eprint = {astro-ph/0206508},
 primaryClass = {astro-ph},
       adsurl = {https://ui.adsabs.harvard.edu/abs/2002PhR...372....1C}
}

@article{zehavi2005,
doi = {10.1086/427495},
url = {https://dx.doi.org/10.1086/427495},
year = {2005},
month = {mar},
publisher = {},
volume = {621},
number = {1},
pages = {22},
author = {Zehavi, Idit and Eisenstein, Daniel J. and Nichol, Robert C. and Blanton, Michael R. and Hogg, David W. and Brinkmann, Jon and Loveday, Jon and Meiksin, Avery and Schneider, Donald P. and Tegmark, Max},
title = {The Intermediate-Scale Clustering of Luminous Red Galaxies},
journal = {The Astrophysical Journal}
}

@article{mo1996,
    author = {Mo, H. J. and White, S. D. M.},
    title = {An analytic model for the spatial clustering of dark matter haloes},
    journal = {\mnras},
    volume = {282},
    number = {2},
    pages = {347-361},
    year = {1996},
    month = {09},
    issn = {0035-8711},
    doi = {10.1093/mnras/282.2.347},
    url = {https://doi.org/10.1093/mnras/282.2.347},
    eprint = {https://academic.oup.com/mnras/article-pdf/282/2/347/3510677/282-2-347.pdf},
}

@BOOK{peacock1999,
       author = {{Peacock}, John A.},
        title = "{Cosmological Physics}",
         year = 1999,
       adsurl = {https://ui.adsabs.harvard.edu/abs/1999coph.book.....P}
}

@ARTICLE{kaiser1984,
       author = {{Kaiser}, N.},
        title = "{On the spatial correlations of Abell clusters.}",
      journal = {\apjl},
         year = 1984,
        month = sep,
       volume = {284},
        pages = {L9-L12},
          doi = {10.1086/184341},
       adsurl = {https://ui.adsabs.harvard.edu/abs/1984ApJ...284L...9K}
}

@article{best2005,
    author = {Best, P. N. and Kauffmann, G. and Heckman, T. M. and Brinchmann, J. and Charlot, S. and Ivezi\'{c}, \v{Z}. and White, S. D. M.},
    title = {The host galaxies of radio-loud active galactic nuclei: mass dependences, gas cooling and active galactic nuclei feedback},
    journal = {\mnras},
    volume = {362},
    number = {1},
    pages = {25-40},
    year = {2005},
    month = {09},
    issn = {0035-8711},
    doi = {10.1111/j.1365-2966.2005.09192.x},
    url = {https://doi.org/10.1111/j.1365-2966.2005.09192.x},
    eprint = {https://academic.oup.com/mnras/article-pdf/362/1/25/6029587/362-1-25.pdf},
}

@article{best2007,
    author = {Best, P. N. and Von Der Linden, A. and Kauffmann, G. and Heckman, T. M. and Kaiser, C. R.},
    title = {On the prevalence of radio-loud active galactic nuclei in brightest cluster galaxies: implications for AGN heating of cooling flows},
    journal = {\mnras},
    volume = {379},
    number = {3},
    pages = {894-908},
    year = {2007},
    month = {07},
    issn = {0035-8711},
    doi = {10.1111/j.1365-2966.2007.11937.x},
    url = {https://doi.org/10.1111/j.1365-2966.2007.11937.x},
    eprint = {https://academic.oup.com/mnras/article-pdf/379/3/894/3496171/mnras0379-0894.pdf},
}

@article{kauffmann2008,
    author = {Kauffmann, Guinevere and Heckman, Timothy M. and Best, Philip N.},
    title = {Radio jets in galaxies with actively accreting black holes: new insights from the SDSS},
    journal = {\mnras},
    volume = {384},
    number = {3},
    pages = {953-971},
    year = {2008},
    month = {07},
    issn = {0035-8711},
    doi = {10.1111/j.1365-2966.2007.12752.x},
    url = {https://doi.org/10.1111/j.1365-2966.2007.12752.x},
    eprint = {https://academic.oup.com/mnras/article-pdf/384/3/953/5785269/mnras0384-0953.pdf},
}

@article{donoso2010,
    author = {Donoso, E. and Li, Cheng and Kauffmann, G. and Best, P. N. and Heckman, T. M.},
    title = {Clustering of radio galaxies and quasars},
    journal = {\mnras},
    volume = {407},
    number = {2},
    pages = {1078-1089},
    year = {2010},
    month = {09},
    issn = {0035-8711},
    doi = {10.1111/j.1365-2966.2010.16907.x},
    url = {https://doi.org/10.1111/j.1365-2966.2010.16907.x},
    eprint = {https://academic.oup.com/mnras/article-pdf/407/2/1078/3900690/mnras0407-1078.pdf},
}

@article{retana2017,
   title={Probing the radio loud/quiet AGN dichotomy with quasar clustering},
   volume={600},
   ISSN={1432-0746},
   url={http://dx.doi.org/10.1051/0004-6361/201526433},
   DOI={10.1051/0004-6361/201526433},
   journal={Astronomy &amp; Astrophysics},
   publisher={EDP Sciences},
   author={Retana-Montenegro, E. and Röttgering, H. J. A.},
   year={2017},
   month=apr, pages={A97} }

@ARTICLE{sabater2019,
       author = {{Sabater}, J. and {Best}, P.~N. and {Hardcastle}, M.~J. and {Shimwell}, T.~W. and {Tasse}, C. and {Williams}, W.~L. and {Br{\"u}ggen}, M. and {Cochrane}, R.~K. and {Croston}, J.~H. and {de Gasperin}, F. and {Duncan}, K.~J. and {G{\"u}rkan}, G. and {Mechev}, A.~P. and {Morabito}, L.~K. and {Prandoni}, I. and {R{\"o}ttgering}, H.~J.~A. and {Smith}, D.~J.~B. and {Harwood}, J.~J. and {Mingo}, B. and {Mooney}, S. and {Saxena}, A.},
        title = "{The LoTSS view of radio AGN in the local Universe. The most massive galaxies are always switched on}",
      journal = {\aap},
         year = 2019,
        month = feb,
       volume = {622},
          eid = {A17},
        pages = {A17},
          doi = {10.1051/0004-6361/201833883},
archivePrefix = {arXiv},
       eprint = {1811.05528},
 primaryClass = {astro-ph.GA},
       adsurl = {https://ui.adsabs.harvard.edu/abs/2019A&A...622A..17S}
}

@ARTICLE{capetti2022,
       author = {{Capetti}, A. and {Brienza}, M. and {Balmaverde}, B. and {Best}, P.~N. and {Baldi}, R.~D. and {Drabent}, A. and {G{\"u}rkan}, G. and {Rottgering}, H.~J.~A. and {Tasse}, C. and {Webster}, B.},
        title = "{The LOFAR view of giant, early-type galaxies: Radio emission from active nuclei and star formation}",
      journal = {\aap},
         year = 2022,
        month = apr,
       volume = {660},
          eid = {A93},
        pages = {A93},
          doi = {10.1051/0004-6361/202142911},
archivePrefix = {arXiv},
       eprint = {2202.08593},
 primaryClass = {astro-ph.GA},
       adsurl = {https://ui.adsabs.harvard.edu/abs/2022A&A...660A..93C}
}

@article{gurkan2014,
    author = {Gürkan, G. and Hardcastle, M. J. and Jarvis, M. J.},
    title = {The Wide-field Infrared Survey Explorer properties of complete samples of radio-loud active galactic nucleus},
    journal = {\mnras},
    volume = {438},
    number = {2},
    pages = {1149-1161},
    year = {2014},
    month = {01},
    issn = {0035-8711},
    doi = {10.1093/mnras/stt2264},
    url = {https://doi.org/10.1093/mnras/stt2264},
    eprint = {https://academic.oup.com/mnras/article-pdf/438/2/1149/18752763/stt2264.pdf},
}

@ARTICLE{smolcic2009,
       author = {{Smol{\v{c}}i{\'c}}, V. and {Zamorani}, G. and {Schinnerer}, E. and {Bardelli}, S. and {Bondi}, M. and {B{\^\i}rzan}, L. and {Carilli}, C.~L. and {Ciliegi}, P. and {Elvis}, M. and {Impey}, C.~D. and {Koekemoer}, A.~M. and {Merloni}, A. and {Paglione}, T. and {Salvato}, M. and {Scodeggio}, M. and {Scoville}, N. and {Trump}, J.~R.},
        title = "{Cosmic Evolution of Radio Selected Active Galactic Nuclei in the Cosmos Field}",
      journal = {\apj},
         year = 2009,
        month = may,
       volume = {696},
       number = {1},
        pages = {24-39},
          doi = {10.1088/0004-637X/696/1/24},
archivePrefix = {arXiv},
       eprint = {0901.3372},
 primaryClass = {astro-ph.CO},
       adsurl = {https://ui.adsabs.harvard.edu/abs/2009ApJ...696...24S}
}

@ARTICLE{mandelbaum2009,
       author = {{Mandelbaum}, Rachel and {Li}, Cheng and {Kauffmann}, Guinevere and {White}, Simon D.~M.},
        title = "{Halo masses for optically selected and for radio-loud AGN from clustering and galaxy-galaxy lensing}",
      journal = {\mnras},
         year = 2009,
        month = feb,
       volume = {393},
       number = {2},
        pages = {377-392},
          doi = {10.1111/j.1365-2966.2008.14235.x},
archivePrefix = {arXiv},
       eprint = {0806.4089},
 primaryClass = {astro-ph},
       adsurl = {https://ui.adsabs.harvard.edu/abs/2009MNRAS.393..377M}
}

@article{magliocchetti2020,
    author = {Magliocchetti, M and Pentericci, L and Cirasuolo, M and Zamorani, G and Amorin, R and Bongiorno, A and Cimatti, A and Fontana, A and Garilli, B and Gargiulo, A and Hathi, N P and McLeod, D J and McLure, R J and Brusa, M and Saxena, A and Talia, M},
    title = {The role of galaxy mass on AGN emission: a view from the VANDELS survey},
    journal = {\mnras},
    volume = {493},
    number = {3},
    pages = {3838-3853},
    year = {2020},
    month = {02},
    issn = {0035-8711},
    doi = {10.1093/mnras/staa410},
    url = {https://doi.org/10.1093/mnras/staa410},
    eprint = {https://academic.oup.com/mnras/article-pdf/493/3/3838/32902902/staa410.pdf},
}

@ARTICLE{brown2011,
       author = {{Brown}, Michael J.~I. and {Jannuzi}, Buell T. and {Floyd}, David J.~E. and {Mould}, Jeremy R.},
        title = "{The Ubiquitous Radio Continuum Emission from the Most Massive Early-type Galaxies}",
      journal = {\apjl},
         year = 2011,
        month = apr,
       volume = {731},
       number = {2},
          eid = {L41},
        pages = {L41},
          doi = {10.1088/2041-8205/731/2/L41},
archivePrefix = {arXiv},
       eprint = {1103.2828},
 primaryClass = {astro-ph.CO},
       adsurl = {https://ui.adsabs.harvard.edu/abs/2011ApJ...731L..41B}
}

@article{grossova2022,
author = {Grossová, Romana and Werner, Norbert and Massaro, Francesco and Lakhchaura, Kiran and Plšek, Tomáš and Gabányi, Krisztina and Rajpurohit, Kamlesh and Canning, Rebecca and Nulsen, Paul and O’Sullivan, Ewan and Allen, Steven and Fabian, Andrew},
year = {2022},
month = {02},
pages = {30},
title = {Very Large Array Radio Study of a Sample of Nearby X-Ray and Optically Bright Early-type Galaxies},
volume = {258},
journal = {The Astrophysical Journal Supplement Series},
doi = {10.3847/1538-4365/ac366c}
}

@ARTICLE{jarvis2002,
       author = {{Jarvis}, Matt J. and {McLure}, Ross J.},
        title = "{On the black hole mass-radio luminosity relation for flat-spectrum radio-loud quasars}",
      journal = {\mnras},
         year = 2002,
        month = oct,
       volume = {336},
       number = {2},
        pages = {L38-L42},
          doi = {10.1046/j.1365-8711.2002.05997.x},
archivePrefix = {arXiv},
       eprint = {astro-ph/0208390},
 primaryClass = {astro-ph},
       adsurl = {https://ui.adsabs.harvard.edu/abs/2002MNRAS.336L..38J}
}

@article{metcalf2006,
    author = {Metcalf, R. B. and Magliocchetti, M.},
    title = {The role of black hole mass in quasar radio activity},
    journal = {\mnras},
    volume = {365},
    number = {1},
    pages = {101-109},
    year = {2006},
    month = {01},
    issn = {0035-8711},
    doi = {10.1111/j.1365-2966.2005.09649.x},
    url = {https://doi.org/10.1111/j.1365-2966.2005.09649.x},
    eprint = {https://academic.oup.com/mnras/article-pdf/365/1/101/5529890/365-1-101.pdf},
}

@ARTICLE{wylezalek2013,
       author = {{Wylezalek}, Dominika and {Galametz}, Audrey and {Stern}, Daniel and {Vernet}, Jo{\"e}l and {De Breuck}, Carlos and {Seymour}, Nick and {Brodwin}, Mark and {Eisenhardt}, Peter R.~M. and {Gonzalez}, Anthony H. and {Hatch}, Nina and {Jarvis}, Matt and {Rettura}, Alessandro and {Stanford}, Spencer A. and {Stevens}, Jason A.},
        title = "{Galaxy Clusters around Radio-loud Active Galactic Nuclei at 1.3 < z < 3.2 as Seen by Spitzer}",
      journal = {\apj},
         year = 2013,
        month = may,
       volume = {769},
       number = {1},
          eid = {79},
        pages = {79},
          doi = {10.1088/0004-637X/769/1/79},
archivePrefix = {arXiv},
       eprint = {1304.0770},
 primaryClass = {astro-ph.CO},
       adsurl = {https://ui.adsabs.harvard.edu/abs/2013ApJ...769...79W}
}

@article{castignani2014,
doi = {10.1088/0004-637X/792/2/114},
url = {https://dx.doi.org/10.1088/0004-637X/792/2/114},
year = {2014},
month = {aug},
publisher = {The American Astronomical Society},
volume = {792},
number = {2},
pages = {114},
author = {Castignani, G. and Chiaberge, M. and Celotti, A. and Norman, C. and De Zotti, G.},
title = {CLUSTER CANDIDATES AROUND LOW-POWER RADIO GALAXIES AT z ∼ 1–2 IN COSMOS},
journal = {The Astrophysical Journal}
}

@ARTICLE{kolwa2019,
       author = {{Kolwa}, S. and {Jarvis}, M.~J. and {McAlpine}, K. and {Heywood}, I.},
        title = "{The relation between galaxy density and radio jet power for 1.4 GHz VLA selected AGNs in Stripe 82}",
      journal = {\mnras},
         year = 2019,
        month = feb,
       volume = {482},
       number = {4},
        pages = {5156-5166},
          doi = {10.1093/mnras/sty3019},
archivePrefix = {arXiv},
       eprint = {1811.01822},
 primaryClass = {astro-ph.GA},
       adsurl = {https://ui.adsabs.harvard.edu/abs/2019MNRAS.482.5156K}
}

@ARTICLE{croston2019,
       author = {{Croston}, J.~H. and {Hardcastle}, M.~J. and {Mingo}, B. and {Best}, P.~N. and {Sabater}, J. and {Shimwell}, T.~M. and {Williams}, W.~L. and {Duncan}, K.~J. and {R{\"o}ttgering}, H.~J.~A. and {Brienza}, M. and {G{\"u}rkan}, G. and {Ineson}, J. and {Miley}, G.~K. and {Morabito}, L.~M. and {O'Sullivan}, S.~P. and {Prandoni}, I.},
        title = "{The environments of radio-loud AGN from the LOFAR Two-Metre Sky Survey (LoTSS)}",
      journal = {\aap},
         year = 2019,
        month = feb,
       volume = {622},
          eid = {A10},
        pages = {A10},
          doi = {10.1051/0004-6361/201834019},
archivePrefix = {arXiv},
       eprint = {1811.07949},
 primaryClass = {astro-ph.GA},
       adsurl = {https://ui.adsabs.harvard.edu/abs/2019A&A...622A..10C}
}

@ARTICLE{bardelli2010,
       author = {{Bardelli}, S. and {Schinnerer}, E. and {Smol{\v{c}}ic}, V. and {Zamorani}, G. and {Zucca}, E. and {Mignoli}, M. and {Halliday}, C. and {Kova{\v{c}}}, K. and {Ciliegi}, P. and {Caputi}, K. and {Koekemoer}, A.~M. and {Bongiorno}, A. and {Bondi}, M. and {Bolzonella}, M. and {Vergani}, D. and {Pozzetti}, L. and {Carollo}, C.~M. and {Contini}, T. and {Kneib}, J. -P. and {Le F{\`e}vre}, O. and {Lilly}, S. and {Mainieri}, V. and {Renzini}, A. and {Scodeggio}, M. and {Coppa}, G. and {Cucciati}, O. and {de la Torre}, S. and {de Ravel}, L. and {Franzetti}, P. and {Garilli}, B. and {Iovino}, A. and {Kampczyk}, P. and {Knobel}, C. and {Lamareille}, F. and {Le Borgne}, J. -F. and {Le Brun}, V. and {Maier}, C. and {Pell{\`o}}, R. and {Peng}, Y. and {Perez-Montero}, E. and {Ricciardelli}, E. and {Silverman}, J.~D. and {Tanaka}, M. and {Tasca}, L. and {Tresse}, L. and {Abbas}, U. and {Bottini}, D. and {Cappi}, A. and {Cassata}, P. and {Cimatti}, A. and {Guzzo}, L. and {Leauthaud}, A. and {Maccagni}, D. and {Marinoni}, C. and {McCracken}, H.~J. and {Memeo}, P. and {Meneux}, B. and {Oesch}, P. and {Porciani}, C. and {Scaramella}, R. and {Capak}, P. and {Sanders}, D. and {Scoville}, N. and {Taniguchi}, Y. and {Jahnke}, K.},
        title = "{Properties and environment of radio-emitting galaxies in the VLA-zCOSMOS survey}",
      journal = {\aap},
         year = 2010,
        month = feb,
       volume = {511},
          eid = {A1},
        pages = {A1},
          doi = {10.1051/0004-6361/200912809},
archivePrefix = {arXiv},
       eprint = {0911.0523},
 primaryClass = {astro-ph.CO},
       adsurl = {https://ui.adsabs.harvard.edu/abs/2010A&A...511A...1B}
}

@article{mo2020,
doi = {10.3847/1538-4357/abb08d},
url = {https://dx.doi.org/10.3847/1538-4357/abb08d},
year = {2020},
month = {sep},
publisher = {The American Astronomical Society},
volume = {901},
number = {2},
pages = {131},
author = {Mo, Wenli and Gonzalez, Anthony and Brodwin, Mark and Decker, Bandon and Eisenhardt, Peter and Moravec, Emily and Stanford, S. A. and Stern, Daniel and Wylezalek, Dominika},
title = {The Massive and Distant Clusters of WISE Survey. VIII. Radio Activity in Massive Galaxy Clusters},
journal = {The Astrophysical Journal}
}

@ARTICLE{magliocchetti2022,
       author = {{Magliocchetti}, Manuela},
        title = "{Hosts and environments: a (large-scale) radio history of AGN and star-forming galaxies}",
      journal = {\aapr},
         year = 2022,
        month = dec,
       volume = {30},
       number = {1},
          eid = {6},
        pages = {6},
          doi = {10.1007/s00159-022-00142-1},
archivePrefix = {arXiv},
       eprint = {2206.15286},
 primaryClass = {astro-ph.CO},
       adsurl = {https://ui.adsabs.harvard.edu/abs/2022A&ARv..30....6M}
}

@ARTICLE{brand2005,
       author = {{Brand}, Kate and {Rawlings}, Steve and {Hill}, Gary J. and {Tufts}, Joseph R.},
        title = "{The three-dimensional clustering of radio galaxies in the Texas-Oxford NVSS structure survey}",
      journal = {\mnras},
         year = 2005,
        month = mar,
       volume = {357},
       number = {4},
        pages = {1231-1254},
          doi = {10.1111/j.1365-2966.2005.08719.x},
archivePrefix = {arXiv},
       eprint = {astro-ph/0412237},
 primaryClass = {astro-ph},
       adsurl = {https://ui.adsabs.harvard.edu/abs/2005MNRAS.357.1231B}
}

@ARTICLE{magliocchetti2004,
       author = {{Magliocchetti}, Manuela and {Maddox}, Steve J. and {Hawkins}, Ed and {Peacock}, John A. and {Bland-Hawthorn}, Joss and {Bridges}, Terry and {Cannon}, Russell and {Cole}, Shaun and {Colless}, Matthew and {Collins}, Chris and {Couch}, Warrick and {Dalton}, Gavin and {de Propris}, Roberto and {Driver}, Simon P. and {Efstathiou}, George and {Ellis}, Richard S. and {Frenk}, Carlos S. and {Glazebrook}, Karl and {Jackson}, Carole A. and {Jones}, Bryn and {Lahav}, Ofer and {Lewis}, Ian and {Lumsden}, Stuart and {Norberg}, Peder and {Peterson}, Bruce A. and {Sutherland}, Will and {Taylor}, Keith and {2dFGRS Team}},
        title = "{The 2dF galaxy redshift survey: clustering properties of radio galaxies}",
      journal = {\mnras},
         year = 2004,
        month = jun,
       volume = {350},
       number = {4},
        pages = {1485-1494},
          doi = {10.1111/j.1365-2966.2004.07751.x},
archivePrefix = {arXiv},
       eprint = {astro-ph/0312160},
 primaryClass = {astro-ph},
       adsurl = {https://ui.adsabs.harvard.edu/abs/2004MNRAS.350.1485M}
}

@ARTICLE{wake2008,
       author = {{Wake}, David A. and {Croom}, Scott M. and {Sadler}, Elaine M. and {Johnston}, Helen M.},
        title = "{The clustering of radio galaxies at z \raisebox{-0.5ex}\textasciitilde= 0.55 from the 2SLAQ LRG survey}",
      journal = {\mnras},
         year = 2008,
        month = dec,
       volume = {391},
       number = {4},
        pages = {1674-1684},
          doi = {10.1111/j.1365-2966.2008.14039.x},
archivePrefix = {arXiv},
       eprint = {0810.1050},
 primaryClass = {astro-ph},
       adsurl = {https://ui.adsabs.harvard.edu/abs/2008MNRAS.391.1674W}
}

@ARTICLE{fine2011,
       author = {{Fine}, S. and {Shanks}, T. and {Nikoloudakis}, N. and {Sawangwit}, U.},
        title = "{Evolution in the clustering strength of radio galaxies}",
      journal = {\mnras},
         year = 2011,
        month = dec,
       volume = {418},
       number = {4},
        pages = {2251-2259},
          doi = {10.1111/j.1365-2966.2011.19527.x},
archivePrefix = {arXiv},
       eprint = {1107.5666},
 primaryClass = {astro-ph.CO},
       adsurl = {https://ui.adsabs.harvard.edu/abs/2011MNRAS.418.2251F}
}

@ARTICLE{lindsay2014a,
       author = {{Lindsay}, S.~N. and {Jarvis}, M.~J. and {McAlpine}, K.},
        title = "{Evolution in the bias of faint radio sources to z {\ensuremath{\sim}} 2.2}",
      journal = {\mnras},
         year = 2014,
        month = may,
       volume = {440},
       number = {3},
        pages = {2322-2332},
          doi = {10.1093/mnras/stu453},
archivePrefix = {arXiv},
       eprint = {1403.0882},
 primaryClass = {astro-ph.CO},
       adsurl = {https://ui.adsabs.harvard.edu/abs/2014MNRAS.440.2322L}
}

@ARTICLE{bertin1996,
       author = {{Bertin}, E. and {Arnouts}, S.},
        title = "{SExtractor: Software for source extraction.}",
      journal = {\aaps},
         year = 1996,
        month = jun,
       volume = {117},
        pages = {393-404},
          doi = {10.1051/aas:1996164},
       adsurl = {https://ui.adsabs.harvard.edu/abs/1996A&AS..117..393B}
}

@article{mo2018,
doi = {10.3847/1538-4357/aaef83},
url = {https://dx.doi.org/10.3847/1538-4357/aaef83},
year = {2018},
month = {dec},
publisher = {The American Astronomical Society},
volume = {869},
number = {2},
pages = {131},
author = {Mo, Wenli and Gonzalez, Anthony and Stern, Daniel and Brodwin, Mark and Decker, Bandon and Eisenhardt, Peter and Moravec, Emily and Stanford, S. A. and Wylezalek, Dominika},
title = {The Massive and Distant Clusters of WISE Survey. IV. The Distribution of Active Galactic Nuclei in Galaxy Clusters at z ∼ 1},
journal = {The Astrophysical Journal}
}

@ARTICLE{smolcic2011,
       author = {{Smol{\v{c}}i{\'c}}, V. and {Finoguenov}, A. and {Zamorani}, G. and {Schinnerer}, E. and {Tanaka}, M. and {Giodini}, S. and {Scoville}, N.},
        title = "{On the occupation of X-ray-selected galaxy groups by radio active galactic nuclei since z = 1.3}",
      journal = {\mnras},
         year = 2011,
        month = sep,
       volume = {416},
       number = {1},
        pages = {L31-L35},
          doi = {10.1111/j.1745-3933.2011.01092.x},
archivePrefix = {arXiv},
       eprint = {1108.2952},
 primaryClass = {astro-ph.CO},
       adsurl = {https://ui.adsabs.harvard.edu/abs/2011MNRAS.416L..31S}
}

@ARTICLE{antonucci1993,
       author = {{Antonucci}, Robert},
        title = "{Unified models for active galactic nuclei and quasars.}",
      journal = {\araa},
         year = 1993,
        month = jan,
       volume = {31},
        pages = {473-521},
          doi = {10.1146/annurev.aa.31.090193.002353},
       adsurl = {https://ui.adsabs.harvard.edu/abs/1993ARA&A..31..473A}
}

@ARTICLE{rawlings2004,
       author = {{Rawlings}, Steve and {Jarvis}, Matt J.},
        title = "{Evidence that powerful radio jets have a profound influence on the evolution of galaxies}",
      journal = {\mnras},
         year = 2004,
        month = dec,
       volume = {355},
       number = {3},
        pages = {L9-L12},
          doi = {10.1111/j.1365-2966.2004.08234.x},
archivePrefix = {arXiv},
       eprint = {astro-ph/0409687},
 primaryClass = {astro-ph},
       adsurl = {https://ui.adsabs.harvard.edu/abs/2004MNRAS.355L...9R}
}

@ARTICLE{aihara2022,
       author = {{Aihara}, Hiroaki and {AlSayyad}, Yusra and {Ando}, Makoto and {Armstrong}, Robert and {Bosch}, James and {Egami}, Eiichi and {Furusawa}, Hisanori and {Furusawa}, Junko and {Harasawa}, Sumiko and {Harikane}, Yuichi and {Hsieh}, Bau-Ching and {Ikeda}, Hiroyuki and {Ito}, Kei and {Iwata}, Ikuru and {Kodama}, Tadayuki and {Koike}, Michitaro and {Kokubo}, Mitsuru and {Komiyama}, Yutaka and {Li}, Xiangchong and {Liang}, Yongming and {Lin}, Yen-Ting and {Lupton}, Robert H. and {Lust}, Nate B. and {MacArthur}, Lauren A. and {Mawatari}, Ken and {Mineo}, Sogo and {Miyatake}, Hironao and {Miyazaki}, Satoshi and {More}, Surhud and {Morishima}, Takahiro and {Murayama}, Hitoshi and {Nakajima}, Kimihiko and {Nakata}, Fumiaki and {Nishizawa}, Atsushi J. and {Oguri}, Masamune and {Okabe}, Nobuhiro and {Okura}, Yuki and {Ono}, Yoshiaki and {Osato}, Ken and {Ouchi}, Masami and {Pan}, Yen-Chen and {Plazas Malag{\'o}n}, Andr{\'e}s A. and {Price}, Paul A. and {Reed}, Sophie L. and {Rykoff}, Eli S. and {Shibuya}, Takatoshi and {Simunovic}, Mirko and {Strauss}, Michael A. and {Sugimori}, Kanako and {Suto}, Yasushi and {Suzuki}, Nao and {Takada}, Masahiro and {Takagi}, Yuhei and {Takata}, Tadafumi and {Takita}, Satoshi and {Tanaka}, Masayuki and {Tang}, Shenli and {Taranu}, Dan S. and {Terai}, Tsuyoshi and {Toba}, Yoshiki and {Turner}, Edwin L. and {Uchiyama}, Hisakazu and {Vijarnwannaluk}, Bovornpratch and {Waters}, Christopher Z. and {Yamada}, Yoshihiko and {Yamamoto}, Naoaki and {Yamashita}, Takuji},
        title = "{Third data release of the Hyper Suprime-Cam Subaru Strategic Program}",
      journal = {\pasj},
         year = 2022,
        month = apr,
       volume = {74},
       number = {2},
        pages = {247-272},
          doi = {10.1093/pasj/psab122},
archivePrefix = {arXiv},
       eprint = {2108.13045},
 primaryClass = {astro-ph.IM},
       adsurl = {https://ui.adsabs.harvard.edu/abs/2022PASJ...74..247A}
}

@INPROCEEDINGS{cuillandre2012,
       author = {{Cuillandre}, Jean-Charles J. and {Withington}, Kanoa and {Hudelot}, Patrick and {Goranova}, Yuliana and {McCracken}, Henry and {Magnard}, Fr{\'e}d{\'e}ric and {Mellier}, Yannick and {Regnault}, Nicolas and {B{\'e}toule}, Marc and {Aussel}, Herv{\'e} and {Kavelaars}, J.~J. and {Fernique}, Pierre and {Bonnarel}, Fran{\c{c}}ois and {Ochsenbein}, Francois and {Ilbert}, Olivier},
        title = "{Introduction to the CFHT Legacy Survey final release (CFHTLS T0007)}",
    booktitle = {Observatory Operations: Strategies, Processes, and Systems IV},
         year = 2012,
       editor = {{Peck}, Alison B. and {Seaman}, Robert L. and {Comeron}, Fernando},
       series = {Society of Photo-Optical Instrumentation Engineers (SPIE) Conference Series},
       volume = {8448},
        month = sep,
          eid = {84480M},
        pages = {84480M},
          doi = {10.1117/12.925584},
       adsurl = {https://ui.adsabs.harvard.edu/abs/2012SPIE.8448E..0MC}
}

@INPROCEEDINGS{vaccari2016,
       author = {{Vaccari}, M. and {Covone}, G. and {Radovich}, M. and {Grado}, A. and {Limatola}, L. and {Botticella}, M.~T. and {Cappellaro}, E. and {Paolillo}, M. and {Pignata}, G. and {De Cicco}, D. and {Falocco}, S. and {Marchetti}, L. and {Brescia}, M. and {Cavuoti}, S. and {Longo}, G. and {Capaccioli}, M. and {Napolitano}, N. and {Schipani}, P.},
        title = "{The VOICE Survey : VST Optical Imaging of the CDFS and ES1 Fields}",
    booktitle = {The 4th Annual Conference on High Energy Astrophysics in Southern Africa (HEASA 2016)},
         year = 2016,
        month = jan,
          eid = {26},
        pages = {26},
          doi = {10.22323/1.275.0026},
archivePrefix = {arXiv},
       eprint = {1704.01495},
 primaryClass = {astro-ph.GA},
       adsurl = {https://ui.adsabs.harvard.edu/abs/2016heas.confE..26V}
}

@ARTICLE{uchiyama2022,
       author = {{Uchiyama}, Hisakazu and {Yamashita}, Takuji and {Toshikawa}, Jun and {Kashikawa}, Nobunari and {Ichikawa}, Kohei and {Kubo}, Mariko and {Ito}, Kei and {Kawakatu}, Nozomu and {Nagao}, Tohru and {Toba}, Yoshiki and {Ono}, Yoshiaki and {Harikane}, Yuichi and {Imanishi}, Masatoshi and {Kajisawa}, Masaru and {Lee}, Chien-Hsiu and {Liang}, Yongming},
        title = "{A Wide and Deep Exploration of Radio Galaxies with Subaru HSC (WERGS). VI. Distant Filamentary Structures Pointed Out by High-z Radio Galaxies at z 4}",
      journal = {\apj},
         year = 2022,
        month = feb,
       volume = {926},
       number = {1},
          eid = {76},
        pages = {76},
          doi = {10.3847/1538-4357/ac441c},
archivePrefix = {arXiv},
       eprint = {2112.01684},
 primaryClass = {astro-ph.GA},
       adsurl = {https://ui.adsabs.harvard.edu/abs/2022ApJ...926...76U}
}

@ARTICLE{mccracken2012,
       author = {{McCracken}, H.~J. and {Milvang-Jensen}, B. and {Dunlop}, J. and {Franx}, M. and {Fynbo}, J.~P.~U. and {Le F{\`e}vre}, O. and {Holt}, J. and {Caputi}, K.~I. and {Goranova}, Y. and {Buitrago}, F. and {Emerson}, J.~P. and {Freudling}, W. and {Hudelot}, P. and {L{\'o}pez-Sanjuan}, C. and {Magnard}, F. and {Mellier}, Y. and {M{\o}ller}, P. and {Nilsson}, K.~K. and {Sutherland}, W. and {Tasca}, L. and {Zabl}, J.},
        title = "{UltraVISTA: a new ultra-deep near-infrared survey in COSMOS}",
      journal = {\aap},
         year = 2012,
        month = aug,
       volume = {544},
          eid = {A156},
        pages = {A156},
          doi = {10.1051/0004-6361/201219507},
archivePrefix = {arXiv},
       eprint = {1204.6586},
 primaryClass = {astro-ph.CO},
       adsurl = {https://ui.adsabs.harvard.edu/abs/2012A&A...544A.156M}
}

@ARTICLE{moster2011,
       author = {{Moster}, Benjamin P. and {Somerville}, Rachel S. and {Newman}, Jeffrey A. and {Rix}, Hans-Walter},
        title = "{A Cosmic Variance Cookbook}",
      journal = {\apj},
         year = 2011,
        month = apr,
       volume = {731},
       number = {2},
          eid = {113},
        pages = {113},
          doi = {10.1088/0004-637X/731/2/113},
archivePrefix = {arXiv},
       eprint = {1001.1737},
 primaryClass = {astro-ph.CO},
       adsurl = {https://ui.adsabs.harvard.edu/abs/2011ApJ...731..113M}
}

@ARTICLE{despali2016,
       author = {{Despali}, Giulia and {Giocoli}, Carlo and {Angulo}, Raul E. and {Tormen}, Giuseppe and {Sheth}, Ravi K. and {Baso}, Giacomo and {Moscardini}, Lauro},
        title = "{The universality of the virial halo mass function and models for non-universality of other halo definitions}",
      journal = {\mnras},
         year = 2016,
        month = mar,
       volume = {456},
       number = {3},
        pages = {2486-2504},
          doi = {10.1093/mnras/stv2842},
archivePrefix = {arXiv},
       eprint = {1507.05627},
 primaryClass = {astro-ph.CO},
       adsurl = {https://ui.adsabs.harvard.edu/abs/2016MNRAS.456.2486D}
}

@INCOLLECTION{fortson2012,
       author = {{Fortson}, Lucy and {Masters}, Karen and {Nichol}, Robert and {Borne}, Kirk D. and {Edmondson}, Edward M. and {Lintott}, Chris and {Raddick}, Jordan and {Schawinski}, Kevin and {Wallin}, John},
        title = "{Galaxy Zoo: Morphological Classification and Citizen Science}",
    booktitle = {Advances in Machine Learning and Data Mining for Astronomy},
         year = 2012,
       editor = {{Way}, Michael J. and {Scargle}, Jeffrey D. and {Ali}, Kamal M. and {Srivastava}, Ashok N.},
        pages = {213-236},
          doi = {10.48550/arXiv.1104.5513},
       adsurl = {https://ui.adsabs.harvard.edu/abs/2012amld.book..213F}
}

@ARTICLE{mcalpine2012,
       author = {{McAlpine}, K. and {Smith}, D.~J.~B. and {Jarvis}, M.~J. and {Bonfield}, D.~G. and {Fleuren}, S.},
        title = "{The likelihood ratio as a tool for radio continuum surveys with Square Kilometre Array precursor telescopes}",
      journal = {\mnras},
         year = 2012,
        month = jun,
       volume = {423},
       number = {1},
        pages = {132-140},
          doi = {10.1111/j.1365-2966.2012.20715.x},
archivePrefix = {arXiv},
       eprint = {1202.1958},
 primaryClass = {astro-ph.IM},
       adsurl = {https://ui.adsabs.harvard.edu/abs/2012MNRAS.423..132M}
}

@ARTICLE{kondapally2021,
       author = {{Kondapally}, R. and {Best}, P.~N. and {Hardcastle}, M.~J. and {Nisbet}, D. and {Bonato}, M. and {Sabater}, J. and {Duncan}, K.~J. and {McCheyne}, I. and {Cochrane}, R.~K. and {Bowler}, R.~A.~A. and {Williams}, W.~L. and {Shimwell}, T.~W. and {Tasse}, C. and {Croston}, J.~H. and {Goyal}, A. and {Jamrozy}, M. and {Jarvis}, M.~J. and {Mahatma}, V.~H. and {R{\"o}ttgering}, H.~J.~A. and {Smith}, D.~J.~B. and {Wo{\l}owska}, A. and {Bondi}, M. and {Brienza}, M. and {Brown}, M.~J.~I. and {Br{\"u}ggen}, M. and {Chambers}, K. and {Garrett}, M.~A. and {G{\"u}rkan}, G. and {Huber}, M. and {Kunert-Bajraszewska}, M. and {Magnier}, E. and {Mingo}, B. and {Mostert}, R. and {Nikiel-Wroczy{\'n}ski}, B. and {O'Sullivan}, S.~P. and {Paladino}, R. and {Ploeckinger}, T. and {Prandoni}, I. and {Rosenthal}, M.~J. and {Schwarz}, D.~J. and {Shulevski}, A. and {Wagenveld}, J.~D. and {Wang}, L.},
        title = "{The LOFAR Two-meter Sky Survey: Deep Fields Data Release 1. III. Host-galaxy identifications and value added catalogues}",
      journal = {\aap},
         year = 2021,
        month = apr,
       volume = {648},
          eid = {A3},
        pages = {A3},
          doi = {10.1051/0004-6361/202038813},
archivePrefix = {arXiv},
       eprint = {2011.08201},
 primaryClass = {astro-ph.GA},
       adsurl = {https://ui.adsabs.harvard.edu/abs/2021A&A...648A...3K}
}

@ARTICLE{chakraborty2020,
       author = {{Chakraborty}, Arnab and {Dutta}, Prasun and {Datta}, Abhirup and {Roy}, Nirupam},
        title = "{The study of the angular and spatial distribution of radio-selected AGNs and star-forming galaxies in the ELAIS N1 field}",
      journal = {\mnras},
         year = 2020,
        month = may,
       volume = {494},
       number = {3},
        pages = {3392-3404},
          doi = {10.1093/mnras/staa945},
archivePrefix = {arXiv},
       eprint = {2002.12383},
 primaryClass = {astro-ph.GA},
       adsurl = {https://ui.adsabs.harvard.edu/abs/2020MNRAS.494.3392C}
}

@ARTICLE{berlind2002,
       author = {{Berlind}, Andreas A. and {Weinberg}, David H.},
        title = "{The Halo Occupation Distribution: Toward an Empirical Determination of the Relation between Galaxies and Mass}",
      journal = {\apj},
         year = 2002,
        month = aug,
       volume = {575},
       number = {2},
        pages = {587-616},
          doi = {10.1086/341469},
archivePrefix = {arXiv},
       eprint = {astro-ph/0109001},
 primaryClass = {astro-ph},
       adsurl = {https://ui.adsabs.harvard.edu/abs/2002ApJ...575..587B}
}

@article{quenouille1956,
 ISSN = {00063444, 14643510},
 URL = {http://www.jstor.org/stable/2332914},
 author = {M. H. Quenouille},
 journal = {Biometrika},
 number = {3/4},
 pages = {353--360},
 publisher = {[Oxford University Press, Biometrika Trust]},
 title = {Notes on Bias in Estimation},
 urldate = {2025-05-06},
 volume = {43},
 year = {1956}
}

@ARTICLE{salcedo2020,
       author = {{Salcedo}, Andr{\'e}s N. and {Wibking}, Benjamin D. and {Weinberg}, David H. and {Wu}, Hao-Yi and {Ferrer}, Douglas and {Eisenstein}, Daniel and {Pinto}, Philip},
        title = "{Cosmology with stacked cluster weak lensing and cluster-galaxy cross-correlations}",
      journal = {\mnras},
         year = 2020,
        month = jan,
       volume = {491},
       number = {3},
        pages = {3061-3081},
          doi = {10.1093/mnras/stz2963},
archivePrefix = {arXiv},
       eprint = {1906.06499},
 primaryClass = {astro-ph.CO},
       adsurl = {https://ui.adsabs.harvard.edu/abs/2020MNRAS.491.3061S}
}

@ARTICLE{arnouts2002,
       author = {{Arnouts}, S. and {Moscardini}, L. and {Vanzella}, E. and {Colombi}, S. and {Cristiani}, S. and {Fontana}, A. and {Giallongo}, E. and {Matarrese}, S. and {Saracco}, P.},
        title = "{Measuring the redshift evolution of clustering: the Hubble Deep Field South}",
      journal = {\mnras},
         year = 2002,
        month = jan,
       volume = {329},
       number = {2},
        pages = {355-366},
          doi = {10.1046/j.1365-8711.2002.04988.x},
archivePrefix = {arXiv},
       eprint = {astro-ph/0109453},
 primaryClass = {astro-ph},
       adsurl = {https://ui.adsabs.harvard.edu/abs/2002MNRAS.329..355A}
}

@ARTICLE{allison2015,
       author = {{Allison}, Rupert and {Lindsay}, Sam N. and {Sherwin}, Blake D. and {de Bernardis}, Francesco and {Bond}, J. Richard and {Calabrese}, Erminia and {Devlin}, Mark J. and {Dunkley}, Joanna and {Gallardo}, Patricio and {Henderson}, Shawn and {Hincks}, Adam D. and {Hlozek}, Ren{\'e}e and {Jarvis}, Matt and {Kosowsky}, Arthur and {Louis}, Thibaut and {Madhavacheril}, Mathew and {McMahon}, Jeff and {Moodley}, Kavilan and {Naess}, Sigurd and {Newburgh}, Laura and {Niemack}, Michael D. and {Page}, Lyman A. and {Partridge}, Bruce and {Sehgal}, Neelima and {Spergel}, David N. and {Staggs}, Suzanne T. and {van Engelen}, Alexander and {Wollack}, Edward J.},
        title = "{The Atacama Cosmology Telescope: measuring radio galaxy bias through cross-correlation with lensing}",
      journal = {\mnras},
         year = 2015,
        month = jul,
       volume = {451},
       number = {1},
        pages = {849-858},
          doi = {10.1093/mnras/stv991},
archivePrefix = {arXiv},
       eprint = {1502.06456},
 primaryClass = {astro-ph.CO},
       adsurl = {https://ui.adsabs.harvard.edu/abs/2015MNRAS.451..849A}
}

@INPROCEEDINGS{cress1996,
       author = {{Cress}, C.~M. and {Helfand}, D.~J. and {Becker}, R.~H. and {Gregg}, M.~D. and {White}, R.~L.},
        title = "{The Angular Two Point Correlation Function for the FIRST Radio Survey}",
    booktitle = {Clusters, Lensing, and the Future of the Universe},
         year = 1996,
       editor = {{Trimble}, Virginia and {Reisenegger}, Andreas},
       series = {Astronomical Society of the Pacific Conference Series},
       volume = {88},
        month = jan,
        pages = {193},
          doi = {10.48550/arXiv.astro-ph/9509043},
archivePrefix = {arXiv},
       eprint = {astro-ph/9509043},
 primaryClass = {astro-ph},
       adsurl = {https://ui.adsabs.harvard.edu/abs/1996ASPC...88..193C}
}

@article{mazumder2022,
    author = {Mazumder, Aishrila and Chakraborty, Arnab and Datta, Abhirup},
    title = {A study on the clustering properties of radio-selected sources in the Lockman Hole region at 325 MHz},
    journal = {\mnras},
    volume = {517},
    number = {3},
    pages = {3407-3422},
    year = {2022},
    month = {10},
    issn = {0035-8711},
    doi = {10.1093/mnras/stac2801},
    url = {https://doi.org/10.1093/mnras/stac2801},
    eprint = {https://academic.oup.com/mnras/article-pdf/517/3/3407/46649687/stac2801.pdf},
}

@ARTICLE{magliocchetti2014,
       author = {{Magliocchetti}, M. and {Lutz}, D. and {Rosario}, D. and {Berta}, S. and {Le Floc'h}, E. and {Magnelli}, B. and {Pozzi}, F. and {Riguccini}, L. and {Santini}, P.},
        title = "{The PEP survey: infrared properties of radio-selected AGN}",
      journal = {\mnras},
         year = 2014,
        month = jul,
       volume = {442},
       number = {1},
        pages = {682-693},
          doi = {10.1093/mnras/stu863},
archivePrefix = {arXiv},
       eprint = {1404.7014},
 primaryClass = {astro-ph.GA},
       adsurl = {https://ui.adsabs.harvard.edu/abs/2014MNRAS.442..682M}
}

@ARTICLE{blake2002,
       author = {{Blake}, Chris and {Wall}, Jasper},
        title = "{Quantifying angular clustering in wide-area radio surveys}",
      journal = {\mnras},
         year = 2002,
        month = dec,
       volume = {337},
       number = {3},
        pages = {993-1003},
          doi = {10.1046/j.1365-8711.2002.05979.x},
archivePrefix = {arXiv},
       eprint = {astro-ph/0208350},
 primaryClass = {astro-ph},
       adsurl = {https://ui.adsabs.harvard.edu/abs/2002MNRAS.337..993B}
}

@ARTICLE{overzier2003,
       author = {{Overzier}, R.~A. and {R{\"o}ttgering}, H.~J.~A. and {Rengelink}, R.~B. and {Wilman}, R.~J.},
        title = "{The spatial clustering of radio sources in NVSS and FIRST; implications for galaxy clustering evolution}",
      journal = {\aap},
         year = 2003,
        month = jul,
       volume = {405},
        pages = {53-72},
          doi = {10.1051/0004-6361:20030527},
archivePrefix = {arXiv},
       eprint = {astro-ph/0304160},
 primaryClass = {astro-ph},
       adsurl = {https://ui.adsabs.harvard.edu/abs/2003A&A...405...53O}
}

@ARTICLE{gilli2019,
       author = {{Gilli}, R. and {Mignoli}, M. and {Peca}, A. and {Nanni}, R. and {Prandoni}, I. and {Liuzzo}, E. and {D'Amato}, Q. and {Brusa}, M. and {Calura}, F. and {Caminha}, G.~B. and {Chiaberge}, M. and {Comastri}, A. and {Cucciati}, O. and {Cusano}, F. and {Grandi}, P. and {Decarli}, R. and {Lanzuisi}, G. and {Mannucci}, F. and {Pinna}, E. and {Tozzi}, P. and {Vanzella}, E. and {Vignali}, C. and {Vito}, F. and {Balmaverde}, B. and {Citro}, A. and {Cappelluti}, N. and {Zamorani}, G. and {Norman}, C.},
        title = "{Discovery of a galaxy overdensity around a powerful, heavily obscured FRII radio galaxy at z = 1.7: star formation promoted by large-scale AGN feedback?}",
      journal = {\aap},
         year = 2019,
        month = dec,
       volume = {632},
          eid = {A26},
        pages = {A26},
          doi = {10.1051/0004-6361/201936121},
archivePrefix = {arXiv},
       eprint = {1909.00814},
 primaryClass = {astro-ph.GA},
       adsurl = {https://ui.adsabs.harvard.edu/abs/2019A&A...632A..26G}
}

@ARTICLE{mcnamara2012,
       author = {{McNamara}, B.~R. and {Nulsen}, P.~E.~J.},
        title = "{Mechanical feedback from active galactic nuclei in galaxies, groups and clusters}",
      journal = {New Journal of Physics},
         year = 2012,
        month = may,
       volume = {14},
       number = {5},
          eid = {055023},
        pages = {055023},
          doi = {10.1088/1367-2630/14/5/055023},
archivePrefix = {arXiv},
       eprint = {1204.0006},
 primaryClass = {astro-ph.CO},
       adsurl = {https://ui.adsabs.harvard.edu/abs/2012NJPh...14e5023M}
}

@ARTICLE{hale2024,
       author = {{Hale}, C.~L. and {Schwarz}, D.~J. and {Best}, P.~N. and {Nakoneczny}, S.~J. and {Alonso}, D. and {Bacon}, D. and {B{\"o}hme}, L. and {Bhardwaj}, N. and {Bilicki}, M. and {Camera}, S. and {Heneka}, C.~S. and {Pashapour-Ahmadabadi}, M. and {Tiwari}, P. and {Zheng}, J. and {Duncan}, K.~J. and {Jarvis}, M.~J. and {Kondapally}, R. and {Magliocchetti}, M. and {Rottgering}, H.~J.~A. and {Shimwell}, T.~W.},
        title = "{Cosmology from LOFAR Two-metre Sky Survey Data Release 2: angular clustering of radio sources}",
      journal = {\mnras},
         year = 2024,
        month = jan,
       volume = {527},
       number = {3},
        pages = {6540-6568},
          doi = {10.1093/mnras/stad3088},
archivePrefix = {arXiv},
       eprint = {2310.07627},
 primaryClass = {astro-ph.CO},
       adsurl = {https://ui.adsabs.harvard.edu/abs/2024MNRAS.527.6540H}
}

@ARTICLE{best2023,
       author = {{Best}, P.~N. and {Kondapally}, R. and {Williams}, W.~L. and {Cochrane}, R.~K. and {Duncan}, K.~J. and {Hale}, C.~L. and {Haskell}, P. and {Ma{\l}ek}, K. and {McCheyne}, I. and {Smith}, D.~J.~B. and {Wang}, L. and {Botteon}, A. and {Bonato}, M. and {Bondi}, M. and {Calistro Rivera}, G. and {Gao}, F. and {G{\"u}rkan}, G. and {Hardcastle}, M.~J. and {Jarvis}, M.~J. and {Mingo}, B. and {Miraghaei}, H. and {Morabito}, L.~K. and {Nisbet}, D. and {Prandoni}, I. and {R{\"o}ttgering}, H.~J.~A. and {Sabater}, J. and {Shimwell}, T. and {Tasse}, C. and {van Weeren}, R.},
        title = "{The LOFAR Two-metre Sky Survey: Deep Fields data release 1. V. Survey description, source classifications, and host galaxy properties}",
      journal = {\mnras},
         year = 2023,
        month = aug,
       volume = {523},
       number = {2},
        pages = {1729-1755},
          doi = {10.1093/mnras/stad1308},
archivePrefix = {arXiv},
       eprint = {2305.05782},
 primaryClass = {astro-ph.GA},
       adsurl = {https://ui.adsabs.harvard.edu/abs/2023MNRAS.523.1729B}
}

@ARTICLE{hatch2014,
       author = {{Hatch}, N.~A. and {Wylezalek}, D. and {Kurk}, J.~D. and {Stern}, D. and {De Breuck}, C. and {Jarvis}, M.~J. and {Galametz}, A. and {Gonzalez}, A.~H. and {Hartley}, W.~G. and {Mortlock}, A. and {Seymour}, N. and {Stevens}, J.~A.},
        title = "{Why z > 1 radio-loud galaxies are commonly located in protoclusters}",
      journal = {\mnras},
         year = 2014,
        month = nov,
       volume = {445},
       number = {1},
        pages = {280-289},
          doi = {10.1093/mnras/stu1725},
archivePrefix = {arXiv},
       eprint = {1409.1218},
 primaryClass = {astro-ph.GA},
       adsurl = {https://ui.adsabs.harvard.edu/abs/2014MNRAS.445..280H}
}

@ARTICLE{calderon2019,
       author = {{Calderon}, Victor F. and {Berlind}, Andreas A.},
        title = "{Prediction of galaxy halo masses in SDSS DR7 via a machine learning approach}",
      journal = {\mnras},
         year = 2019,
        month = dec,
       volume = {490},
       number = {2},
        pages = {2367-2379},
          doi = {10.1093/mnras/stz2775},
archivePrefix = {arXiv},
       eprint = {1902.02680},
 primaryClass = {astro-ph.GA},
       adsurl = {https://ui.adsabs.harvard.edu/abs/2019MNRAS.490.2367C}
}

@ARTICLE{lim2017,
       author = {{Lim}, S.~H. and {Mo}, H.~J. and {Lu}, Yi and {Wang}, Huiyuan and {Yang}, Xiaohu},
        title = "{Galaxy groups in the low-redshift Universe}",
      journal = {\mnras},
         year = 2017,
        month = sep,
       volume = {470},
       number = {3},
        pages = {2982-3005},
          doi = {10.1093/mnras/stx1462},
archivePrefix = {arXiv},
       eprint = {1706.02307},
 primaryClass = {astro-ph.GA},
       adsurl = {https://ui.adsabs.harvard.edu/abs/2017MNRAS.470.2982L}
}

@ARTICLE{jarvis2001,
       author = {{Jarvis}, Matt J. and {Rawlings}, Steve and {Willott}, Chris J. and {Blundell}, Katherine M. and {Eales}, Steve and {Lacy}, Mark},
        title = "{On the redshift cut-off for steep-spectrum radio sources}",
      journal = {\mnras},
         year = 2001,
        month = nov,
       volume = {327},
       number = {3},
        pages = {907-917},
          doi = {10.1046/j.1365-8711.2001.04778.x},
archivePrefix = {arXiv},
       eprint = {astro-ph/0106473},
 primaryClass = {astro-ph},
       adsurl = {https://ui.adsabs.harvard.edu/abs/2001MNRAS.327..907J}
}

@ARTICLE{dunlop1990,
       author = {{Dunlop}, J.~S. and {Peacock}, J.~A.},
        title = "{The redshift cut-off in the luminosity function of radio galaxies and quasars.}",
      journal = {\mnras},
         year = 1990,
        month = nov,
       volume = {247},
        pages = {19},
       adsurl = {https://ui.adsabs.harvard.edu/abs/1990MNRAS.247...19D}
}

@ARTICLE{rigby2011,
       author = {{Rigby}, E.~E. and {Best}, P.~N. and {Brookes}, M.~H. and {Peacock}, J.~A. and {Dunlop}, J.~S. and {R{\"o}ttgering}, H.~J.~A. and {Wall}, J.~V. and {Ker}, L.},
        title = "{The luminosity-dependent high-redshift turnover in the steep spectrum radio luminosity function: clear evidence for downsizing in the radio-AGN population}",
      journal = {\mnras},
         year = 2011,
        month = sep,
       volume = {416},
       number = {3},
        pages = {1900-1915},
          doi = {10.1111/j.1365-2966.2011.19167.x},
archivePrefix = {arXiv},
       eprint = {1104.5020},
 primaryClass = {astro-ph.CO},
       adsurl = {https://ui.adsabs.harvard.edu/abs/2011MNRAS.416.1900R}
}

@ARTICLE{magliocchetti2016,
       author = {{Magliocchetti}, M. and {Lutz}, D. and {Santini}, P. and {Salvato}, M. and {Popesso}, P. and {Berta}, S. and {Pozzi}, F.},
        title = "{The PEP survey: evidence for intense star-forming activity in the majority of radio-selected AGN at z {\ensuremath{\gtrsim}} 1}",
      journal = {\mnras},
         year = 2016,
        month = feb,
       volume = {456},
       number = {1},
        pages = {431-447},
          doi = {10.1093/mnras/stv2645},
archivePrefix = {arXiv},
       eprint = {1511.03032},
 primaryClass = {astro-ph.GA},
       adsurl = {https://ui.adsabs.harvard.edu/abs/2016MNRAS.456..431M}
}

@ARTICLE{petter2024,
       author = {{Petter}, Grayson C. and {Hickox}, Ryan C. and {Morabito}, Leah K. and {Alexander}, David M.},
        title = "{Environments of Luminous Low-frequency Radio Galaxies Since Cosmic Noon: Jet-mode Feedback Dominates in Groups}",
      journal = {\apj},
         year = 2024,
        month = sep,
       volume = {972},
       number = {2},
          eid = {184},
        pages = {184},
          doi = {10.3847/1538-4357/ad6849},
archivePrefix = {arXiv},
       eprint = {2407.18744},
 primaryClass = {astro-ph.GA},
       adsurl = {https://ui.adsabs.harvard.edu/abs/2024ApJ...972..184P}
}

@ARTICLE{deruiter1977,
       author = {{de Ruiter}, H.~R. and {Willis}, A.~G. and {Arp}, H.~C.},
        title = "{A Westerbork 1415 MHz survey of background radio sources. II. Optical identifications with deep IIIa-J plates.}",
      journal = {\aaps},
         year = 1977,
        month = may,
       volume = {28},
        pages = {211-293},
       adsurl = {https://ui.adsabs.harvard.edu/abs/1977A&AS...28..211D}
}

@ARTICLE{lintott2008,
       author = {{Lintott}, Chris J. and {Schawinski}, Kevin and {Slosar}, An{\v{z}}e and {Land}, Kate and {Bamford}, Steven and {Thomas}, Daniel and {Raddick}, M. Jordan and {Nichol}, Robert C. and {Szalay}, Alex and {Andreescu}, Dan and {Murray}, Phil and {Vandenberg}, Jan},
        title = "{Galaxy Zoo: morphologies derived from visual inspection of galaxies from the Sloan Digital Sky Survey}",
      journal = {\mnras},
         year = 2008,
        month = sep,
       volume = {389},
       number = {3},
        pages = {1179-1189},
          doi = {10.1111/j.1365-2966.2008.13689.x},
archivePrefix = {arXiv},
       eprint = {0804.4483},
 primaryClass = {astro-ph},
       adsurl = {https://ui.adsabs.harvard.edu/abs/2008MNRAS.389.1179L}
}

@ARTICLE{swarup1991,
       author = {{Swarup}, G. and {Ananthakrishnan}, S. and {Kapahi}, V.~K. and {Rao}, A.~P. and {Subrahmanya}, C.~R. and {Kulkarni}, V.~K.},
        title = "{The Giant Metre-Wave Radio Telescope}",
      journal = {Current Science},
         year = 1991,
        month = jan,
       volume = {60},
        pages = {95},
       adsurl = {https://ui.adsabs.harvard.edu/abs/1991CSci...60...95S}
}

@ARTICLE{hickox2009,
       author = {{Hickox}, Ryan C. and {Jones}, Christine and {Forman}, William R. and {Murray}, Stephen S. and {Kochanek}, Christopher S. and {Eisenstein}, Daniel and {Jannuzi}, Buell T. and {Dey}, Arjun and {Brown}, Michael J.~I. and {Stern}, Daniel and {Eisenhardt}, Peter R. and {Gorjian}, Varoujan and {Brodwin}, Mark and {Narayan}, Ramesh and {Cool}, Richard J. and {Kenter}, Almus and {Caldwell}, Nelson and {Anderson}, Michael E.},
        title = "{Host Galaxies, Clustering, Eddington Ratios, and Evolution of Radio, X-Ray, and Infrared-Selected AGNs}",
      journal = {\apj},
         year = 2009,
        month = may,
       volume = {696},
       number = {1},
        pages = {891-919},
          doi = {10.1088/0004-637X/696/1/891},
archivePrefix = {arXiv},
       eprint = {0901.4121},
 primaryClass = {astro-ph.GA},
       adsurl = {https://ui.adsabs.harvard.edu/abs/2009ApJ...696..891H}
}

@ARTICLE{devries2002,
       author = {{de Vries}, W.~H. and {Morganti}, R. and {R{\"o}ttgering}, H.~J.~A. and {Vermeulen}, R. and {van Breugel}, W. and {Rengelink}, R. and {Jarvis}, M.~J.},
        title = "{Deep Westerbork 1.4 GHz Imaging of the Bootes Field}",
      journal = {\aj},
         year = 2002,
        month = mar,
       volume = {123},
       number = {3},
        pages = {1784-1800},
          doi = {10.1086/338906},
archivePrefix = {arXiv},
       eprint = {astro-ph/0111543},
 primaryClass = {astro-ph},
       adsurl = {https://ui.adsabs.harvard.edu/abs/2002AJ....123.1784D}
}

@ARTICLE{chakraborty2019,
       author = {{Chakraborty}, Arnab and {Datta}, Abhirup and {Choudhuri}, Samir and {Roy}, Nirupam and {Intema}, Huib and {Choudhury}, Madhurima and {Datta}, Kanan K. and {Pal}, Srijita and {Bharadwaj}, Somnath and {Dutta}, Prasun and {Choudhury}, Tirthankar Roy},
        title = "{Detailed study of the ELAIS N1 field with the uGMRT - I. Characterizing the 325 MHz foreground for redshifted 21 cm observations}",
      journal = {\mnras},
         year = 2019,
        month = aug,
       volume = {487},
       number = {3},
        pages = {4102-4113},
          doi = {10.1093/mnras/stz1580},
archivePrefix = {arXiv},
       eprint = {1906.01655},
 primaryClass = {astro-ph.CO},
       adsurl = {https://ui.adsabs.harvard.edu/abs/2019MNRAS.487.4102C}
}

@ARTICLE{limber1954,
       author = {{Limber}, D. Nelson},
        title = "{The Analysis of Counts of the Extragalactic Nebulae in Terms of a Fluctuating Density Field. II.}",
      journal = {\apj},
         year = 1954,
        month = may,
       volume = {119},
        pages = {655},
          doi = {10.1086/145870},
       adsurl = {https://ui.adsabs.harvard.edu/abs/1954ApJ...119..655L}
}

@ARTICLE{afshordi2004,
       author = {{Afshordi}, Niayesh and {Loh}, Yeong-Shang and {Strauss}, Michael A.},
        title = "{Cross-correlation of the cosmic microwave background with the 2MASS galaxy survey: Signatures of dark energy, hot gas, and point sources}",
      journal = {\prd},
         year = 2004,
        month = apr,
       volume = {69},
       number = {8},
          eid = {083524},
        pages = {083524},
          doi = {10.1103/PhysRevD.69.083524},
archivePrefix = {arXiv},
       eprint = {astro-ph/0308260},
 primaryClass = {astro-ph},
       adsurl = {https://ui.adsabs.harvard.edu/abs/2004PhRvD..69h3524A}
}

@ARTICLE{totsuji1969,
       author = {{Totsuji}, H. and {Kihara}, T.},
        title = "{The Correlation Function for the Distribution of Galaxies}",
      journal = {\pasj},
         year = 1969,
        month = jan,
       volume = {21},
        pages = {221},
       adsurl = {https://ui.adsabs.harvard.edu/abs/1969PASJ...21..221T}
}

@ARTICLE{wang2013,
       author = {{Wang}, Y. and {Brunner}, R.~J. and {Dolence}, J.~C.},
        title = "{The SDSS galaxy angular two-point correlation function}",
      journal = {\mnras},
         year = 2013,
        month = jul,
       volume = {432},
       number = {3},
        pages = {1961-1979},
          doi = {10.1093/mnras/stt450},
archivePrefix = {arXiv},
       eprint = {1303.2432},
 primaryClass = {astro-ph.CO},
       adsurl = {https://ui.adsabs.harvard.edu/abs/2013MNRAS.432.1961W}
}

@INCOLLECTION{coil2013,
       author = {{Coil}, Alison L.},
        title = "{The Large-Scale Structure of the Universe}",
    booktitle = {Planets, Stars and Stellar Systems. Volume 6: Extragalactic Astronomy and Cosmology},
         year = 2013,
       editor = {{Oswalt}, Terry D. and {Keel}, William C.},
       volume = {6},
        pages = {387},
          doi = {10.1007/978-94-007-5609-0_8},
       adsurl = {https://ui.adsabs.harvard.edu/abs/2013pss6.book..387C}
}

@ARTICLE{peebles1983,
       author = {{Davis}, M. and {Peebles}, P.~J.~E.},
        title = "{A survey of galaxy redshifts. V. The two-point position and velocity correlations.}",
      journal = {\apj},
         year = 1983,
        month = apr,
       volume = {267},
        pages = {465-482},
          doi = {10.1086/160884},
       adsurl = {https://ui.adsabs.harvard.edu/abs/1983ApJ...267..465D}
}

@ARTICLE{harrison2018,
       author = {{Harrison}, C.~M. and {Costa}, T. and {Tadhunter}, C.~N. and {Fl{\"u}tsch}, A. and {Kakkad}, D. and {Perna}, M. and {Vietri}, G.},
        title = "{AGN outflows and feedback twenty years on}",
      journal = {Nature Astronomy},
         year = 2018,
        month = feb,
       volume = {2},
        pages = {198-205},
          doi = {10.1038/s41550-018-0403-6},
archivePrefix = {arXiv},
       eprint = {1802.10306},
 primaryClass = {astro-ph.GA},
       adsurl = {https://ui.adsabs.harvard.edu/abs/2018NatAs...2..198H}
}

@ARTICLE{kodama2003,
       author = {{Kodama}, Tadayuki and {Bower}, Richard},
        title = "{The K$_{s}$-band luminosity and stellar mass functions of galaxies in z\raisebox{-0.5ex}\textasciitilde 1 clusters}",
      journal = {\mnras},
         year = 2003,
        month = nov,
       volume = {346},
       number = {1},
        pages = {1-12},
          doi = {10.1046/j.1365-2966.2003.07093.x},
archivePrefix = {arXiv},
       eprint = {astro-ph/0308130},
 primaryClass = {astro-ph},
       adsurl = {https://ui.adsabs.harvard.edu/abs/2003MNRAS.346....1K}
}

@ARTICLE{bower2006,
       author = {{Bower}, R.~G. and {Benson}, A.~J. and {Malbon}, R. and {Helly}, J.~C. and {Frenk}, C.~S. and {Baugh}, C.~M. and {Cole}, S. and {Lacey}, C.~G.},
        title = "{Breaking the hierarchy of galaxy formation}",
      journal = {\mnras},
         year = 2006,
        month = aug,
       volume = {370},
       number = {2},
        pages = {645-655},
          doi = {10.1111/j.1365-2966.2006.10519.x},
archivePrefix = {arXiv},
       eprint = {astro-ph/0511338},
 primaryClass = {astro-ph},
       adsurl = {https://ui.adsabs.harvard.edu/abs/2006MNRAS.370..645B}
}

@ARTICLE{hopkins2006,
       author = {{Hopkins}, Philip F. and {Hernquist}, Lars and {Cox}, Thomas J. and {Di Matteo}, Tiziana and {Robertson}, Brant and {Springel}, Volker},
        title = "{A Unified, Merger-driven Model of the Origin of Starbursts, Quasars, the Cosmic X-Ray Background, Supermassive Black Holes, and Galaxy Spheroids}",
      journal = {\apjs},
         year = 2006,
        month = mar,
       volume = {163},
       number = {1},
        pages = {1-49},
          doi = {10.1086/499298},
archivePrefix = {arXiv},
       eprint = {astro-ph/0506398},
 primaryClass = {astro-ph},
       adsurl = {https://ui.adsabs.harvard.edu/abs/2006ApJS..163....1H}
}

@ARTICLE{krumpe2015,
       author = {{Krumpe}, Mirko and {Miyaji}, Takamitsu and {Husemann}, Bernd and {Fanidakis}, Nikos and {Coil}, Alison L. and {Aceves}, Hector},
        title = "{The Spatial Clustering of ROSAT All-Sky Survey Active Galactic Nuclei. IV. More Massive Black Holes Reside in More Massive Dark Matter Halos}",
      journal = {\apj},
         year = 2015,
        month = dec,
       volume = {815},
       number = {1},
          eid = {21},
        pages = {21},
          doi = {10.1088/0004-637X/815/1/21},
archivePrefix = {arXiv},
       eprint = {1509.01261},
 primaryClass = {astro-ph.GA},
       adsurl = {https://ui.adsabs.harvard.edu/abs/2015ApJ...815...21K}
}

@ARTICLE{miyaji2011,
       author = {{Miyaji}, Takamitsu and {Krumpe}, Mirko and {Coil}, Alison L. and {Aceves}, Hector},
        title = "{The Spatial Clustering of ROSAT All-sky Survey AGNs. II. Halo Occupation Distribution Modeling of the Cross-correlation Function}",
      journal = {\apj},
         year = 2011,
        month = jan,
       volume = {726},
       number = {2},
          eid = {83},
        pages = {83},
          doi = {10.1088/0004-637X/726/2/83},
archivePrefix = {arXiv},
       eprint = {1010.5498},
 primaryClass = {astro-ph.CO},
       adsurl = {https://ui.adsabs.harvard.edu/abs/2011ApJ...726...83M}
}

@ARTICLE{blundell1999,
       author = {{Blundell}, Katherine M. and {Rawlings}, Steve},
        title = "{The inevitable youthfulness of known high-redshift radio galaxies}",
      journal = {\nat},
         year = 1999,
        month = may,
       volume = {399},
       number = {6734},
        pages = {330-332},
          doi = {10.1038/20612},
archivePrefix = {arXiv},
       eprint = {astro-ph/9905333},
 primaryClass = {astro-ph},
       adsurl = {https://ui.adsabs.harvard.edu/abs/1999Natur.399..330B}
}

@ARTICLE{kondapally2023,
       author = {{Kondapally}, Rohit and {Best}, Philip N. and {Raouf}, Mojtaba and {Thomas}, Nicole L. and {Dav{\'e}}, Romeel and {Shabala}, Stanislav S. and {R{\"o}ttgering}, Huub J.~A. and {Hardcastle}, Martin J. and {Bonato}, Matteo and {Cochrane}, Rachel K. and {Ma{\l}ek}, Katarzyna and {Morabito}, Leah K. and {Prandoni}, Isabella and {Smith}, Daniel J.~B.},
        title = "{Cosmic evolution of radio-AGN feedback: confronting models with data}",
      journal = {\mnras},
         year = 2023,
        month = aug,
       volume = {523},
       number = {4},
        pages = {5292-5305},
          doi = {10.1093/mnras/stad1813},
archivePrefix = {arXiv},
       eprint = {2306.11795},
 primaryClass = {astro-ph.GA},
       adsurl = {https://ui.adsabs.harvard.edu/abs/2023MNRAS.523.5292K}
}

@ARTICLE{kondapally2022,
       author = {{Kondapally}, Rohit and {Best}, Philip N. and {Cochrane}, Rachel K. and {Sabater}, Jos{\'e} and {Duncan}, Kenneth J. and {Hardcastle}, Martin J. and {Haskell}, Paul and {Mingo}, Beatriz and {R{\"o}ttgering}, Huub J.~A. and {Smith}, Daniel J.~B. and {Williams}, Wendy L. and {Bonato}, Matteo and {Calistro Rivera}, Gabriela and {Gao}, Fangyou and {Hale}, Catherine L. and {Ma{\l}ek}, Katarzyna and {Miley}, George K. and {Prandoni}, Isabella and {Wang}, Lingyu},
        title = "{Cosmic evolution of low-excitation radio galaxies in the LOFAR two-metre sky survey deep fields}",
      journal = {\mnras},
         year = 2022,
        month = jul,
       volume = {513},
       number = {3},
        pages = {3742-3767},
          doi = {10.1093/mnras/stac1128},
archivePrefix = {arXiv},
       eprint = {2204.07588},
 primaryClass = {astro-ph.GA},
       adsurl = {https://ui.adsabs.harvard.edu/abs/2022MNRAS.513.3742K}
}

@ARTICLE{heckman2014,
       author = {{Heckman}, Timothy M. and {Best}, Philip N.},
        title = "{The Coevolution of Galaxies and Supermassive Black Holes: Insights from Surveys of the Contemporary Universe}",
      journal = {\araa},
         year = 2014,
        month = aug,
       volume = {52},
        pages = {589-660},
          doi = {10.1146/annurev-astro-081913-035722},
archivePrefix = {arXiv},
       eprint = {1403.4620},
 primaryClass = {astro-ph.GA},
       adsurl = {https://ui.adsabs.harvard.edu/abs/2014ARA&A..52..589H}
}

@ARTICLE{birzan2008,
       author = {{B{\^\i}rzan}, L. and {McNamara}, B.~R. and {Nulsen}, P.~E.~J. and {Carilli}, C.~L. and {Wise}, M.~W.},
        title = "{Radiative Efficiency and Content of Extragalactic Radio Sources: Toward a Universal Scaling Relation between Jet Power and Radio Power}",
      journal = {\apj},
         year = 2008,
        month = oct,
       volume = {686},
       number = {2},
        pages = {859-880},
          doi = {10.1086/591416},
archivePrefix = {arXiv},
       eprint = {0806.1929},
 primaryClass = {astro-ph},
       adsurl = {https://ui.adsabs.harvard.edu/abs/2008ApJ...686..859B}
}

@ARTICLE{cavagnolo2010,
       author = {{Cavagnolo}, K.~W. and {McNamara}, B.~R. and {Nulsen}, P.~E.~J. and {Carilli}, C.~L. and {Jones}, C. and {B{\^\i}rzan}, L.},
        title = "{A Relationship Between AGN Jet Power and Radio Power}",
      journal = {\apj},
         year = 2010,
        month = sep,
       volume = {720},
       number = {2},
        pages = {1066-1072},
          doi = {10.1088/0004-637X/720/2/1066},
archivePrefix = {arXiv},
       eprint = {1006.5699},
 primaryClass = {astro-ph.CO},
       adsurl = {https://ui.adsabs.harvard.edu/abs/2010ApJ...720.1066C}
}

@ARTICLE{popesso2024,
       author = {{Popesso}, P. and {Biviano}, A. and {Marini}, I. and {Dolag}, K. and {Vladutescu-Zopp}, S. and {Csizi}, B. and {Biffi}, V. and {Lamer}, G. and {Robothan}, A. and {Bravo}, M. and {Lovisari}, L. and {Ettori}, S. and {Angelinelli}, M. and {Driver}, S. and {Toptun}, V. and {Dev}, A. and {Mazengo}, D. and {Merloni}, A. and {Comparat}, J. and {Ponti}, G. and {Mroczkowski}, T. and {Bulbul}, E. and {Grandis}, S. and {Bahar}, E.},
        title = "{The hot gas mass fraction in halos. From Milky Way-like groups to massive clusters}",
      journal = {arXiv e-prints},
         year = 2024,
        month = nov,
          eid = {arXiv:2411.16555},
        pages = {arXiv:2411.16555},
          doi = {10.48550/arXiv.2411.16555},
archivePrefix = {arXiv},
       eprint = {2411.16555},
 primaryClass = {astro-ph.GA},
       adsurl = {https://ui.adsabs.harvard.edu/abs/2024arXiv241116555P}
}

@ARTICLE{borgani2002,
       author = {{Borgani}, S. and {Governato}, F. and {Wadsley}, J. and {Menci}, N. and {Tozzi}, P. and {Quinn}, T. and {Stadel}, J. and {Lake}, G.},
        title = "{The effect of non-gravitational gas heating in groups and clusters of galaxies}",
      journal = {\mnras},
         year = 2002,
        month = oct,
       volume = {336},
       number = {2},
        pages = {409-424},
          doi = {10.1046/j.1365-8711.2002.05746.x},
archivePrefix = {arXiv},
       eprint = {astro-ph/0205471},
 primaryClass = {astro-ph},
       adsurl = {https://ui.adsabs.harvard.edu/abs/2002MNRAS.336..409B}
}

@ARTICLE{kravtsov2012,
       author = {{Kravtsov}, Andrey V. and {Borgani}, Stefano},
        title = "{Formation of Galaxy Clusters}",
      journal = {\araa},
         year = 2012,
        month = sep,
       volume = {50},
        pages = {353-409},
          doi = {10.1146/annurev-astro-081811-125502},
archivePrefix = {arXiv},
       eprint = {1205.5556},
 primaryClass = {astro-ph.CO},
       adsurl = {https://ui.adsabs.harvard.edu/abs/2012ARA&A..50..353K}
}

@ARTICLE{croton2007,
       author = {{Croton}, Darren J. and {Gao}, Liang and {White}, Simon D.~M.},
        title = "{Halo assembly bias and its effects on galaxy clustering}",
      journal = {\mnras},
         year = 2007,
        month = feb,
       volume = {374},
       number = {4},
        pages = {1303-1309},
          doi = {10.1111/j.1365-2966.2006.11230.x},
archivePrefix = {arXiv},
       eprint = {astro-ph/0605636},
 primaryClass = {astro-ph},
       adsurl = {https://ui.adsabs.harvard.edu/abs/2007MNRAS.374.1303C}
}

@ARTICLE{gao2007,
       author = {{Gao}, Liang and {White}, Simon D.~M.},
        title = "{Assembly bias in the clustering of dark matter haloes}",
      journal = {\mnras},
         year = 2007,
        month = apr,
       volume = {377},
       number = {1},
        pages = {L5-L9},
          doi = {10.1111/j.1745-3933.2007.00292.x},
archivePrefix = {arXiv},
       eprint = {astro-ph/0611921},
 primaryClass = {astro-ph},
       adsurl = {https://ui.adsabs.harvard.edu/abs/2007MNRAS.377L...5G}
}

@ARTICLE{gao2005,
       author = {{Gao}, Liang and {Springel}, Volker and {White}, Simon D.~M.},
        title = "{The age dependence of halo clustering}",
      journal = {\mnras},
         year = 2005,
        month = oct,
       volume = {363},
       number = {1},
        pages = {L66-L70},
          doi = {10.1111/j.1745-3933.2005.00084.x},
archivePrefix = {arXiv},
       eprint = {astro-ph/0506510},
 primaryClass = {astro-ph},
       adsurl = {https://ui.adsabs.harvard.edu/abs/2005MNRAS.363L..66G}
}

@ARTICLE{wechsler2006,
       author = {{Wechsler}, Risa H. and {Zentner}, Andrew R. and {Bullock}, James S. and {Kravtsov}, Andrey V. and {Allgood}, Brandon},
        title = "{The Dependence of Halo Clustering on Halo Formation History, Concentration, and Occupation}",
      journal = {\apj},
         year = 2006,
        month = nov,
       volume = {652},
       number = {1},
        pages = {71-84},
          doi = {10.1086/507120},
archivePrefix = {arXiv},
       eprint = {astro-ph/0512416},
 primaryClass = {astro-ph},
       adsurl = {https://ui.adsabs.harvard.edu/abs/2006ApJ...652...71W}
}

@ARTICLE{mahatma2023,
       author = {{Mahatma}, Vijay H.},
        title = "{The Dynamics and Energetics of Remnant and Restarting RLAGN}",
      journal = {Galaxies},
         year = 2023,
        month = jun,
       volume = {11},
       number = {3},
          eid = {74},
        pages = {74},
          doi = {10.3390/galaxies11030074},
archivePrefix = {arXiv},
       eprint = {2306.08471},
 primaryClass = {astro-ph.GA},
       adsurl = {https://ui.adsabs.harvard.edu/abs/2023Galax..11...74M}
}

@ARTICLE{gao2020,
       author = {{Gao}, F. and {Wang}, L. and {Pearson}, W.~J. and {Gordon}, Y.~A. and {Holwerda}, B.~W. and {Hopkins}, A.~M. and {Brown}, M.~J.~I. and {Bland-Hawthorn}, J. and {Owers}, M.~S.},
        title = "{Mergers trigger active galactic nuclei out to z {\ensuremath{\sim}} 0.6}",
      journal = {\aap},
         year = 2020,
        month = may,
       volume = {637},
          eid = {A94},
        pages = {A94},
          doi = {10.1051/0004-6361/201937178},
archivePrefix = {arXiv},
       eprint = {2004.00680},
 primaryClass = {astro-ph.GA},
       adsurl = {https://ui.adsabs.harvard.edu/abs/2020A&A...637A..94G}
}

@ARTICLE{goulding2018,
       author = {{Goulding}, Andy D. and {Greene}, Jenny E. and {Bezanson}, Rachel and {Greco}, Johnny and {Johnson}, Sean and {Leauthaud}, Alexie and {Matsuoka}, Yoshiki and {Medezinski}, Elinor and {Price-Whelan}, Adrian M.},
        title = "{Galaxy interactions trigger rapid black hole growth: An unprecedented view from the Hyper Suprime-Cam survey}",
      journal = {\pasj},
         year = 2018,
        month = jan,
       volume = {70},
          eid = {S37},
        pages = {S37},
          doi = {10.1093/pasj/psx135},
archivePrefix = {arXiv},
       eprint = {1706.07436},
 primaryClass = {astro-ph.GA},
       adsurl = {https://ui.adsabs.harvard.edu/abs/2018PASJ...70S..37G}
}

@ARTICLE{pierce2022,
       author = {{Pierce}, J.~C.~S. and {Tadhunter}, C.~N. and {Gordon}, Y. and {Ramos Almeida}, C. and {Ellison}, S.~L. and {O'Dea}, C. and {Grimmett}, L. and {Makrygianni}, L. and {Bessiere}, P.~S. and {Do{\~n}a Gir{\'o}n}, P.},
        title = "{Do AGN triggering mechanisms vary with radio power? - II. The importance of mergers as a function of radio power and optical luminosity}",
      journal = {\mnras},
         year = 2022,
        month = feb,
       volume = {510},
       number = {1},
        pages = {1163-1183},
          doi = {10.1093/mnras/stab3231},
archivePrefix = {arXiv},
       eprint = {2111.03075},
 primaryClass = {astro-ph.GA},
       adsurl = {https://ui.adsabs.harvard.edu/abs/2022MNRAS.510.1163P}
}

@ARTICLE{mccarthy2010,
       author = {{McCarthy}, I.~G. and {Schaye}, J. and {Ponman}, T.~J. and {Bower}, R.~G. and {Booth}, C.~M. and {Dalla Vecchia}, C. and {Crain}, R.~A. and {Springel}, V. and {Theuns}, T. and {Wiersma}, R.~P.~C.},
        title = "{The case for AGN feedback in galaxy groups}",
      journal = {\mnras},
         year = 2010,
        month = aug,
       volume = {406},
       number = {2},
        pages = {822-839},
          doi = {10.1111/j.1365-2966.2010.16750.x},
archivePrefix = {arXiv},
       eprint = {0911.2641},
 primaryClass = {astro-ph.CO},
       adsurl = {https://ui.adsabs.harvard.edu/abs/2010MNRAS.406..822M}
}

@ARTICLE{gitti2012,
       author = {{Gitti}, Myriam and {Brighenti}, Fabrizio and {McNamara}, Brian R.},
        title = "{Evidence for AGN Feedback in Galaxy Clusters and Groups}",
      journal = {Advances in Astronomy},
         year = 2012,
        month = jan,
       volume = {2012},
          eid = {950641},
        pages = {950641},
          doi = {10.1155/2012/950641},
archivePrefix = {arXiv},
       eprint = {1109.3334},
 primaryClass = {astro-ph.CO},
       adsurl = {https://ui.adsabs.harvard.edu/abs/2012AdAst2012E...6G}
}

@ARTICLE{lian2016,
       author = {{Lian}, Jianhui and {Yan}, Renbin and {Zhang}, Kai and {Kong}, Xu},
        title = "{The Quenching Timescale and Quenching Rate of Galaxies}",
      journal = {\apj},
         year = 2016,
        month = nov,
       volume = {832},
       number = {1},
          eid = {29},
        pages = {29},
          doi = {10.3847/0004-637X/832/1/29},
archivePrefix = {arXiv},
       eprint = {1609.04805},
 primaryClass = {astro-ph.GA},
       adsurl = {https://ui.adsabs.harvard.edu/abs/2016ApJ...832...29L}
}

@ARTICLE{best2004,
       author = {{Best}, P.~N.},
        title = "{The environmental dependence of radio-loud AGN activity and star formation in the 2dFGRS}",
      journal = {\mnras},
         year = 2004,
        month = jun,
       volume = {351},
       number = {1},
        pages = {70-82},
          doi = {10.1111/j.1365-2966.2004.07752.x},
archivePrefix = {arXiv},
       eprint = {astro-ph/0402523},
 primaryClass = {astro-ph},
       adsurl = {https://ui.adsabs.harvard.edu/abs/2004MNRAS.351...70B}
}

@ARTICLE{saikia2009,
       author = {{Saikia}, D.~J. and {Jamrozy}, M.},
        title = "{Recurrent activity in Active Galactic Nuclei}",
      journal = {Bulletin of the Astronomical Society of India},
         year = 2009,
        month = sep,
       volume = {37},
        pages = {63-89},
          doi = {10.48550/arXiv.1002.1841},
archivePrefix = {arXiv},
       eprint = {1002.1841},
 primaryClass = {astro-ph.CO},
       adsurl = {https://ui.adsabs.harvard.edu/abs/2009BASI...37...63S}
}

@ARTICLE{lara1999,
       author = {{Lara}, L. and {M{\'a}rquez}, I. and {Cotton}, W.~D. and {Feretti}, L. and {Giovannini}, G. and {Marcaide}, J.~M. and {Venturi}, T.},
        title = "{Restarting activity in the giant radio galaxy J1835+620}",
      journal = {\aap},
         year = 1999,
        month = aug,
       volume = {348},
        pages = {699-704},
          doi = {10.48550/arXiv.astro-ph/9906287},
archivePrefix = {arXiv},
       eprint = {astro-ph/9906287},
 primaryClass = {astro-ph},
       adsurl = {https://ui.adsabs.harvard.edu/abs/1999A&A...348..699L}
}

@ARTICLE{nandi2012,
       author = {{Nandi}, S. and {Saikia}, D.~J.},
        title = "{Double-double radio galaxies from the FIRST survey}",
      journal = {Bulletin of the Astronomical Society of India},
         year = 2012,
        month = jun,
       volume = {40},
       number = {2},
        pages = {121-137},
          doi = {10.48550/arXiv.1208.1941},
archivePrefix = {arXiv},
       eprint = {1208.1941},
 primaryClass = {astro-ph.CO},
       adsurl = {https://ui.adsabs.harvard.edu/abs/2012BASI...40..121N}
}

@ARTICLE{schoenmakers2000,
       author = {{Schoenmakers}, Arno P. and {de Bruyn}, A.~G. and {R{\"o}ttgering}, H.~J.~A. and {van der Laan}, H. and {Kaiser}, C.~R.},
        title = "{Radio galaxies with a `double-double morphology' - I. Analysis of the radio properties and evidence for interrupted activity in active galactic nuclei}",
      journal = {\mnras},
         year = 2000,
        month = jun,
       volume = {315},
       number = {2},
        pages = {371-380},
          doi = {10.1046/j.1365-8711.2000.03430.x},
archivePrefix = {arXiv},
       eprint = {astro-ph/9912141},
 primaryClass = {astro-ph},
       adsurl = {https://ui.adsabs.harvard.edu/abs/2000MNRAS.315..371S}
}

@INPROCEEDINGS{smith2016,
       author = {{Smith}, D.~J.~B. and {Best}, P.~N. and {Duncan}, K.~J. and {Hatch}, N.~A. and {Jarvis}, M.~J. and {R{\"o}ttgering}, H.~J.~A. and {Simpson}, C.~J. and {Stott}, J.~P. and {Cochrane}, R.~K. and {Coppin}, K.~E. and {Dannerbauer}, H. and {Davis}, T.~A. and {Geach}, J.~E. and {Hale}, C.~L. and {Hardcastle}, M.~J. and {Hatfield}, P.~W. and {Houghton}, R.~C.~W. and {Maddox}, N. and {McGee}, S.~L. and {Morabito}, L. and {Nisbet}, D. and {Pandey-Pommier}, M. and {Prandoni}, I. and {Saxena}, A. and {Shimwell}, T.~W. and {Tarr}, M. and {van Bemmel}, I. and {Verma}, A. and {White}, G.~J. and {Williams}, W.~L.},
        title = "{The WEAVE-LOFAR Survey}",
    booktitle = {SF2A-2016: Proceedings of the Annual meeting of the French Society of Astronomy and Astrophysics},
         year = 2016,
       editor = {{Reyl{\'e}}, C. and {Richard}, J. and {Cambr{\'e}sy}, L. and {Deleuil}, M. and {P{\'e}contal}, E. and {Tresse}, L. and {Vauglin}, I.},
        month = dec,
        pages = {271-280},
          doi = {10.48550/arXiv.1611.02706},
archivePrefix = {arXiv},
       eprint = {1611.02706},
 primaryClass = {astro-ph.GA},
       adsurl = {https://ui.adsabs.harvard.edu/abs/2016sf2a.conf..271S}
}

@INPROCEEDINGS{haarlem2005,
       author = {{van Haarlem}, M.~P.},
        title = "{LOFAR: The Low Frequency Array}",
    booktitle = {EAS Publications Series},
         year = 2005,
       editor = {{Gurvits}, Leonid I. and {Frey}, Sandor and {Rawlings}, Steve},
       series = {EAS Publications Series},
       volume = {15},
        month = jan,
        pages = {431-444},
          doi = {10.1051/eas:2005169},
       adsurl = {https://ui.adsabs.harvard.edu/abs/2005EAS....15..431V}
}

@ARTICLE{norris2011,
       author = {{Norris}, Ray P. and {Hopkins}, A.~M. and {Afonso}, J. and {Brown}, S. and {Condon}, J.~J. and {Dunne}, L. and {Feain}, I. and {Hollow}, R. and {Jarvis}, M. and {Johnston-Hollitt}, M. and {Lenc}, E. and {Middelberg}, E. and {Padovani}, P. and {Prandoni}, I. and {Rudnick}, L. and {Seymour}, N. and {Umana}, G. and {Andernach}, H. and {Alexander}, D.~M. and {Appleton}, P.~N. and {Bacon}, D. and {Banfield}, J. and {Becker}, W. and {Brown}, M.~J.~I. and {Ciliegi}, P. and {Jackson}, C. and {Eales}, S. and {Edge}, A.~C. and {Gaensler}, B.~M. and {Giovannini}, G. and {Hales}, C.~A. and {Hancock}, P. and {Huynh}, M.~T. and {Ibar}, E. and {Ivison}, R.~J. and {Kennicutt}, R. and {Kimball}, Amy E. and {Koekemoer}, A.~M. and {Koribalski}, B.~S. and {L{\'o}pez-S{\'a}nchez}, {\'A}. R. and {Mao}, M.~Y. and {Murphy}, T. and {Messias}, H. and {Pimbblet}, K.~A. and {Raccanelli}, A. and {Randall}, K.~E. and {Reiprich}, T.~H. and {Roseboom}, I.~G. and {R{\"o}ttgering}, H. and {Saikia}, D.~J. and {Sharp}, R.~G. and {Slee}, O.~B. and {Smail}, Ian and {Thompson}, M.~A. and {Urquhart}, J.~S. and {Wall}, J.~V. and {Zhao}, G. -B.},
        title = "{EMU: Evolutionary Map of the Universe}",
      journal = {\pasa},
         year = 2011,
        month = aug,
       volume = {28},
       number = {3},
        pages = {215-248},
          doi = {10.1071/AS11021},
archivePrefix = {arXiv},
       eprint = {1106.3219},
 primaryClass = {astro-ph.CO},
       adsurl = {https://ui.adsabs.harvard.edu/abs/2011PASA...28..215N}
}

@ARTICLE{euclid2025,
       author = {{Euclid Collaboration} and {Mellier}, Y. and {Abdurro'uf} and {Acevedo Barroso}, J.~A. and {Ach{\'u}carro}, A. and {Adamek}, J. and {Adam}, R. and {Addison}, G.~E. and {Aghanim}, N. and {Aguena}, M. and {Ajani}, V. and {Akrami}, Y. and {Al-Bahlawan}, A. and {Alavi}, A. and {Albuquerque}, I.~S. and {Alestas}, G. and {Alguero}, G. and {Allaoui}, A. and {Allen}, S.~W. and {Allevato}, V. and {Alonso-Tetilla}, A.~V. and {Altieri}, B. and {Alvarez-Candal}, A. and {Alvi}, S. and {Amara}, A. and {Amendola}, L. and {Amiaux}, J. and {Andika}, I.~T. and {Andreon}, S. and {Andrews}, A. and {Angora}, G. and {Angulo}, R.~E. and {Annibali}, F. and {Anselmi}, A. and {Anselmi}, S. and {Arcari}, S. and {Archidiacono}, M. and {Aric{\`o}}, G. and {Arnaud}, M. and {Arnouts}, S. and {Asgari}, M. and {Asorey}, J. and {Atayde}, L. and {Atek}, H. and {Atrio-Barandela}, F. and {Aubert}, M. and {Aubourg}, E. and {Auphan}, T. and {Auricchio}, N. and {Aussel}, B. and {Aussel}, H. and {Avelino}, P.~P. and {Avgoustidis}, A. and {Avila}, S. and {Awan}, S. and {Azzollini}, R. and {Baccigalupi}, C. and {Bachelet}, E. and {Bacon}, D. and {Baes}, M. and {Bagley}, M.~B. and {Bahr-Kalus}, B. and {Balaguera-Antolinez}, A. and {Balbinot}, E. and {Balcells}, M. and {Baldi}, M. and {Baldry}, I. and {Balestra}, A. and {Ballardini}, M. and {Ballester}, O. and {Balogh}, M. and {Ba{\~n}ados}, E. and {Barbier}, R. and {Bardelli}, S. and {Baron}, M. and {Barreiro}, T. and {Barrena}, R. and {Barriere}, J. -C. and {Barros}, B.~J. and {Barthelemy}, A. and {Bartolo}, N. and {Basset}, A. and {Battaglia}, P. and {Battisti}, A.~J. and {Baugh}, C.~M. and {Baumont}, L. and {Bazzanini}, L. and {Beaulieu}, J. -P. and {Beckmann}, V. and {Belikov}, A.~N. and {Bel}, J. and {Bellagamba}, F. and {Bella}, M. and {Bellini}, E. and {Benabed}, K. and {Bender}, R. and {Benevento}, G. and {Bennett}, C.~L. and {Benson}, K. and {Bergamini}, P. and {Bermejo-Climent}, J.~R. and {Bernardeau}, F. and {Bertacca}, D. and {Berthe}, M. and {Berthier}, J. and {Bethermin}, M. and {Beutler}, F. and {Bevillon}, C. and {Bhargava}, S. and {Bhatawdekar}, R. and {Bianchi}, D. and {Bisigello}, L. and {Biviano}, A. and {Blake}, R.~P. and {Blanchard}, A. and {Blazek}, J. and {Blot}, L. and {Bosco}, A. and {Bodendorf}, C. and {Boenke}, T. and {B{\"o}hringer}, H. and {Boldrini}, P. and {Bolzonella}, M. and {Bonchi}, A. and {Bonici}, M. and {Bonino}, D. and {Bonino}, L. and {Bonvin}, C. and {Bon}, W. and {Booth}, J.~T. and {Borgani}, S. and {Borlaff}, A.~S. and {Borsato}, E. and {Bose}, B. and {Botticella}, M.~T. and {Boucaud}, A. and {Bouche}, F. and {Boucher}, J.~S. and {Boutigny}, D. and {Bouvard}, T. and {Bouwens}, R. and {Bouy}, H. and {Bowler}, R.~A.~A. and {Bozza}, V. and {Bozzo}, E. and {Branchini}, E. and {Brando}, G. and {Brau-Nogue}, S. and {Brekke}, P. and {Bremer}, M.~N. and {Brescia}, M. and {Breton}, M. -A. and {Brinchmann}, J. and {Brinckmann}, T. and {Brockley-Blatt}, C. and {Brodwin}, M. and {Brouard}, L. and {Brown}, M.~L. and {Bruton}, S. and {Bucko}, J. and {Buddelmeijer}, H. and {Buenadicha}, G. and {Buitrago}, F. and {Burger}, P. and {Burigana}, C. and {Busillo}, V. and {Busonero}, D. and {Cabanac}, R. and {Cabayol-Garcia}, L. and {Cagliari}, M.~S. and {Caillat}, A. and {Caillat}, L. and {Calabrese}, M. and {Calabro}, A. and {Calderone}, G. and {Calura}, F. and {Camacho Quevedo}, B. and {Camera}, S. and {Campos}, L. and {Ca{\~n}as-Herrera}, G. and {Candini}, G.~P. and {Cantiello}, M. and {Capobianco}, V. and {Cappellaro}, E. and {Cappelluti}, N. and {Cappi}, A. and {Caputi}, K.~I. and {Cara}, C. and {Carbone}, C. and {Cardone}, V.~F. and {Carella}, E. and {Carlberg}, R.~G. and {Carle}, M. and {Carminati}, L. and {Caro}, F. and {Carrasco}, J.~M. and {Carretero}, J. and {Carrilho}, P. and {Carron Duque}, J. and {Carry}, B.},
        title = "{Euclid: I. Overview of the Euclid mission}",
      journal = {\aap},
         year = 2025,
        month = may,
       volume = {697},
          eid = {A1},
        pages = {A1},
          doi = {10.1051/0004-6361/202450810},
archivePrefix = {arXiv},
       eprint = {2405.13491},
 primaryClass = {astro-ph.CO},
       adsurl = {https://ui.adsabs.harvard.edu/abs/2025A&A...697A...1E}
}

@ARTICLE{lsst2009,
       author = {{LSST Science Collaboration} and {Abell}, Paul A. and {Allison}, Julius and {Anderson}, Scott F. and {Andrew}, John R. and {Angel}, J. Roger P. and {Armus}, Lee and {Arnett}, David and {Asztalos}, S.~J. and {Axelrod}, Tim S. and {Bailey}, Stephen and {Ballantyne}, D.~R. and {Bankert}, Justin R. and {Barkhouse}, Wayne A. and {Barr}, Jeffrey D. and {Barrientos}, L. Felipe and {Barth}, Aaron J. and {Bartlett}, James G. and {Becker}, Andrew C. and {Becla}, Jacek and {Beers}, Timothy C. and {Bernstein}, Joseph P. and {Biswas}, Rahul and {Blanton}, Michael R. and {Bloom}, Joshua S. and {Bochanski}, John J. and {Boeshaar}, Pat and {Borne}, Kirk D. and {Bradac}, Marusa and {Brandt}, W.~N. and {Bridge}, Carrie R. and {Brown}, Michael E. and {Brunner}, Robert J. and {Bullock}, James S. and {Burgasser}, Adam J. and {Burge}, James H. and {Burke}, David L. and {Cargile}, Phillip A. and {Chandrasekharan}, Srinivasan and {Chartas}, George and {Chesley}, Steven R. and {Chu}, You-Hua and {Cinabro}, David and {Claire}, Mark W. and {Claver}, Charles F. and {Clowe}, Douglas and {Connolly}, A.~J. and {Cook}, Kem H. and {Cooke}, Jeff and {Cooray}, Asantha and {Covey}, Kevin R. and {Culliton}, Christopher S. and {de Jong}, Roelof and {de Vries}, Willem H. and {Debattista}, Victor P. and {Delgado}, Francisco and {Dell'Antonio}, Ian P. and {Dhital}, Saurav and {Di Stefano}, Rosanne and {Dickinson}, Mark and {Dilday}, Benjamin and {Djorgovski}, S.~G. and {Dobler}, Gregory and {Donalek}, Ciro and {Dubois-Felsmann}, Gregory and {Durech}, Josef and {Eliasdottir}, Ardis and {Eracleous}, Michael and {Eyer}, Laurent and {Falco}, Emilio E. and {Fan}, Xiaohui and {Fassnacht}, Christopher D. and {Ferguson}, Harry C. and {Fernandez}, Yanga R. and {Fields}, Brian D. and {Finkbeiner}, Douglas and {Figueroa}, Eduardo E. and {Fox}, Derek B. and {Francke}, Harold and {Frank}, James S. and {Frieman}, Josh and {Fromenteau}, Sebastien and {Furqan}, Muhammad and {Galaz}, Gaspar and {Gal-Yam}, A. and {Garnavich}, Peter and {Gawiser}, Eric and {Geary}, John and {Gee}, Perry and {Gibson}, Robert R. and {Gilmore}, Kirk and {Grace}, Emily A. and {Green}, Richard F. and {Gressler}, William J. and {Grillmair}, Carl J. and {Habib}, Salman and {Haggerty}, J.~S. and {Hamuy}, Mario and {Harris}, Alan W. and {Hawley}, Suzanne L. and {Heavens}, Alan F. and {Hebb}, Leslie and {Henry}, Todd J. and {Hileman}, Edward and {Hilton}, Eric J. and {Hoadley}, Keri and {Holberg}, J.~B. and {Holman}, Matt J. and {Howell}, Steve B. and {Infante}, Leopoldo and {Ivezic}, Zeljko and {Jacoby}, Suzanne H. and {Jain}, Bhuvnesh and {R} and {Jedicke} and {Jee}, M. James and {Garrett Jernigan}, J. and {Jha}, Saurabh W. and {Johnston}, Kathryn V. and {Jones}, R. Lynne and {Juric}, Mario and {Kaasalainen}, Mikko and {Styliani} and {Kafka} and {Kahn}, Steven M. and {Kaib}, Nathan A. and {Kalirai}, Jason and {Kantor}, Jeff and {Kasliwal}, Mansi M. and {Keeton}, Charles R. and {Kessler}, Richard and {Knezevic}, Zoran and {Kowalski}, Adam and {Krabbendam}, Victor L. and {Krughoff}, K. Simon and {Kulkarni}, Shrinivas and {Kuhlman}, Stephen and {Lacy}, Mark and {Lepine}, Sebastien and {Liang}, Ming and {Lien}, Amy and {Lira}, Paulina and {Long}, Knox S. and {Lorenz}, Suzanne and {Lotz}, Jennifer M. and {Lupton}, R.~H. and {Lutz}, Julie and {Macri}, Lucas M. and {Mahabal}, Ashish A. and {Mandelbaum}, Rachel and {Marshall}, Phil and {May}, Morgan and {McGehee}, Peregrine M. and {Meadows}, Brian T. and {Meert}, Alan and {Milani}, Andrea and {Miller}, Christopher J. and {Miller}, Michelle and {Mills}, David and {Minniti}, Dante and {Monet}, David and {Mukadam}, Anjum S. and {Nakar}, Ehud and {Neill}, Douglas R. and {Newman}, Jeffrey A. and {Nikolaev}, Sergei and {Nordby}, Martin and {O'Connor}, Paul and {Oguri}, Masamune and {Oliver}, John and {Olivier}, Scot S. and {Olsen}, Julia K. and {Olsen}, Knut and {Olszewski}, Edward W. and {Oluseyi}, Hakeem and {Padilla}, Nelson D. and {Parker}, Alex and {Pepper}, Joshua and {Peterson}, John R. and {Petry}, Catherine and {Pinto}, Philip A. and {Pizagno}, James L. and {Popescu}, Bogdan and {Prsa}, Andrej and {Radcka}, Veljko and {Raddick}, M. Jordan and {Rasmussen}, Andrew and {Rau}, Arne and {Rho}, Jeonghee and {Rhoads}, James E. and {Richards}, Gordon T. and {Ridgway}, Stephen T. and {Robertson}, Brant E. and {Roskar}, Rok and {Saha}, Abhijit and {Sarajedini}, Ata and {Scannapieco}, Evan and {Schalk}, Terry and {Schindler}, Rafe and {Schmidt}, Samuel},
        title = "{LSST Science Book, Version 2.0}",
      journal = {arXiv e-prints},
         year = 2009,
        month = dec,
          eid = {arXiv:0912.0201},
        pages = {arXiv:0912.0201},
          doi = {10.48550/arXiv.0912.0201},
archivePrefix = {arXiv},
       eprint = {0912.0201},
 primaryClass = {astro-ph.IM},
       adsurl = {https://ui.adsabs.harvard.edu/abs/2009arXiv0912.0201L}
}

@ARTICLE{levi2013,
       author = {{Levi}, Michael and {Bebek}, Chris and {Beers}, Timothy and {Blum}, Robert and {Cahn}, Robert and {Eisenstein}, Daniel and {Flaugher}, Brenna and {Honscheid}, Klaus and {Kron}, Richard and {Lahav}, Ofer and {McDonald}, Patrick and {Roe}, Natalie and {Schlegel}, David and {representing the DESI collaboration}},
        title = "{The DESI Experiment, a whitepaper for Snowmass 2013}",
      journal = {arXiv e-prints},
         year = 2013,
        month = aug,
          eid = {arXiv:1308.0847},
        pages = {arXiv:1308.0847},
          doi = {10.48550/arXiv.1308.0847},
archivePrefix = {arXiv},
       eprint = {1308.0847},
 primaryClass = {astro-ph.CO},
       adsurl = {https://ui.adsabs.harvard.edu/abs/2013arXiv1308.0847L}
}

@ARTICLE{pozetti2010,
       author = {{Pozzetti}, L. and {Bolzonella}, M. and {Zucca}, E. and {Zamorani}, G. and {Lilly}, S. and {Renzini}, A. and {Moresco}, M. and {Mignoli}, M. and {Cassata}, P. and {Tasca}, L. and {Lamareille}, F. and {Maier}, C. and {Meneux}, B. and {Halliday}, C. and {Oesch}, P. and {Vergani}, D. and {Caputi}, K. and {Kova{\v{c}}}, K. and {Cimatti}, A. and {Cucciati}, O. and {Iovino}, A. and {Peng}, Y. and {Carollo}, M. and {Contini}, T. and {Kneib}, J. -P. and {Le F{\'e}vre}, O. and {Mainieri}, V. and {Scodeggio}, M. and {Bardelli}, S. and {Bongiorno}, A. and {Coppa}, G. and {de la Torre}, S. and {de Ravel}, L. and {Franzetti}, P. and {Garilli}, B. and {Kampczyk}, P. and {Knobel}, C. and {Le Borgne}, J. -F. and {Le Brun}, V. and {Pell{\`o}}, R. and {Perez Montero}, E. and {Ricciardelli}, E. and {Silverman}, J.~D. and {Tanaka}, M. and {Tresse}, L. and {Abbas}, U. and {Bottini}, D. and {Cappi}, A. and {Guzzo}, L. and {Koekemoer}, A.~M. and {Leauthaud}, A. and {Maccagni}, D. and {Marinoni}, C. and {McCracken}, H.~J. and {Memeo}, P. and {Porciani}, C. and {Scaramella}, R. and {Scarlata}, C. and {Scoville}, N.},
        title = "{zCOSMOS - 10k-bright spectroscopic sample. The bimodality in the galaxy stellar mass function: exploring its evolution with redshift}",
      journal = {\aap},
         year = 2010,
        month = nov,
       volume = {523},
          eid = {A13},
        pages = {A13},
          doi = {10.1051/0004-6361/200913020},
archivePrefix = {arXiv},
       eprint = {0907.5416},
 primaryClass = {astro-ph.CO},
       adsurl = {https://ui.adsabs.harvard.edu/abs/2010A&A...523A..13P}
}

@ARTICLE{kin1999,
       author = {{Ng}, Kin-Wang and {Liu}, Guo-Chin},
        title = "{Correlation Functions of CMB Anisotropy and Polarization}",
      journal = {International Journal of Modern Physics D},
         year = 1999,
        month = jan,
       volume = {8},
       number = {1},
        pages = {61-83},
          doi = {10.1142/S0218271899000079},
archivePrefix = {arXiv},
       eprint = {astro-ph/9710012},
 primaryClass = {astro-ph},
       adsurl = {https://ui.adsabs.harvard.edu/abs/1999IJMPD...8...61N}
}

@ARTICLE{chon2004,
       author = {{Chon}, Gayoung and {Challinor}, Anthony and {Prunet}, Simon and {Hivon}, Eric and {Szapudi}, Istv{\'a}n},
        title = "{Fast estimation of polarization power spectra using correlation functions}",
      journal = {\mnras},
         year = 2004,
        month = may,
       volume = {350},
       number = {3},
        pages = {914-926},
          doi = {10.1111/j.1365-2966.2004.07737.x},
archivePrefix = {arXiv},
       eprint = {astro-ph/0303414},
 primaryClass = {astro-ph},
       adsurl = {https://ui.adsabs.harvard.edu/abs/2004MNRAS.350..914C}
}

@ARTICLE{bryan1998,
       author = {{Bryan}, Greg L. and {Norman}, Michael L.},
        title = "{Statistical Properties of X-Ray Clusters: Analytic and Numerical Comparisons}",
      journal = {\apj},
         year = 1998,
        month = mar,
       volume = {495},
       number = {1},
        pages = {80-99},
          doi = {10.1086/305262},
archivePrefix = {arXiv},
       eprint = {astro-ph/9710107},
 primaryClass = {astro-ph},
       adsurl = {https://ui.adsabs.harvard.edu/abs/1998ApJ...495...80B}
}

@ARTICLE{fedeli2014,
       author = {{Fedeli}, C. and {Semboloni}, E. and {Velliscig}, M. and {Daalen}, M. Van and {Schaye}, J. and {Hoekstra}, H.},
        title = "{The clustering of baryonic matter. II: halo model and hydrodynamic simulations}",
      journal = {\jcap},
         year = 2014,
        month = aug,
       volume = {2014},
       number = {8},
        pages = {028-028},
          doi = {10.1088/1475-7516/2014/08/028},
archivePrefix = {arXiv},
       eprint = {1406.5013},
 primaryClass = {astro-ph.CO},
       adsurl = {https://ui.adsabs.harvard.edu/abs/2014JCAP...08..028F}
}

@ARTICLE{mead2015,
       author = {{Mead}, A.~J. and {Peacock}, J.~A. and {Heymans}, C. and {Joudaki}, S. and {Heavens}, A.~F.},
        title = "{An accurate halo model for fitting non-linear cosmological power spectra and baryonic feedback models}",
      journal = {\mnras},
         year = 2015,
        month = dec,
       volume = {454},
       number = {2},
        pages = {1958-1975},
          doi = {10.1093/mnras/stv2036},
archivePrefix = {arXiv},
       eprint = {1505.07833},
 primaryClass = {astro-ph.CO},
       adsurl = {https://ui.adsabs.harvard.edu/abs/2015MNRAS.454.1958M}
}

@INPROCEEDINGS{vaccari2015,
       author = {{Vaccari}, M.},
        title = "{The Spitzer Data Fusion: Contents, Construction and Applications to Galaxy Evolution Studies}",
    booktitle = {The Many Facets of Extragalactic Radio Surveys: Towards New Scientific Challenges},
         year = 2015,
        month = oct,
          eid = {27},
        pages = {27},
          doi = {https://pos.sissa.it/267/027},
archivePrefix = {arXiv},
       eprint = {1604.02353},
 primaryClass = {astro-ph.GA},
       adsurl = {https://ui.adsabs.harvard.edu/abs/2015fers.confE..27V}
}

@ARTICLE{wu1999,
       author = {{Wu}, Xiang-Ping and {Xue}, Yan-Jie and {Fang}, Li-Zhi},
        title = "{The L$_{X}$-T and L$_{X}$-{\ensuremath{\sigma}} Relationships for Galaxy Clusters Revisited}",
      journal = {\apj},
         year = 1999,
        month = oct,
       volume = {524},
       number = {1},
        pages = {22-30},
          doi = {10.1086/307791},
archivePrefix = {arXiv},
       eprint = {astro-ph/9905106},
 primaryClass = {astro-ph},
       adsurl = {https://ui.adsabs.harvard.edu/abs/1999ApJ...524...22W}
}

@ARTICLE{mclure2004,
       author = {{McLure}, Ross J. and {Jarvis}, Matt J.},
        title = "{The relationship between radio luminosity and black hole mass in optically selected quasars}",
      journal = {\mnras},
         year = 2004,
        month = oct,
       volume = {353},
       number = {4},
        pages = {L45-L49},
          doi = {10.1111/j.1365-2966.2004.08305.x},
archivePrefix = {arXiv},
       eprint = {astro-ph/0408203},
 primaryClass = {astro-ph},
       adsurl = {https://ui.adsabs.harvard.edu/abs/2004MNRAS.353L..45M}
}

@ARTICLE{magorrian1998,
       author = {{Magorrian}, John and {Tremaine}, Scott and {Richstone}, Douglas and {Bender}, Ralf and {Bower}, Gary and {Dressler}, Alan and {Faber}, S.~M. and {Gebhardt}, Karl and {Green}, Richard and {Grillmair}, Carl and {Kormendy}, John and {Lauer}, Tod},
        title = "{The Demography of Massive Dark Objects in Galaxy Centers}",
      journal = {\aj},
         year = 1998,
        month = jun,
       volume = {115},
       number = {6},
        pages = {2285-2305},
          doi = {10.1086/300353},
archivePrefix = {arXiv},
       eprint = {astro-ph/9708072},
 primaryClass = {astro-ph},
       adsurl = {https://ui.adsabs.harvard.edu/abs/1998AJ....115.2285M}
}

@ARTICLE{haring2004,
       author = {{H{\"a}ring}, Nadine and {Rix}, Hans-Walter},
        title = "{On the Black Hole Mass-Bulge Mass Relation}",
      journal = {\apjl},
         year = 2004,
        month = apr,
       volume = {604},
       number = {2},
        pages = {L89-L92},
          doi = {10.1086/383567},
archivePrefix = {arXiv},
       eprint = {astro-ph/0402376},
 primaryClass = {astro-ph},
       adsurl = {https://ui.adsabs.harvard.edu/abs/2004ApJ...604L..89H}
}

@ARTICLE{mcconnell2013,
       author = {{McConnell}, Nicholas J. and {Ma}, Chung-Pei},
        title = "{Revisiting the Scaling Relations of Black Hole Masses and Host Galaxy Properties}",
      journal = {\apj},
         year = 2013,
        month = feb,
       volume = {764},
       number = {2},
          eid = {184},
        pages = {184},
          doi = {10.1088/0004-637X/764/2/184},
archivePrefix = {arXiv},
       eprint = {1211.2816},
 primaryClass = {astro-ph.CO},
       adsurl = {https://ui.adsabs.harvard.edu/abs/2013ApJ...764..184M}
}

@ARTICLE{kormendy2013,
       author = {{Kormendy}, John and {Ho}, Luis C.},
        title = "{Coevolution (Or Not) of Supermassive Black Holes and Host Galaxies}",
      journal = {\araa},
         year = 2013,
        month = aug,
       volume = {51},
       number = {1},
        pages = {511-653},
          doi = {10.1146/annurev-astro-082708-101811},
archivePrefix = {arXiv},
       eprint = {1304.7762},
 primaryClass = {astro-ph.CO},
       adsurl = {https://ui.adsabs.harvard.edu/abs/2013ARA&A..51..511K}
}

@ARTICLE{kondapally2025,
       author = {{Kondapally}, Rohit and {Best}, Philip N. and {Duncan}, Kenneth J. and {R{\"o}ttgering}, Huub J.~A. and {Smith}, Daniel J.~B. and {Prandoni}, Isabella and {Hardcastle}, Martin J. and {Holc}, Tanja and {Patrick}, Abigail L. and {Arnaudova}, Marina I. and {Mingo}, Beatriz and {Cochrane}, Rachel K. and {Das}, Soumyadeep and {Haskell}, Paul and {Magliocchetti}, Manuela and {Ma{\l}ek}, Katarzyna and {Miley}, George K. and {Tasse}, Cyril and {Williams}, Wendy L.},
        title = "{Radio-AGN activity across the galaxy population: dependence on stellar mass, star formation rate, and redshift}",
      journal = {\mnras},
         year = 2025,
        month = jan,
       volume = {536},
       number = {1},
        pages = {554-571},
          doi = {10.1093/mnras/stae2567},
archivePrefix = {arXiv},
       eprint = {2411.08104},
 primaryClass = {astro-ph.GA},
       adsurl = {https://ui.adsabs.harvard.edu/abs/2025MNRAS.536..554K}
}




\appendix

\section{Cross-correlation model}\label{sec:cross_correlation_model}

In this section, we discuss the halo model for the cross-correlation between the AGN (or matched galaxy) sample, labelled with superscript $\rm{A}$, and the full optical/NIR population, $\rm{B}$. It is assumed that sample $\rm{A}$ is a subset of sample $\rm{B}$, so while all central galaxies in $\rm{A}$ are also in $\rm{B}$, only a fraction $f_{\rm{c}}(M)$ of central galaxies in $\rm{B}$ are also in $\rm{A}$. Similarly, only a fraction $f_{\rm{s}}(M)$ of satellite galaxies in $\rm{B}$ are also in $\rm{A}$, and all satellites in $\rm{A}$ are in $\rm{B}$.

The full halo model power spectrum is the sum of the one- and two-halo contributions \citep{cooray2002}:
\begin{equation} \label{eqn:total_power_spectrum}
    P_{\rm{hm}}(k \mid z)=P_{\rm{1h}}(k \mid z)+P_{\rm{2h}}(k \mid z).
\end{equation}

The two-halo term measures the clustering between galaxies that reside in different haloes, so depends only on the large-scale bias of each sample. It therefore remains unchanged from the standard form:
\begin{equation} \label{eqn:two_halo_power_spectrum}
    P_{\rm{2h}}(k \mid z)=\frac{1}{\overline{n}_{\rm{g}}^{\rm{A}}\overline{n}_{\rm{g}}^{\rm{B}}}I(k,z \mid U^{\rm{A}})I(k,z \mid U^{\rm{B}})P_{\rm{lin}}(k \mid z),
\end{equation}
where $\overline{n}_{\rm{g}}^{\rm{A}}$ and $\overline{n}_{\rm{g}}^{\rm{B}}$ are the predicted mean number densities of galaxies in each sample given by Equation~\ref{eqn:mean_number_density}, $P_{\rm{lin}}(k \mid z)$ is the linear matter power spectrum, and the integral
\begin{equation} \label{eqn:two_halo_integral}
    I(k,z \mid U)=\int{\rm{d}}M\,n(M,z) b_{\rm{h}}(M,z) \langle U(k \mid M) \rangle.
\end{equation}
$\langle U(k \mid M) \rangle$ is the Fourier transform of the mean galaxy density profile for a halo of mass $M$:
\begin{equation} \label{eqn:first_fourier_moment}
\begin{split}
    \langle n_{\rm{g}}(r) \mid M) \rangle&=\overline{N}_{\rm{c}}(M) \left[ 1+\overline{n}_{\rm{s}}(r \mid M) \right]\\
    &=\overline{N}_{\rm{c}}(M) \left[ 1+\overline{N}_{\rm{s}}(M)u_{\rm{s}}(r \mid M) \right],
\end{split}
\end{equation}
where the normalized distribution of satellites, $u_{\rm{s}}(r \mid M)$, is a function of radial distance from the halo centre, $r$. The distribution of satellites in a halo is assumed to follow that of the dark matter, which is modelled as a truncated NFW \citep{navarro1995} profile (see \citealt{nicola2020} for the full parametrization). Then
\begin{equation}
\begin{split}
    \langle U(k \mid M) \rangle &= \overline{N}_{\rm{c}}(M) \left[ 1+\overline{n}_{\rm{s}}(k \mid M) \right]\\
    &=\overline{N}_{\rm{c}}(M) \left[ 1+\overline{N}_{\rm{s}}(M)\tilde{u}_{\rm{s}}(k \mid M) \right],
\end{split}
\end{equation}
where $\tilde{u}_{\rm{s}}(k \mid M)$ is the Fourier transform of $u_{\rm{s}}(r \mid M)$.

The one-halo term of the cross-correlation power spectrum accounts for pairs of galaxies that reside in the same halo. It is given by
\begin{equation} \label{eqn:one_halo_power_spectrum}
    P_{\rm{1h}}(k \mid z)=\frac{1}{\overline{n}_{\rm{g}}^{\rm{A}}\overline{n}_{\rm{g}}^{\rm{B}}}\int{\rm{d}}M\,n(M,z)\langle U^{\rm{A}}(k \mid M) U^{\rm{B}}(k \mid M) \rangle. 
\end{equation}
If the two samples were drawn completely independently of each other, then their joint two-point Fourier moment would factorize: 
\begin{equation} \label{eqn:two_point_fourier_moment_independent}
\begin{split}
    &\langle U^{\rm{A}}(k \mid M) U^{\rm{B}}(k \mid M) \rangle=\langle U^{\rm{A}}(k \mid M) \rangle \langle U^{\rm{B}}(k \mid M) \rangle\\
    =& \ \overline{N}_{\rm{c}}^{\rm{A}}(M) \overline{N}_{\rm{c}}^{\rm{B}}(M)\big[\big(\overline{N}_{\rm{s}}^{\rm{A}}(M)+\overline{N}_{\rm{s}}^{\rm{B}}(M)\big)\tilde{u}_{\rm{s}}(k \mid M) \\
    &+\overline{N}_{\rm{s}}^{\rm{A}}(M)\overline{N}_{\rm{s}}^{\rm{B}}(M) \tilde{u}_{\rm{s}}^2(k \mid M)\big].
\end{split}
\end{equation}
However, since the samples overlap, this cannot be assumed to be the case.

To derive $\langle U^{\rm{A}} U^{\rm{B}} \rangle$ for the case where $\rm{A}$ is a subset of $\rm{B}$, we divide the halo into small volume elements, $\Delta V$, with coordinates $X$. In the following, we drop the explicit $M$ notation for brevity. The Fourier transforms then become discrete sums:
\begin{equation} \label{eqn:two_point_fourier_moment}
    \langle U^{\rm{A}}(k) U^{\rm{B}}(k) \rangle=\frac{1}{(\Delta V)^2}\sum_{X,X'}e^{ik(X-X')}(\Delta V)^2\langle N_{X}^{\rm{A}} N_{X'}^{\rm{B}} \rangle.
\end{equation}
$\langle N_{X}^{\rm{A}} N_{X'}^{\rm{B}} \rangle$ can be decomposed into the contributions from central-central, central-satellite, satellite-central and satellite-satellite galaxy pairs:
\begin{equation}\label{eqn:NANB_decomposed}
\begin{split}
    &\langle N_{X}^{\rm{A}} N_{X'}^{\rm{B}} \rangle\\ &= \langle N_{\rm{c}X}^{{\rm{A}}} N_{\rm{c}X'}^{{\rm{B}}} \rangle + \langle N_{\rm{c}X}^{{\rm{A}}} N_{\rm{s}X'}^{{\rm{B}}} \rangle + \langle N_{\rm{s}X}^{{\rm{A}}} N_{\rm{c}X'}^{{\rm{B}}} \rangle + \langle N_{\rm{s}X}^{{\rm{A}}} N_{\rm{s}X'}^{{\rm{B}}} \rangle,
\end{split}
\end{equation}
where, for example, $N_{{\rm{c}}X}^{\rm{A}} \equiv N_{\rm{c}}^{\rm{A}}(X)\in\{0,1\}$ denotes the number of central galaxies from sample $\rm{A}$ in the volume element at position $X$, with analogous definitions for satellite galaxies and for sample $\rm{B}$. 

To evaluate these four terms, we introduce $\mathcal{P}(N_X)$, the probability distribution for finding $N_X$ galaxies in the volume element at position $X$. For central galaxies, the expectation value is
\begin{equation}
    \sum_{N_{{\rm{c}}X}=0}^1 N_{{\rm{c}}X} \mathcal{P}(N_{{\rm{c}}X}) = \mathcal{P}(N_{{\rm{c}}X}=1) = \delta_{X0}^{\rm{K}} \overline{N}_{\rm{c}},
\end{equation}
where $\overline{N}_{\rm{c}}$ is the mean number of central galaxies per halo from Equation~\ref{eqn:num_centrals}, and the Kronecker-delta function, $\delta^{\rm{K}}$, reflects the assumption that central galaxies reside at the centre of their halo, $X=0$.

For satellite galaxies, the expectation value,
\begin{equation}
    \sum_{N_{{\rm{s}}X}=0}^\infty N_{{\rm{s}}X} \mathcal{P}(N_{{\rm{s}}X}) = \langle N_{{\rm{s}}X} \rangle = \overline{N}_{\rm{c}} \overline{n}_{\rm{s}}(X) = \overline{N}_{\rm{c}} \overline{N}_{\rm{s}} u_{\rm{s}}(X),
\end{equation}
where $\overline{N}_{\rm{s}}$ is the mean number of satellite galaxies in a halo with a central galaxy from Equation~\ref{eqn:num_satellites}, $u_{\rm{s}}(X)$ is the normalized spatial profile of satellite galaxies within the halo, and $\overline{n}_{\rm{s}}(X) \equiv \overline{N}_{\rm{s}} u_{\rm{s}}(X)$.

The central-central term is given by:
\begin{equation}
\begin{split}
    \langle N_{\rm{c}X}^{{\rm{A}}} N_{\rm{c}X'}^{{\rm{B}}} \rangle&=\delta_{XX'}^{\rm{K}}\delta_{X0}^{\rm{K}} \sum_{N_{\rm{c}}^{{\rm{A}}},N_{\rm{c}}^{\rm{B}}} N_{\rm{c}}^{{\rm{A}}} N_{\rm{c}}^{\rm{B}} \mathcal{P}(N_{\rm{c}}^{{\rm{A}}},N_{\rm{c}}^{\rm{B}})\\
    &=\delta_{XX'}^{\rm{K}}\delta_{X0}^{\rm{K}}\sum_{N_{\rm{c}}^{\rm{B}}}N_{\rm{c}}^{\rm{B}}\mathcal{P}(N_{\rm{c}}^{\rm{B}})\sum_{N_{\rm{c}}^{{\rm{A}}}}N_{\rm{c}}^{{\rm{A}}}\mathcal{P}(N_{\rm{c}}^{{\rm{A}}} \mid N_{\rm{c}}^{\rm{B}})\\
    &=\delta_{XX'}^{\rm{K}}\delta_{X0}^{\rm{K}}\sum_{N_{\rm{c}}^{\rm{B}}}N_{\rm{c}}^{\rm{B}}\mathcal{P}(N_{\rm{c}}^{\rm{B}})f_{\rm{c}}N_{\rm{c}}^{\rm{B}}\\
    &=\delta_{XX'}^{\rm{K}}\delta_{X0}^{\rm{K}}f_{\rm{c}}\sum_{N_{\rm{c}}^{\rm{B}}}N_{\rm{c}}^{\rm{B}}\mathcal{P}(N_{\rm{c}}^{\rm{B}})\\
    &=\delta_{XX'}^{\rm{K}}\delta_{X0}^{\rm{K}}f_{\rm{c}}\overline{N}_{\rm{c}}^{\rm{B}},
\end{split}
\end{equation}
where between the third and fourth lines the identity ${N_{\rm{c}}}^2=N_{\rm{c}}$ has been used, since $N_{\rm{c}}\in\{0,1\}$.

The satellite-satellite term is given by:
\begin{equation}
\begin{split}
    \langle N_{\rm{s}X}^{{\rm{A}}} N_{\rm{s}X'}^{{\rm{B}}} \rangle&=\sum_{N_{\rm{s}X}^{{\rm{A}}},N_{\rm{s}X'}^{{\rm{B}}}} N_{\rm{s}X}^{{\rm{A}}} N_{\rm{s}X'}^{{\rm{B}}} \mathcal{P}(N_{\rm{s}X}^{{\rm{A}}}, N_{\rm{s}X'}^{{\rm{B}}})\\
    &=\sum_{N_{\rm{s}X}^{{\rm{A}}},N_{\rm{s}X'}^{{\rm{B}}}} N_{\rm{s}X'}^{{\rm{B}}} \mathcal{P}(N_{\rm{s}X'}^{{\rm{B}}}) N_{\rm{s}X}^{{\rm{A}}} \mathcal{P}(N_{\rm{s}X}^{{\rm{A}}} \mid N_{\rm{s}X'}^{{\rm{B}}}).
\end{split}
\end{equation}
If $X \neq X'$ then the two satellites are distinct, and the probability of finding a satellite at $X$ from sample $\rm{A}$ is independent of the probability of finding a satellite at $X'$ from sample $\rm{B}$. However, if $X=X'$ then they are the same galaxy, and in order for there to be a satellite at $X$ in sample $\rm{A}$, there must also be a satellite from sample $\rm{B}$ at $X$. Therefore,
\begin{equation}
    \mathcal{P}(N_{\rm{s}X}^{{\rm{A}}} \mid N_{\rm{s}X'}^{{\rm{B}}})=
\begin{cases}
    \mathcal{P}(N_{\rm{s}X}^{{\rm{A}}}), &\text{if}\ X \neq X'\\
    \mathcal{P}(N_{\rm{s}X}^{{\rm{A}}} \mid N_{{\rm{s}}X}^{\rm{B}}), &\text{if}\ X = X'
\end{cases}.
\end{equation}
Separating these two cases, and making use of the fact that satellite galaxies are Poisson distributed so the variance ${\langle {N_{\rm{s}}}^2 \rangle}-{\langle {N_{\rm{s}}} \rangle}^2={\langle N_{\rm{s}} \rangle}$,
\begin{equation}
\begin{split}
    \langle N_{\rm{s}X}^{{\rm{A}}} N_{\rm{s}X'}^{{\rm{B}}} \rangle =& \sum_{N_{\rm{s}X}^{{\rm{A}}}} N_{\rm{s}X}^{{\rm{A}}} \mathcal{P}(N_{\rm{s}X}^{{\rm{A}}}) \sum_{N_{\rm{s}X'}^{{\rm{B}}}} N_{\rm{s}X'}^{{\rm{B}}} \mathcal{P}(N_{\rm{s}X'}^{{\rm{B}}}) (1-\delta^{\rm{K}}_{XX'}) \\
    &+\delta^{\rm{K}}_{XX'} \sum_{N_{{\rm{s}}X}^{\rm{B}}} N_{{\rm{s}}X}^{\rm{B}} \mathcal{P}(N_{{\rm{s}}X}^{\rm{B}}) \sum_{N_{\rm{s}X}^{{\rm{A}}}} N_{\rm{s}X}^{{\rm{A}}} \mathcal{P}(N_{\rm{s}X}^{{\rm{A}}} \mid N_{{\rm{s}}X}^{\rm{B}}) \\
    =& \ \langle N_{\rm{s}X}^{{\rm{A}}} \rangle \langle N_{\rm{s}X'}^{{\rm{B}}} \rangle (1-\delta^{\rm{K}}_{XX'}) \\
    &+\delta^{\rm{K}}_{XX'} f_{\rm{s}} \sum_{N_{{\rm{s}}X}^{\rm{B}}} (N_{{\rm{s}}X}^{\rm{B}})^2 \mathcal{P}(N_{{\rm{s}}X}^{\rm{B}}) \\
    =& \ f_{\rm{s}} \langle N_{{\rm{s}}X}^{\rm{B}} \rangle \langle N_{\rm{s}X'}^{{\rm{B}}} \rangle (1-\delta^{\rm{K}}_{XX'}) \\
    &+\delta^{\rm{K}}_{XX'} f_{\rm{s}} \left({\langle N_{{\rm{s}}X}^{\rm{B}} \rangle}^2+{\langle N_{{\rm{s}}X}^{\rm{B}} \rangle}\right) \\
    =& \ f_{\rm{s}} \langle N_{{\rm{s}}X}^{\rm{B}} \rangle \langle N_{\rm{s}X'}^{{\rm{B}}} \rangle + \delta^{\rm{K}}_{XX'} f_{\rm{s}} {\langle N_{{\rm{s}}X}^{\rm{B}} \rangle} \\
    =& \ f_{\rm{s}} {\overline{N}_{\rm{c}}^{\rm{B}}}^2 \overline{n}_{\rm{s}}^{\rm{B}}(X) \overline{n}_{\rm{s}}^{\rm{B}}(X') + \delta^{\rm{K}}_{XX'} f_{\rm{s}} \overline{N}_{\rm{c}}^{\rm{B}} \overline{n}_{\rm{s}}^{\rm{B}}(X).
\end{split}
\end{equation}

Next we consider the central-satellite contribution with a central galaxy from sample $\rm{B}$ and a satellite from $\rm{A}$:
\begin{equation}
\begin{split}
    \langle N_{\rm{c}X'}^{{\rm{B}}} N_{\rm{s}X}^{{\rm{A}}} \rangle =& \sum_{N_{\rm{s}X}^{{\rm{A}}}, N_{{\rm{s}}X}^{\rm{B}}, N_{\rm{c}X'}^{{\rm{B}}}} N_{\rm{c}X'}^{{\rm{B}}} N_{\rm{s}X}^{{\rm{A}}} \mathcal{P}(N_{\rm{s}X}^{{\rm{A}}}, N_{{\rm{s}}X}^{\rm{B}}, N_{\rm{c}X'}^{{\rm{B}}}) \\
    =& \sum_{N_{\rm{s}X}^{{\rm{A}}}, N_{{\rm{s}}X}^{\rm{B}}, N_{\rm{c}X'}^{{\rm{B}}}} N_{\rm{c}X'}^{{\rm{B}}} N_{\rm{s}X}^{{\rm{A}}} \\ 
    &\times \mathcal{P}(N_{\rm{s}X}^{{\rm{A}}} \mid N_{{\rm{s}}X}^{\rm{B}}, N_{\rm{c}X'}^{{\rm{B}}}) \mathcal{P}(N_{{\rm{s}}X}^{\rm{B}} \mid N_{\rm{c}X'}^{{\rm{B}}}) \mathcal{P}(N_{\rm{c}X'}^{{\rm{B}}}) \\
    =& \ f_{\rm{s}} \sum_{N_{{\rm{s}}X}^{\rm{B}}, N_{\rm{c}X'}^{{\rm{B}}}} N_{\rm{c}X'}^{{\rm{B}}} N_{{\rm{s}}X}^{\rm{B}} \mathcal{P}(N_{{\rm{s}}X}^{\rm{B}} \mid N_{\rm{c}X'}^{{\rm{B}}}) \mathcal{P}(N_{\rm{c}X'}^{{\rm{B}}}) \\
    =& \ \delta^{\rm{K}}_{X'0} f_{\rm{s}} \overline{N}_{\rm{c}}^{\rm{B}} \langle N_{{\rm{s}}X}^{\rm{B}} \mid N_{\rm{c}X'}^{{\rm{B}}}=1 \rangle \\
    =& \ \delta^{\rm{K}}_{X'0} f_{\rm{s}} \overline{N}_{\rm{c}}^{\rm{B}} \overline{n}_{\rm{s}}^{\rm{B}}(X).
\end{split}
\end{equation}
And similarly for the satellite-central term:
\begin{equation}
\begin{split}
    \langle N_{\rm{c}X}^{{\rm{A}}} N_{\rm{s}X'}^{{\rm{B}}} \rangle =& \sum_{N_{\rm{c}X}^{{\rm{A}}}, N_{\rm{c}X}^{{\rm{B}}}, N_{\rm{s}X'}^{{\rm{B}}}} N_{\rm{c}X}^{{\rm{A}}} N_{\rm{s}X'}^{{\rm{B}}} \mathcal{P}(N_{\rm{c}X}^{{\rm{A}}}, N_{\rm{c}X}^{{\rm{B}}}, N_{\rm{s}X'}^{{\rm{B}}}) \\
    =& \sum_{N_{\rm{c}X}^{{\rm{A}}}, N_{{\rm{c}}X}^{\rm{B}}, N_{\rm{s}X'}^{{\rm{B}}}} N_{\rm{c}X}^{{\rm{A}}} N_{\rm{s}X'}^{{\rm{B}}} \\ 
    &\times \mathcal{P}(N_{\rm{c}X}^{{\rm{A}}} \mid N_{\rm{c}X}^{{\rm{B}}}, N_{\rm{s}X'}^{{\rm{B}}}) \mathcal{P}(N_{\rm{s}X'}^{{\rm{B}}} \mid N_{\rm{c}X}^{{\rm{B}}}) \mathcal{P}(N_{\rm{c}X}^{{\rm{B}}}) \\
    =& \ f_{\rm{c}} \sum_{N_{\rm{c}X}^{{\rm{B}}}, N_{\rm{s}X'}^{{\rm{B}}}} N_{\rm{s}X'}^{{\rm{B}}} N_{\rm{c}X}^{{\rm{B}}} \mathcal{P}(N_{\rm{s}X'}^{{\rm{B}}} \mid N_{\rm{c}X}^{{\rm{B}}}) \mathcal{P}(N_{\rm{c}X}^{{\rm{B}}}) \\
    =& \ \delta^{\rm{K}}_{X0} f_{\rm{c}} \overline{N}_{\rm{c}}^{\rm{B}} \langle N_{\rm{s}X'}^{{\rm{B}}} \mid N_{\rm{c}X}^{{\rm{B}}}=1 \rangle \\
    =& \ \delta^{\rm{K}}_{X0} f_{\rm{c}} \overline{N}_{\rm{c}}^{\rm{B}} \overline{n}_{\rm{s}}^{\rm{B}}(X').
\end{split}
\end{equation}

Putting these terms into Equation~\ref{eqn:NANB_decomposed}:
\begin{equation} \label{eqn:two_point_fourier_moment_final}
\begin{split}
    \langle &U^{\rm{A}}(k\mid M) U^{\rm{B}}(k \mid M) \rangle=\sum_{X,X'}e^{ik(X-X')} \\
    &\times \Big[ \delta_{XX'}^{\rm{K}}\delta_{X0}^{\rm{K}}f_{\rm{c}}\overline{N}_{\rm{c}}^{\rm{B}} + f_{\rm{s}} {\overline{N}_{\rm{c}}^{\rm{B}}}^2 \overline{n}_{\rm{s}}^{\rm{B}}(X) \overline{n}_{\rm{s}}^{\rm{B}}(X') + \delta^{\rm{K}}_{XX'} f_{\rm{s}} \overline{N}_{\rm{c}}^{\rm{B}} \overline{n}_{\rm{s}}^{\rm{B}}(X) \\
    &+\delta^{\rm{K}}_{X'0} f_{\rm{s}} \overline{N}_{\rm{c}}^{\rm{B}} \overline{n}_{\rm{s}}^{\rm{B}}(X) + \delta^{\rm{K}}_{X0} f_{\rm{c}} \overline{N}_{\rm{c}}^{\rm{B}} \overline{n}_{\rm{s}}^{\rm{B}}(X') \Big] \\
    =& \ f_{\rm{c}}\overline{N}_{\rm{c}}^{\rm{B}} + f_{\rm{s}} {\overline{N}_{\rm{c}}^{\rm{B}}}^2 {\overline{n}_{\rm{s}}^{\rm{B}}}^*(k) {\overline{n}_{\rm{s}}^{\rm{B}}}(k) + \sum_X f_{\rm{s}} \overline{N}_{\rm{c}}^{\rm{B}} \overline{n}_{\rm{s}}^{\rm{B}}(X) \\
    &+ f_{\rm{s}} \overline{N}_{\rm{c}}^{\rm{B}} {\overline{n}_{\rm{s}}^{\rm{B}}}^*(k) + f_{\rm{c}} \overline{N}_{\rm{c}}^{\rm{B}} \overline{n}_{\rm{s}}^{\rm{B}}(k) \\
    =& \ f_{\rm{c}}\overline{N}_{\rm{c}}^{\rm{B}} + f_{\rm{s}} {\overline{N}_{\rm{c}}^{\rm{B}}}^2 \left({\overline{N}_{\rm{s}}^{\rm{B}}} \tilde{u}_{\rm{s}}(k) \right)^2 + f_{\rm{s}} \overline{N}_{\rm{c}}^{\rm{B}} \overline{N}_{\rm{s}}^{\rm{B}} \\
    &+ \big( f_{\rm{s}} + f_{\rm{c}} \big) \overline{N}_{\rm{c}}^{\rm{B}} \overline{N}_{\rm{s}}^{\rm{B}} \tilde{u}_{\rm{s}}(k), 
\end{split}
\end{equation}
where we have made use of the normalization $\sum_Xu_{\rm{s}}(X)=1$. 

The first and third terms represent central-central and satellite-satellite self-pairings between galaxies common to both samples. These terms are independent of $k$, so they correspond to the shot-noise contribution to the power spectrum which affects only the zero-lag correlation function ($\theta = 0$). Since we do not include this in our analysis, these terms can be neglected, so
\begin{equation}\label{eqn:one_halo_power_spectrum_final}
\begin{split}
    P_{\rm{1h}}(k \mid z)=&\frac{1}{\overline{n}_{\rm{g}}^{\rm{A}}\overline{n}_{\rm{g}}^{\rm{B}}}\int{\rm{d}}M\,n(M,z) \\ 
    &\times \Big[ \big( f_{\rm{s}}(M) + f_{\rm{c}}(M) \big) \overline{N}_{\rm{c}}^{\rm{B}}(M) \overline{N}_{\rm{s}}^{\rm{B}}(M) \tilde{u}_{\rm{s}}(k \mid M) \\
    &+ f_{\rm{s}}(M) \left({\overline{N}_{\rm{c}}^{\rm{B}}}(M){\overline{N}_{\rm{s}}^{\rm{B}}}(M) \tilde{u}_{\rm{s}}(k \mid M) \right)^2\Big].
\end{split}
\end{equation}

The fractions of central and satellite galaxies in sample $\rm{B}$ that are also in sample $\rm{A}$ for a given set of HOD parameters are given by
\begin{equation}
    f_{\rm{c}}(M)=\frac{\overline{N}_{\rm{c}}^{\rm{A}}(M)}{\overline{N}_{\rm{c}}^{\rm{B}}(M)},
\end{equation}
and
\begin{equation}
    f_{\rm{s}}(M)=\frac{\overline{N}_{\rm{c}}^{\rm{A}}(M)\overline{N}_{\rm{s}}^{\rm{A}}(M)}{\overline{N}_{\rm{c}}^{\rm{B}}(M)\overline{N}_{\rm{s}}^{\rm{B}}(M)}.
\end{equation}

To implement this in \texttt{CCL}, the \texttt{fourier\_2pt} method of the \texttt{Profile2ptHOD} class\footnote{\href{https://github.com/LSSTDESC/CCL/blob/master/pyccl/halos/profiles_2pt.py}{https://github.com/LSSTDESC/CCL/blob/master/pyccl/halos/profiles\_2pt.py}} was modified.

Once the halo model power spectrum has been calculated, the angular clustering can be predicted using \texttt{CCL} \citep[see Section 2.4.1 of][]{chisari2019}. Using the Limber approximation \citep{limber1954, afshordi2004}, the angular power spectrum can be written as
\begin{equation}
    C_\ell=\frac{2}{2\ell+1} \int {\rm{d}}k \, P_{\rm{hm}}(k \mid z_\ell)\Delta_\ell^{\rm{A}}(k)\Delta_\ell^{\rm{B}}(k),
\end{equation}
where a radial distance $\chi_\ell\equiv(\ell+1/2)/k$ has been defined and $z_\ell$ is the corresponding redshift. $\Delta_\ell^{\rm{A}}(k)$ and $\Delta_\ell^{\rm{B}}(k)$ are the transfer functions corresponding to the samples. For galaxy number count tracers, these are given by
\begin{equation}
    \Delta_\ell^{\rm{NC}}(k)=\frac{H(z_\ell)}{c}p_{\rm{z}}(z_\ell),
\end{equation}
where $H(z)$ is the Hubble parameter, $c$ is the speed of light and $p_{\rm{z}}(z)$ is the normalized redshift distribution of the sample.
Finally, the angular correction function, $\omega(\theta)$, is obtained by summing over the multipole moments: 
\begin{equation}
    \omega(\theta)=\frac{1}{4\pi} \sum_\ell (2\ell+1) C_\ell P_\ell(\cos \theta),
\end{equation}
where $P_\ell(x)$ are the zeroth-order associated Legendre polynomials.

Because stellar mass thresholds are applied to the optical/NIR population but not the AGN and matched galaxy samples, the assumption that $\rm{A}$ is a subset of $\rm{B}$ is only approximate. For the two lower redshift bins ($0<z<1$ and $1<z<1.5$), fewer than $9\%$ of galaxies in $\rm{A}$ lie below the stellar mass thresholds, but this rises to $\sim18\%$ for $1.5<z<2.5$. The only difference in the two-point Fourier moment between the cases where samples $\rm{A}$ and $\rm{B}$ are independent (Equation~\ref{eqn:two_point_fourier_moment_independent}) and when $\rm{A}$ is a subset of $\rm{B}$ (Equation~\ref{eqn:two_point_fourier_moment_final}) is in the central-satellite and satellite-central contributions, which are proportional to $\tilde{u}_{\rm{s}}(k \mid M)$. Isolating these terms, we have $\overline{N}_{\rm{c}}^{\rm{A}} \overline{N}_{\rm{c}}^{\rm{B}} \overline{N}_{\rm{s}}^{\rm{B}} + \overline{N}_{\rm{c}}^{\rm{B}} \overline{N}_{\rm{c}}^{\rm{A}} \overline{N}_{\rm{s}}^{\rm{A}}$ when the samples are independent and $\overline{N}_{\rm{c}}^{\rm{A}} \overline{N}_{\rm{s}}^{\rm{B}} + \overline{N}_{\rm{c}}^{\rm{A}} \overline{N}_{\rm{s}}^{\rm{A}}$ when one is a subset of the other. We find that the second term in each of these expressions is much smaller than the first term since $\overline{N}_{\rm{s}}^{\rm{B}} \gg \overline{N}_{\rm{s}}^{\rm{A}}$, so can be neglected, and the independent case differs only by a factor of $\overline{N}_{\rm{c}}^{\rm{B}}$. In a perfect subset, $\overline{N}_{\rm{c}}^{\rm{B}}=1$ whenever $\overline{N}_{\rm{c}}^{\rm{A}}>0$, so the two expressions coincide. When some of $\rm{A}$'s central galaxies fall below $\rm{B}$'s stellar mass limit, $\overline{N}_{\rm{c}}^{\rm{B}}<1$ in those haloes, and exactly that fraction of central-satellite pairs is lost, producing a small, scale-dependent suppression of the one-halo term. However, because $M_1^{\rm{B}} \gg M_{\rm{min}}^{\rm{A}}$ and $M_{\rm{min}}^{\rm{B}}$, both $\overline{N}_{\rm{c}}^{\rm{A}}$ and $\overline{N}_{\rm{c}}^{\rm{B}}$ are very close to 1 at the halo masses where $\rm{B}$'s satellites first appear, so the loss is negligible in practice. 

\section{Posterior probabilities of the HOD parameters}\label{sec:corner_plots}

\begin{figure*}
  \centering
  \begin{subfigure}{\columnwidth}
    \centering
    \includegraphics[width=\columnwidth]{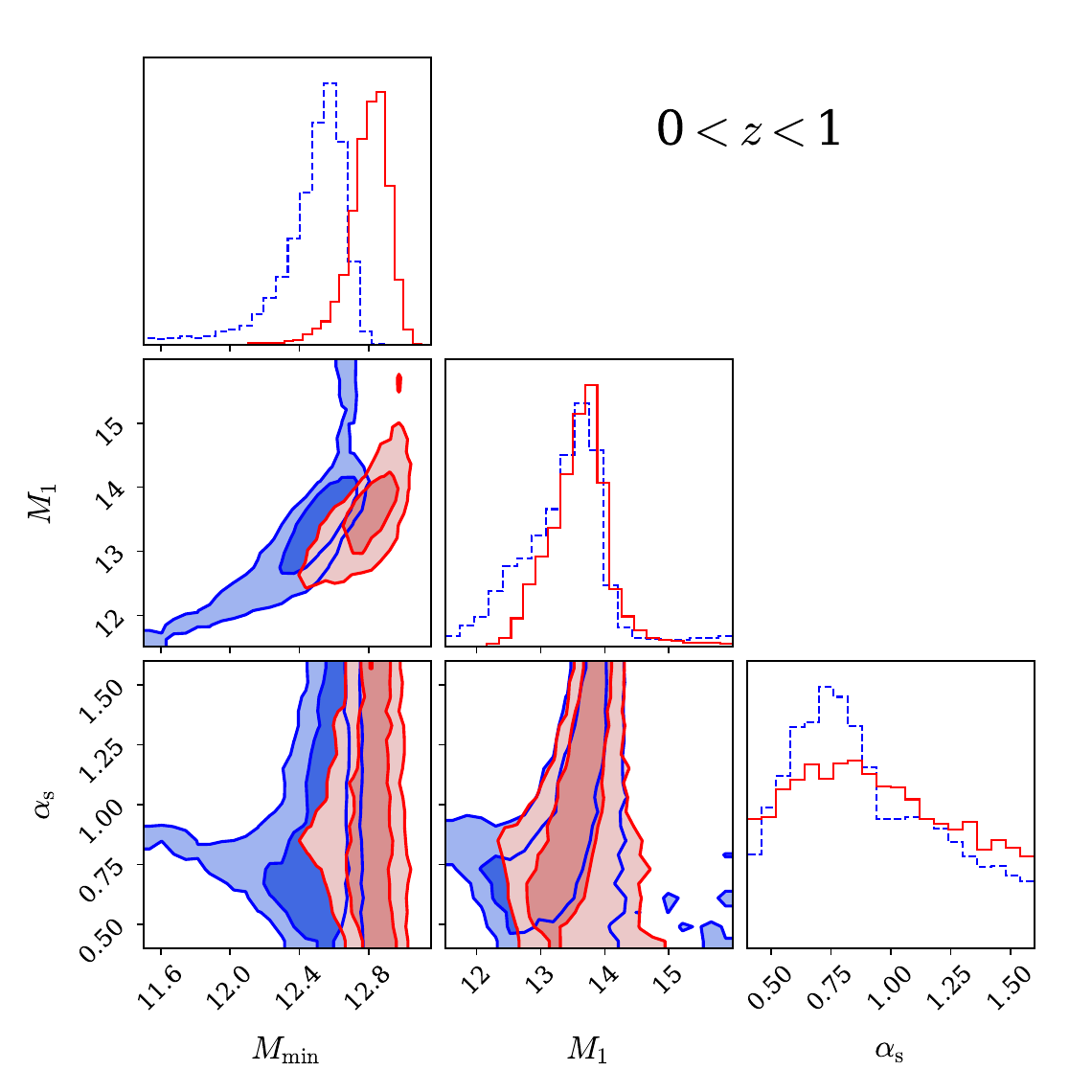}
  \end{subfigure}
  \vspace{1em}
  \begin{subfigure}{\columnwidth}
    \centering
    \includegraphics[width=\columnwidth]{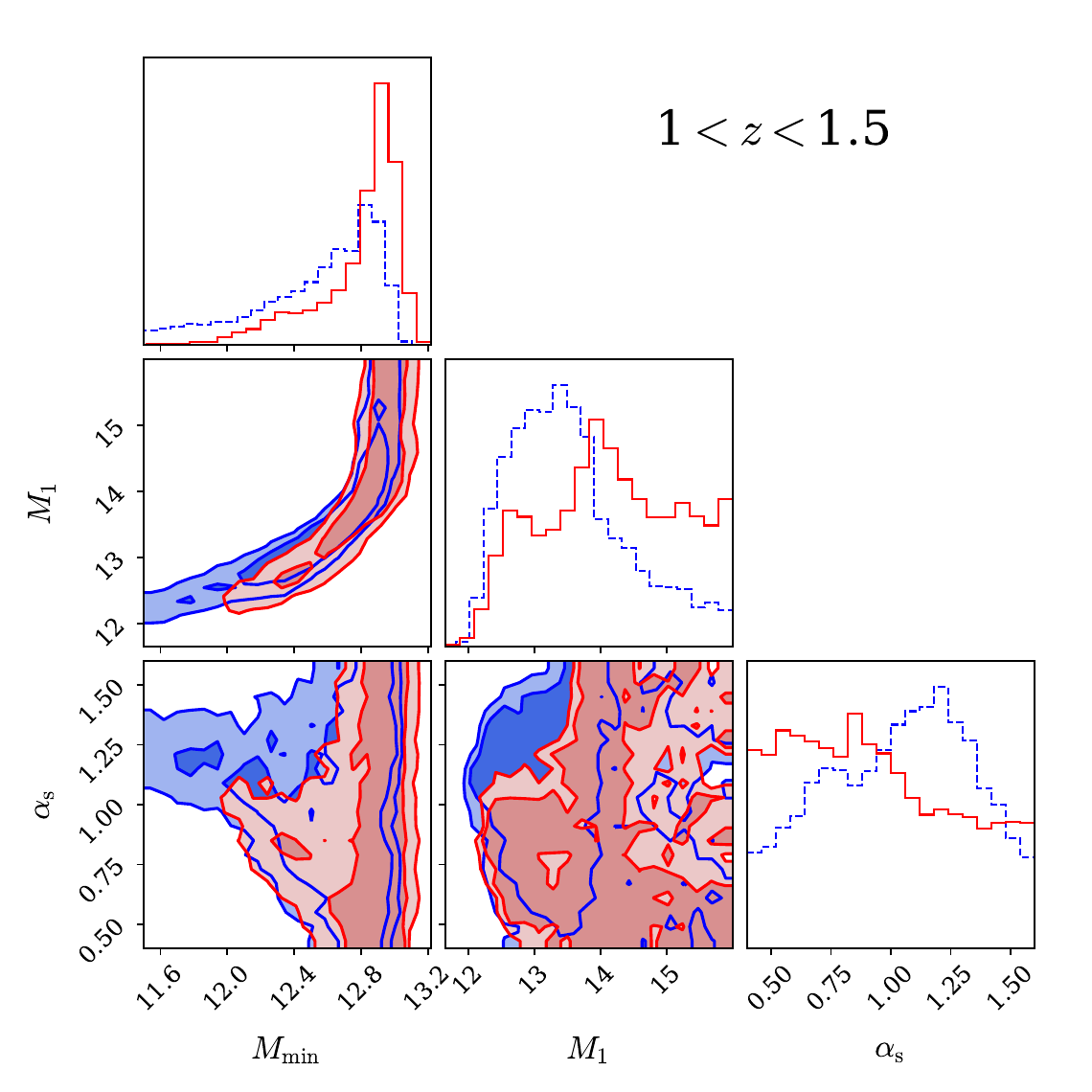}
  \end{subfigure}
  \vspace{1em}
  \begin{subfigure}{\columnwidth}
    \centering
    \includegraphics[width=\columnwidth]{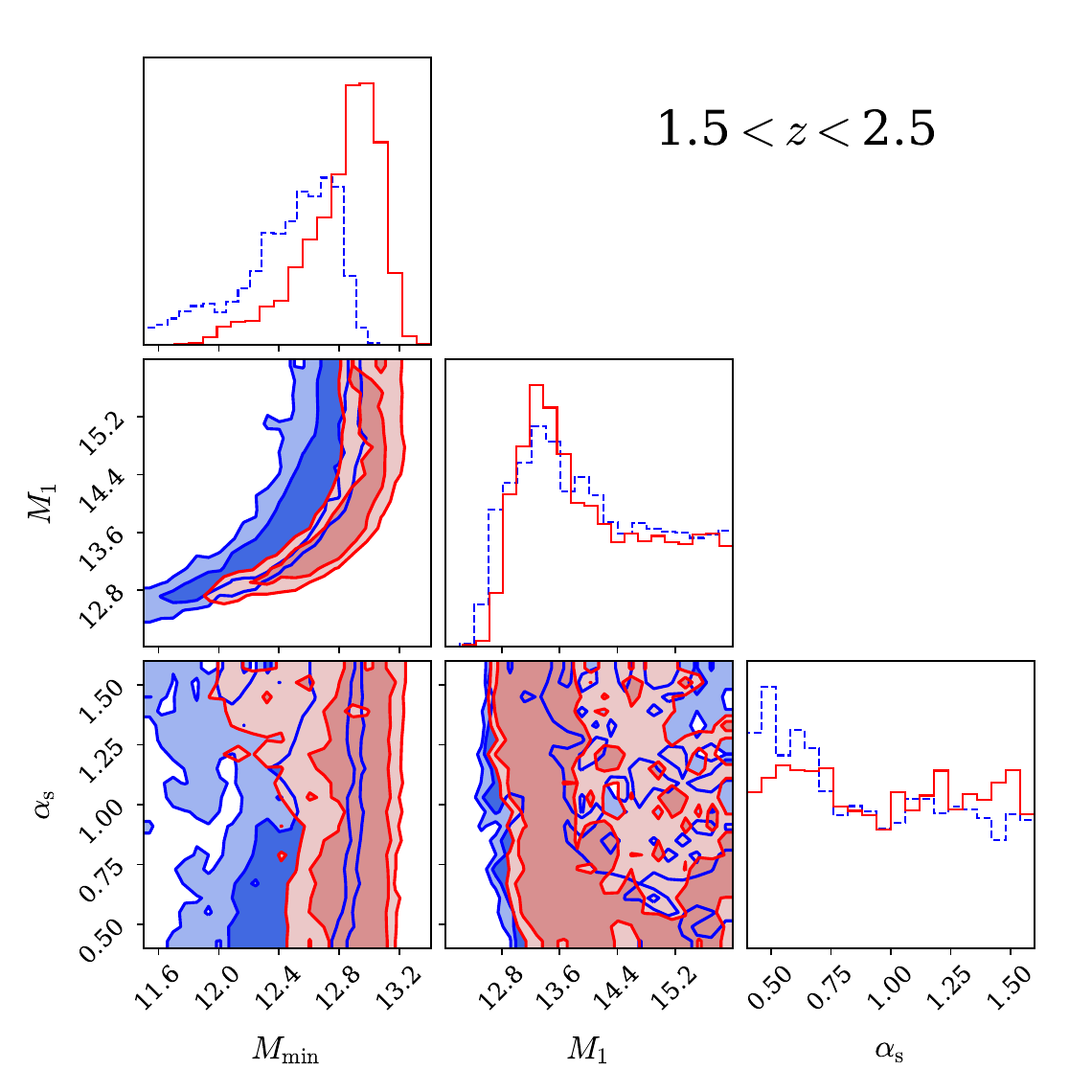}
  \end{subfigure}
  \caption{The posterior distributions and corresponding histograms of the best-fitting HOD model parameters for the AGN (solid red) and matched galaxy (dashed blue) samples for each redshift bin. The contours are the 68\% and 95\% confidence intervals.}
  \label{fig:posteriors_comparison}
\end{figure*}

Figure~\ref{fig:posteriors_comparison} shows the two-dimensional posterior distributions of the best-fitting HOD model parameters for the AGN and matched galaxy samples for each redshift bin.

\bsp	
\label{lastpage}
\end{document}